\documentclass[aps,prx,amsmath,floats,floatfix,twocolumn,
  superscriptaddress,nofootinbib,showpacs]{revtex4-1}

\pdfoutput=1 
\usepackage{diagbox}
\usepackage{mathrsfs}
\usepackage{multirow}
\usepackage{braket}
\usepackage{amsfonts}
\usepackage{amsmath}
\usepackage{amssymb}
\usepackage{amsthm}
\usepackage{bm}
\usepackage{graphicx}
\usepackage{graphics}
\usepackage{latexsym}
\usepackage{rotating}
\usepackage[dvipsnames]{xcolor}
\usepackage{mathrsfs}
\usepackage{microtype}
\usepackage{verbatim}
\usepackage{url}

\usepackage[caption=false]{subfig}
\usepackage[normalem]{ulem}

\usepackage[breaklinks=true]{hyperref}
\hypersetup{
    colorlinks=true,
    linkcolor=NavyBlue,
    filecolor=Magenta,
    urlcolor=NavyBlue,
    citecolor=NavyBlue
}

\usepackage{yfonts}
\usepackage{float}
\usepackage{xspace} 
\usepackage{mathrsfs}
\usepackage{siunitx}
\usepackage{array}

\allowdisplaybreaks[4]

\newcommand{\be}{\begin{equation}}
\newcommand{\ee}{\end{equation}}

\newcommand{\ba}{\begin{align}}
\newcommand{\ea}{\end{align}}

\makeatletter
\newcommand*{\rom}[1]{\expandafter\@slowromancap\romannumeral #1@}
\makeatother

\AtBeginDocument{%
    \newwrite\bibnotes
    \def\bibnotesext{Notes.bib}
    \immediate\openout\bibnotes=\jobname\bibnotesext
    \immediate\write\bibnotes{@CONTROL{REVTEX41Control}}
    \immediate\write\bibnotes{@CONTROL{%
    apsrev41Control,author="08",editor="1",pages="1",title="0",year="1"}}
    \if@filesw
    \immediate\write\@auxout{\string\citation{apsrev41Control}}%
    \fi
}

\allowdisplaybreaks

\begin{document}



\title{Quasinormal mode frequencies and gravitational perturbations of black holes with any subextremal spin in modified gravity through METRICS: \\ The scalar-Gauss-Bonnet gravity case }

\author{Adrian Ka-Wai Chung}
\email{akwchung@illinois.edu}
\affiliation{Illinois Center for Advanced Studies of the Universe \& Department of Physics, University of Illinois Urbana-Champaign, Urbana, Illinois 61801, USA}

\author{Nicol\'as Yunes}
\affiliation{Illinois Center for Advanced Studies of the Universe \& Department of Physics, University of Illinois Urbana-Champaign, Urbana, Illinois 61801, USA}

\date{\today}

\begin{abstract} 

The gravitational waves emitted in the ringdown phase of binary black hole coalescence are a unique probe of strong gravity.
At late times in the ringdown, these waves can be described by quasinormal modes, whose frequencies encode the mass and spin of the remnant, as well as the theory of gravity in play. 
Understanding precisely how deviations from general relativity affect the quasinormal mode frequencies of ringing black holes, however, is extremely challenging, as it requires solving highly coupled and sometimes higher-order partial differential equations.
We here extend a novel approach, \textit{Metric pErTuRbations wIth speCtral methodS} (METRICS), to study the gravitational metric perturbations and the quasinormal mode frequencies of ringing black holes in modified gravity. 
We first derive the asymptotic behavior of gravitational perturbations at the event horizon and spatial infinity for rotating black holes beyond general relativity.
We then extend the eigenvalue perturbation theory approach of METRICS to allow us to compute the leading-order beyond general relativity corrections to the quasinormal-mode frequencies and metric perturbations.
As an example, we apply METRICS to black holes with moderate spins in scalar-Gauss-Bonnet gravity.
Without decoupling or simplifying the linearized field equations in this theory, we compute the leading-order corrections to the quasinormal frequencies of the axial and polar perturbations of the $nlm = 022$, 021 and 033 modes of black holes with dimensionless spin $a \leq 0.85$.
The numerical accuracy of the METRICS frequencies is $\leq 10^{-5}$ when $a \leq 0.6$, $\lesssim 10^{-4}$ when $0.6 < a \leq 0.7 $, and $\lesssim 10^{-2}$ when $0.7 < a \leq 0.85 $ for all modes studied. 
We fit the frequencies with a polynomial in spin, whose coefficients (up to second order in spin) are consistent with those obtained in previous slow-rotating approximations.
These results are the first accurate computations of the gravitational quasinormal-mode frequencies of rapidly rotating black holes (of $a \sim 0.85$) in scalar-Gauss-Bonnet gravity.
\end{abstract}

\maketitle


\section{Introduction}
\label{sec:intro}

The detection of gravitational waves (GWs) allows us to probe the rich physics of the Universe, such as that in play during  
binary black hole (BBH) coalescence~\cite{LIGO_01, LIGO_02, LIGO_03, LIGO_04, LIGO_05, LIGO_06, LIGO_07, LIGO_08, LIGO_09, LIGO_10, LIGO_11, LIGO_GW190412, LIGO_GW190814}. 
Broadly speaking, a BBH coalescence consists of three phases. 
The first phase is the inspiral, during which the BHs orbit around each other with decreasing orbital radius, as a result of GW emission. 
The second phase is the merger, during which the BHs merge to form a single remnant BH. 
The third phase is the ringdown, during which the newly formed, dynamical, and distorted remnant BH relaxes into a stationary configuration by emitting GWs that are characterized by a discrete set of complex quasinormal-mode (QNM) frequencies. 

The validity of general relativity (GR) in the strong-field regime is one aspect about our Universe that GWs can probe. 
GR has passed numerous experimental tests ~\cite{Will2014,Stairs2003,Wex:2020ald,Yunes:2013dva,Will2014,Yagi:2016jml,Berti:2018cxi,Nair:2019iur,Berti:2018vdi, LIGO_07, LIGO_11, LIGOScientific:2021sio, Perkins:2022fhr, Yunes:2013dva} and, thus far, represents our best understanding of spacetime and gravity. 
Nonetheless, GR exhibits theoretical and observational anomalies. 
Theoretically, GR predicts that gravitational collapse inevitably leads to the formation of a spacetime singularity, where GR cannot further describe nature. 
Observationally, GR may require additional parity-violating physics in the early Universe to explain the matter-antimatter asymmetry~\cite{Sakharov:1967dj,Petraki:2013wwa, Gell-Mann:1991kdm, Alexander:2004us}, a cosmological constant that is finely tuned~\cite{Nojiri:2006ri, Tsujikawa:2010zza} to explain the acceleration of the Universe at late times~\cite{late_time_acceleration_01, late_time_acceleration_02}, and cold dark matter to describe the rotation curves of galaxies~\cite{rotation_curve_01, rotation_curve_02}. 
These anomalies have led to the proposal and development of modified gravity theories, each of which amends different aspects of GR in an attempt to remedy these (and other) anomalies. 
To avoid repetition, we refer the reader to \cite{Chung:2023wkd, Chung:2023zdq} for a survey of the motivation of modified gravity theories and relevant GW tests. 

The coalescence of BBHs is a powerful laboratory in which to test these modified theories. 
In many such theories, BHs (or compact objects more generally) still exist, but their spacetime geometry may be different from that of their GR counterpart. 
Moreover, the field equations that describe the dynamics of these compact objects and the behavior of metric perturbations are generically also different from that which arise from the Einstein equations. 
These differences force the GWs emitted during BBH coalescence in these theories to also be different from those in GR. 
By comparing GW observations against GW predictions in GR, we can then test the validity of the latter and probe for GR deviations in the data directly. 

The ringdown phase of BBH coalescence has unique features that make it ideal for testing GR. 
First, the GWs emitted in the ringdown phase usually have the highest frequencies among all three phases. 
Loosely speaking, the GW frequency gives a measure of the relative rate of change in perturbations of spacetime. 
By this standard, the spacetime changes most rapidly during the ringdown phase and might activate effects that could not otherwise be  observed in other phases. 
Second, the QNM frequencies depend only on the properties of the remnant BH and are progenitor-independent, which makes them easy to characterize. 
Third, the QNM response of a BH involves perturbations of the BH geometry near the BH's horizon, the edge of the observable spacetime, where the field strength is the strongest, and where nature is perhaps most likely to be different from GR predictions. 
All these unique features make the ringdown phase a unique probe of the validity of GR in the strong-field regime. 

However, many ringdown tests of gravity have mostly been model agnostic \cite{NHT_Test_01, NHT_Test_02, NHT_Test_03, NHT_Test_04, Isi:2019aib, LIGOScientific:2016src, LIGOScientific:2020tif, LIGOScientific:2021sio}. 
This is understandable because computing the QNM frequencies of a BH in a particular gravity theory has been extremely challenging; this is because of the computational complexity required to solve the dynamical evolution equation of metric perturbations in these theories, which is typically a set of highly coupled partial differential equations. 
These computations are even more challenging when one takes into account BH backgrounds that do not spin slowly, as is the case for BH remnants of BBH coalescence. 

Two approaches have been recently developed to deal with these difficulties. 
The first approach works with curvature perturbations, expressing first the field equations and the Bianchi identities in terms of spinor coefficients, Weyl scalars, and differential operators.
Then, the field equations in terms of these variables are linearized and, if the background spacetime is of Petrov-type D, then the linearized field equations can be simplified into a single master equation, known as the Teukolsky equation, for a certain Teukolsky master function ~\cite{Teukolsky_01_PRL, Teukolsky:1973ha, Press:1973zz, Teukolsky:1974yv}. 
Very recently, this formalism has been extended to modified gravity theories with leading-order deviations from GR \cite{Li:2022pcy,Hussain:2022ins,Wagle:dcsslow,Cano:2023jbk, Cano:2020cao, Cano:2021myl, Cano:2023tmv}, whose spacetime is not Petrov-type D. 
In this modified Teukolsky formalism, one then solves the modified Teukolsky equation, subject to appropriate boundary conditions, to compute the QNM frequencies of the modified Teukolsky function, which can be related to the frequencies of the emitted GWs at future null infinity. 
If one wants to reconstruct the metric perturbations from the Teukolsky master function everywhere outside the perturbed BH, one has to use a Hertz potential in the so-called Chrzanowski-Kogen-Kegeles (CKK) approach~\cite{Chrzanowski:1975wv, Toomani:2021jlo, Whiting_Price_2005, Lousto:2005xu, Ripley:2020xby, Ripley:2022ypi}, which amounts to solving $\sim 10 $ partial differential equations. 

The second approach is the \textit{Metric pErTuRbations wIth speCtral methodS} (METRICS) that we developed in \cite{Chung:2023wkd, Chung:2023zdq}. 
This approach works directly with metric perturbations and perturbations of any other fields that may be present in the theory.
This approach begins by calculating the asymptotic behavior of the field perturbations at the event horizon and at spatial infinity. 
Using this asymptotic behavior, one can then construct an asymptotic factor that regulates the divergent behavior of the field perturbations at the event horizon and spatial infinity. 
The metric perturbations can be power decomposed through the product of the asymptotic factor and a finite but unknown correction function.
Substituting this decomposition of the field perturbations into the field equations, one can then obtain a set of linearized field equations for the unknown functions, which are highly coupled and complicated partial differential equations. 
By performing a spectral expansion of the finite correction functions, and evoking the orthogonality of spectral functions, we transform the linearized field equations into a set of linear, homogeneous, algebraic equations of the spectral coefficients of the finite correction functions. 
The algebraic equations can then be solved as an eigenvalue problem, whose eigenvalues are the QNM frequencies. 
METRICS can accurately compute the QNM frequencies without decoupling and simplifying the linearized field equations into several master equations. 
Since METRICS works directly with metric perturbations, one can swiftly reconstruct the metric perturbations by simply reading the eigenvector corresponding to the QNM frequencies, without undergoing the CKK approach. 

\begin{figure*}[htb!]
\centering  
\subfloat{\includegraphics[width=6cm]{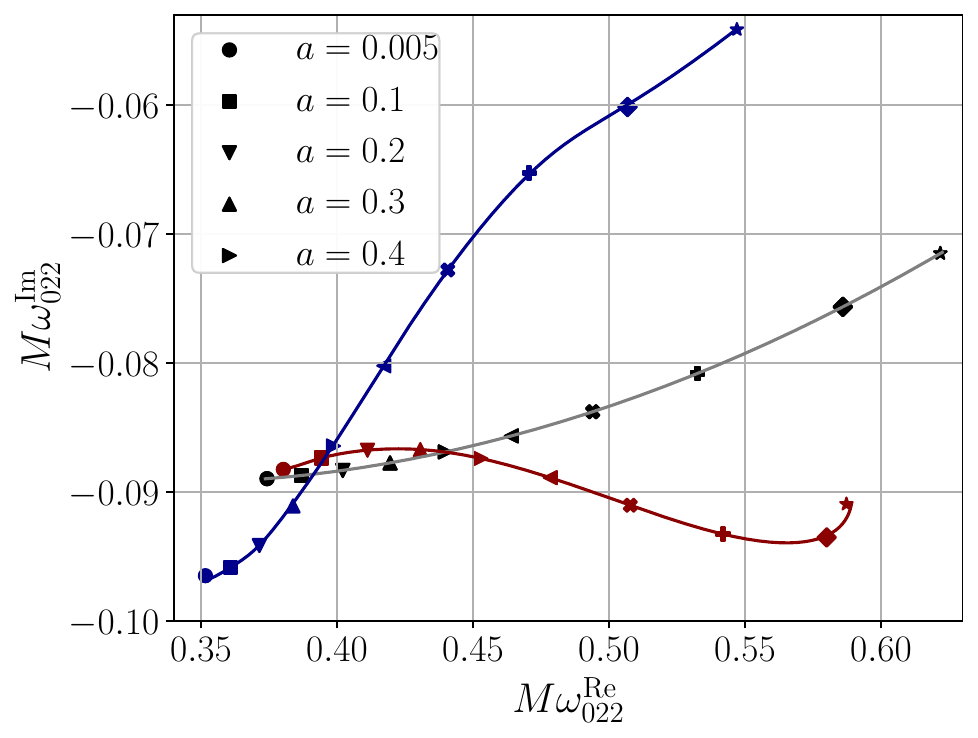}}
\subfloat{\includegraphics[width=6cm]{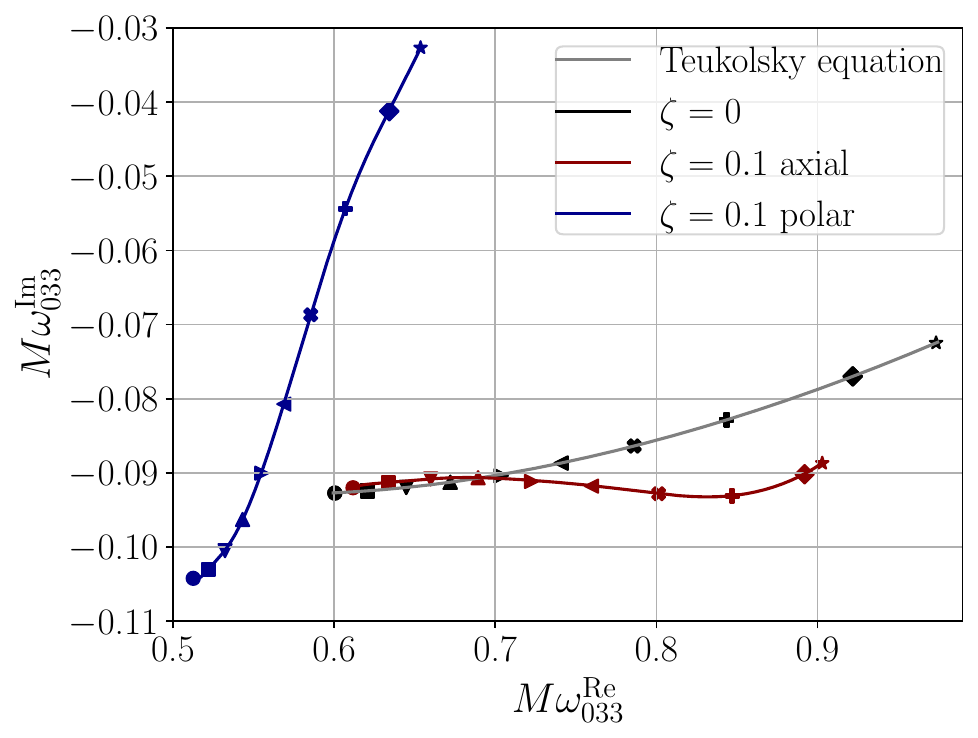}}
\subfloat{\includegraphics[width=6cm]{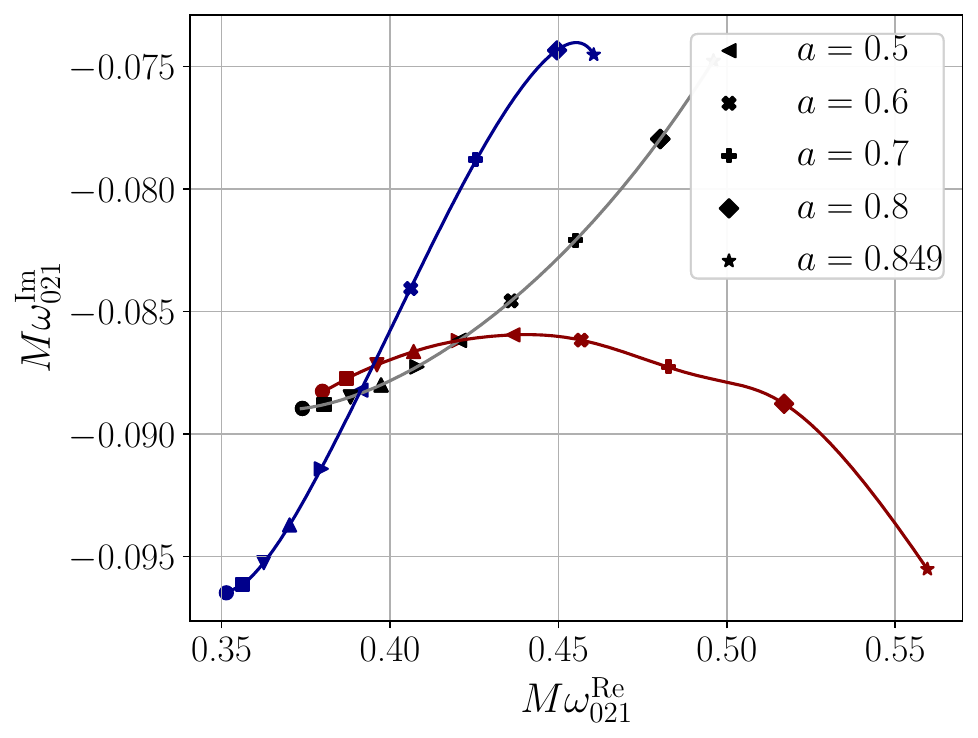}}
\caption{Trajectory in the complex plane of the quasinormal frequencies of the $nlm = 022$, $033$, and $021$ polar (blue) and axial (red) modes in sGB gravity with a dimensionless coupling of $\zeta = 0.1$ for a sequence of black holes with various spins. The symbols correspond to the frequencies computed with METRICS, while the lines correspond to polynomial fits to these frequencies. For comparison, the frequencies in GR are also presented (black symbols and gray lines). Observe how the sGB and GR frequency trajectories are close to each other at small spins, but they separate as the spin increases, specially for the axial modes, which couple more strongly with sGB corrections.    
}
\label{fig:complex_planes}
\end{figure*}

In this paper, we extend METRICS to a wide class of modified gravity theories. 
We begin by reviewing the field equations of this class of theories (Sec.~\ref{sec:FEs}), background metrics that represent BHs in these theories, and the background scalar-field profile (Sec.~\ref{sec:bkgd}).
Then, we specify the perturbation ansatz of the metric tensor and the scalar field in the Regge-Wheeler gauge (Sec.~\ref{sec:metpert}), a gauge that is known to exist in many modified gravity theories.
We then derive the asymptotic behavior of the field perturbations at the event horizon and spatial infinity for BHs in modified gravity (Sec.~\ref{sec:ansatz}). 
The asymptotic behavior at the horizon, derived using the properties of the Killing vector and the horizon, is the first important result of our paper, as it is valid for perturbations that fall through the horizon of a general BH following a null geodesic, regardless of the details of the gravity theory. 
The asymptotic behavior at the horizon and at spatial infinity allows us to construct the asymptotic factor needed to apply METRICS. 

We then continue to apply METRICS to this class of modified gravity theories by product decomposing the field perturbations as a product of the asymptotic factor and a finite correction function. 
We substitute this form of perturbations into the field equations to derive the linearized field equations for the finite correction functions (Sec.~\ref{sec:Linearized_EFEs}). 
By expanding the finite correction functions into a product of Chebyshev polynomials along a compactified spatial coordinate and the associated Legendre polynomial along the azimuthal angle, we obtain a set of linear homogeneous algebraic equations of the spectral coefficients (Sec.~\ref{sec:Conversion}).
Since existing tests of GR indicate that deviations from GR must be small, and since the background metric of BHs in this class of modified gravity theories is constructed only up to leading order in deformations from GR,  we develop an eigenvalue perturbation theory for METRICS (Sec.~\ref{sec:eigenvalue_pert}). This eigenvalue perturbation scheme is the second key result of this paper, as it allows us to accurately and consistently compute the leading-order modifications to the QNM frequencies and the metric and scalar-field perturbations. 

To exemplify the applications of METRICS to a modified gravity theory, we use METRICS to compute the QNM frequencies of rotating BHs in scalar-Gauss-Bonnet (sGB) gravity with zero potential and a shift-symmetric coupling (Sec.~\ref{sec:Application_sGB}). 
We focus on sGB gravity for the following reasons. 
First, it is well-motivated as the low-energy limit of certain string theories \cite{Maeda:2009uy, Moura:2006pz}. 
Moreover, it has recently been shown that sGB gravity is well-posed \cite{AresteSalo:2022hua, Kovacs:2020pns} for small GR deformations, laying the foundation for the first numerical simulations of BBH coalescence in this theory \cite{East:2020hgw, AresteSalo:2023mmd, AresteSalo:2023hcp}. 
Computing the QNM frequencies of BHs in sGB gravity is critical to establish a waveform model in this theory, which ultimately leads to an accurate model-specific test with GW observations. 
Second, sGB gravity belongs to a wide class of modified gravity theories, known as quadratic gravity theories. 
The Lagrangian density of quadratic gravity theories consists of quadratic products of the curvature tensor, which can be viewed as the next leading-order correction to the Lagrangian density if one generalizes it into an infinite power series of the curvature tensor \cite{Cano_Ruiperez_2019}. 
By focusing on sGB gravity, we can learn lessons that will be needed before we apply METRICS to more complicated theories. 

Specifically, we apply METRICS to compute the leading-order modifications to the complex frequency of the $nlm=022$, 033, and 021 modes of rotating BHs of dimensionless spin $a \lesssim 0.85$ in sGB gravity (Sec.~\ref{sec:Rotating_BHs_results}). 
Figure~\ref{fig:complex_planes} presents a summary of our results through the trajectories of the frequencies in the complex plane for a sequence of black holes with different dimensionless spins. We immediately observe the breaking of isospectrality, i.e.~the QNM frequencies of the axial and polar perturbations are different in this theory, as expected~\cite{Manfredi:2017xcv, Chen:2021cts, QNM_dCS_01, QNM_dCS_02, QNM_dCS_03, QNM_dCS_04, QNM_EdGB_01, QNM_EdGB_02, QNM_EdGB_03}. 
We observe also that the sGB frequencies become more and more different from the GR frequencies as the spin increases, specially for the axial modes. 
For all the QNMs we study, the numerical uncertainty of the leading-order correction to the QNM frequencies is 
is smaller than $10^{-5}$ for $a \leq 0.6$, smaller than $10^{-4} $ for $0.6< a \leq 0.7$, and smaller than $10^{-2} $ for $0.7< a \leq 0.85$ (see also Fig.~\ref{fig:delta}). 
We construct a polynomial fitting function to the QNM frequencies we calculate numerically (also shown in Fig.~\ref{fig:complex_planes}), which allows for rapid and accurate evaluation.
The numerical values and the fitting functions are the third key results of this paper. 
These fitting functions can now be used to analyze the detected ringdown signals and test sGB gravity. 
Apart from the QNM frequencies, we also examine the properties of the leading-order modifications to the metric and scalar-field perturbations in sGB gravity, leading to deeper insight into metric perturbations in sGB gravity (Secs.~\ref{sec:PD} and~\ref{sec:SF_role}). 

This work is the first to compute the QNM frequencies of BHs that are not slowly rotating and possess scalar hair to the accuracy described above. 
We note that the QNM frequencies of a rotating BH of $a=0.7$ in sGB gravity were estimated in \cite{Pierini:2022eim} using a slow-spin expansion valid only up to second order in dimensionless spin. 
In what follows, we include sGB corrections to the metric up to 40 orders in dimensionless spin, depending on the dimensionless spin of the BH, to ensure that the effects of the higher spin-order terms in the metric corrections are accurately accounted for when computing the QNM frequencies. 
Also, in \cite{Pierini:2022eim}, Pad\'e resummation is used to help improve the accuracy of the QNM frequencies for $a \geq 0.7$, whereas here we compute the QNM frequencies at $a=0.7$ without using any resummation or extrapolation technique. 
Moreover, QNM frequencies of the axial perturbations are not explicitly computed in \cite{Pierini:2022eim}, whereas, in this work, we also compute the axial-mode frequencies for dimensionless spin $a \leq 0.7$. 
Apart from the semianalytical studies described above, the QNM response of a remnant BH formed by binary BH coalescence in sGB gravity were also simulated in \cite{Okounkova:2019dfo, Okounkova:2019zjf, Okounkova:2020rqw} and beyond an order-reduction scheme \cite{AresteSalo:2022hua, AresteSalo:2023mmd}. 
In principle, one could also extract the QNM frequencies of BHs in sGB gravity from these simulations, but the accuracy of this extraction would be limited by numerical errors in the simulations.
Hence, our work greatly extends the important work of \cite{Pierini:2022eim,Okounkova:2019dfo, Okounkova:2019zjf, Okounkova:2020rqw,AresteSalo:2022hua, AresteSalo:2023mmd}, allowing us to obtain the first accurate computation of the QNM frequencies of rapidly spinning BHs in sGB gravity. 
Our work also extends other important work that focused on QNM frequencies of rapidly rotating BHs using the modified Teukolsky formalism \cite{Cano:2023jbk, Cano:2020cao, Cano:2021myl, Cano:2023tmv} in modified gravity theories without BH scalar hair.

The remainder of this paper describes the computational details of the work summarized above.
In Sec. \ref{sec:MGTs}, we review the field equations (Sec.~\ref{sec:FEs}) and the BH background metric (Sec.~\ref{sec:bkgd}) of the wide class of modified gravity theories we consider. 
We also describe the ansatz of the gravitational and scalar perturbations to BHs in these gravity theories in Sec.~\ref{sec:metpert}. 
In Sec. \ref{sec:ansatz}, we derive the asymptotic behavior of the gravitational and scalar perturbations of modified BHs at spatial infinity and the event horizon. 
The highlight of this section is the derivation of the asymptotic behavior of the perturbations at the event horizon through two separate arguments. 
The asymptotic behavior at the horizon is the first key result of our paper because it is valid for a general BH regardless of the nature of gravity, provided that the BH admits a stationary Killing horizon, and the perturbations follow null geodesics. 
In Sec.~\ref{sec:METRICS}, we review the METRICS approach, which we developed in \cite{Chung:2023wkd, Chung:2023zdq}, and apply it to transform the linearized field equations in sGB gravity into a set of linear homogeneous algebraic equations.
In Sec.~\ref{sec:eigenvalue_pert}, we develop an eigenvalue perturbation theory to solve the algebraic equations for the leading-order modification to the QNM frequencies due to the activation of modifications to gravity. 
The eigenvalue perturbation theory is the second important result of this paper in its own right. 
In Sec.~\ref{sec:Application_sGB}, using METRICS and the eigenvalue perturbation theory, we compute the QNM frequencies of rotating BHs in sGB gravity for dimensionless spin up to 0.85. 
The QNM frequencies of these modes are presented in Table~\ref{tab:omega_1_022},~\ref{tab:omega_1_033} and~\ref{tab:omega_1_021} of Appendix~\ref{sec:Appendix_B}. 
The coefficients (and their uncertainty) of the fitting polynomials of the frequencies are presented in Table~\ref{tab:poly_fit_coeffs} (and~\ref{tab:poly_fit_coeffs_uncertainty}) of Appendix~\ref{sec:Appendix_B}.  

Henceforth, following \cite{Chung:2023zdq, Chung:2023wkd}, we assume the following conventions: 
$x^{\mu} = (x^0, x^1, x^2, x^3) = (t, r, \chi, \phi)$, where $\chi = \cos \theta$ and $\theta$ is the polar angle;
the signature of the metric tensor is $(-, +, +, +)$;
gravitational QNMs are labeled by $n l m$ or $(n, l, m)$, where $n$ is the principal-mode number, $l$ is the azimuthal-mode number \footnote{Note that $l$ is, in general, different from $\ell$, the degree of the associated Legendre polynomials in the product decomposition of the metric perturbation functions. Although these numbers are the same for a Schwarzschild BH background, this is not so for a Kerr BH background.} and $m$ is the magnetic-mode number of the QNM;
Greek letters in index lists stand for spacetime coordinates;
unless otherwise specified, we used geometric units in which $c=G=1$; 
for the convenience of the reader, we have presented a list of all definitions and symbols in Appendix~\ref{sec:Appendix_A}.

\section{Modified gravity theories}
\label{sec:MGTs}

In this section, we define the class of modified gravity theories we will consider. We begin by describing the field equations and their solution for stationary and axisymmetric black hole spacetimes. We then discuss how we perturb these backgrounds.  
\subsection{The field equations}
\label{sec:FEs}

The Lagrangian density of a wide class of modified gravity theories can be written as \cite{Yagi:2015oca, Nair:2019iur, Perkins:2021mhb}
\begin{equation}\label{eq:Lagrangian}
16 \pi \mathscr{L} = R - \frac{1}{2} \nabla_{\mu} \Phi \nabla^{\mu} \Phi- V(\Phi) + \alpha f(\Phi) \mathscr{Q}, 
\end{equation}
where $\Phi$ is a scalar field to which the BH in the modified gravity couples, $V(\Phi)$ is the potential of $\Phi$, $\alpha$ is a coupling constant, which characterizes the strength of the modifications to gravity and has dimensions of length squared in geometric units, and $f(\Phi)$ is a function of $\Phi$ only. 
In the Lagrangian, $\mathscr{Q}$ is a scalar constructed from the curvature tensor. 
For example, in sGB gravity \cite{Ripley:2019irj, Cano_Ruiperez_2019}, $\mathscr{Q}$ is given by 
\begin{equation}
\mathscr{Q}_{\rm sGB} = \mathscr{G} = R^2 - 4 R_{\alpha \beta} R^{\alpha \beta} + R_{\alpha \beta \gamma \delta} R^{\alpha \beta \gamma \delta}, 
\end{equation}
and in dynamical Chern-Simons (dCS) gravity \cite{dCS_01, dCS_02, dCS_03}, 
\begin{equation}
\mathscr{Q}_{\rm dCS} = \mathscr{P} = \frac{1}{4} R_{\nu \mu \rho \sigma}{ }^* R^{\mu \nu \rho \sigma}, 
\end{equation}
where ${}^* R^{\mu \nu \rho \sigma}$ is the dual Riemann tensor
\begin{equation}
{}^* R^{\mu \nu \rho \sigma} = \frac{1}{2} \epsilon^{\rho \sigma \alpha \beta} R^{\mu \nu} {}_{\alpha \beta}, 
\end{equation}
and $\epsilon^{\rho \sigma \alpha \beta}$ is the Levi-Civita tensor, defined as 
\begin{equation}
\epsilon^{\rho \sigma \alpha \beta} = \frac{1}{\sqrt{-g}} \tilde{\epsilon}^{\rho \sigma \alpha \beta}, 
\end{equation}
$g$ is the determinant of $g_{\mu \nu}$, and $\tilde{\epsilon}^{\rho \sigma \alpha \beta}$ is the totally asymmetric Levi-Civita symbol. 

Through this work, to exemplify the application of METRICS to study spinning BHs in modified gravity, we focus on the cases of zero potential and a shift-symmetric coupling function, 
\begin{equation}\label{eq:shift-symmetric}
\begin{split}
V(\Phi) &= 0, \qquad
f(\Phi) = \Phi. 
\end{split}
\end{equation}
We consider a shift-symmetric coupling function because this is the small-coupling approximation (or limit, i.e. when $\alpha \ll 1 $ such that the background $\Phi$ is only slightly displaced from its ground-state value $\Phi_0$) of a general coupling function when $\mathscr{Q}$ is a topological invariant, as in the sGB and dCS cases.  
To see the correspondence, we can Taylor expand a general $f(\Phi)$ around its ground-state potential $\Phi_0$,
\begin{equation}
f(\Phi) = f(\Phi_0) + f'(\Phi_0) (\Phi - \Phi_0) + ... ~ .  
\end{equation}
The constant term $f(\Phi_0) - f'(\Phi_0) \Phi_0 $ has no physical effects when $\mathscr{Q}$ is a topological invariant; this is because $f(\Phi_0) - f'(\Phi_0) \Phi_0 $ pulls out of the integral and $\mathscr{Q}$ has zero variation, so the field equations are not modified. The coefficient $f'(\Phi_0)$ can just be absorbed into the definition of the coupling constant, and one obtains identical field equations as if one had chosen a shift-symmetric coupling function. 

Using the Lagrangian and the least action principle, one can derive the field equations of the modified gravity theory in vacuum, which can be schematically expressed as 
\begin{align}
\label{eq:field_eqs}
& R_{\mu}{}^{\nu} + \zeta \left( \mathscr{A}_{\mu}{}^{\nu} - T_{\mu}{}^{\nu} \right) = 0, \\
\label{eq:field_eqs_scalarfield}
& \square \vartheta + \mathscr{A}_{\vartheta} = 0, 
\end{align}
where $\zeta$ is a dimensionless coupling parameter, 
\begin{equation}\label{eq:zeta}
\zeta = \frac{\alpha^2}{M^4}, 
\end{equation}
where $M$ is the mass of the background BH we consider in this paper. In the small-coupling approximation, $\zeta \ll 1 $ because of existing tests of GR.  
The quantities $\mathscr{A}_{\mu}{}^{\nu}$ and $ \mathscr{A}_{\vartheta}$ are a tensor and a scalar respectively, which represent modifications to GR, 
while $\vartheta$ is a rescaled scalar field, such that $\Phi = \alpha \vartheta$, and 
\begin{equation}\label{eq:Tmunu}
\begin{split}
T_{\mu}{}^{\nu} & \equiv \frac{1}{2}\left(\nabla_{\mu} \vartheta \right)\left(\nabla^{\nu} \vartheta \right) + \frac{1}{2 \zeta} \delta_{\mu} {}^{\nu} V(\Phi), \\
\end{split}
\end{equation}
is the trace-reversed energy-momentum tensor of the rescaled scalar field. 
In general, $\mathscr{A}_{\mu}{}^{\nu}$ satisfies the contracted Bianchi identity $\nabla_{\nu} \mathscr{A}_{\mu}{}^{\nu} = 0 $, and may involve higher-than-second-order derivatives of the metric tensor and derivatives of $\vartheta$.
For example, in sGB gravity \cite{Ripley:2019irj, Cano_Ruiperez_2019}, 
\begin{align}\label{eq:Amunu_sGB}
\mathscr{A}_{\mu}{}^{\nu} & \equiv \delta^{\nu \sigma \alpha \beta}_{\mu \lambda \gamma \delta} R^{\gamma \delta}{}_{\alpha \beta}\nabla^{\lambda} \nabla_{\sigma} \vartheta  -\frac{1}{2} \delta_{\mu}{}^{\nu} \delta^{\eta \sigma \alpha \beta}_{\eta \lambda \gamma \delta} R^{\gamma \delta}{}_{\alpha \beta}\nabla^{\lambda} \nabla_{\sigma} \vartheta, 
\end{align}
where $\delta^{\nu \sigma \alpha \beta}_{\mu \lambda \gamma \delta}$ is the generalized Kronecker $\delta$, defined as 
\begin{equation}
\delta_{\mu_1 \mu_2 \mu_3 \mu_4}^{\nu_1 \nu_2 \nu_3 \nu_4} = \det 
\begin{pmatrix}
\delta_{\mu_1}^{\nu_1} & \delta_{\mu_2}^{\nu_1} & \delta_{\mu_3}^{\nu_1} & \delta_{\mu_4}^{\nu_1} \\
\delta_{\mu_1}^{\nu_2} & \delta_{\mu_2}^{\nu_2} & \delta_{\mu_3}^{\nu_2} & \delta_{\mu_4}^{\nu_2} \\
\delta_{\mu_1}^{\nu_3} & \delta_{\mu_2}^{\nu_3} & \delta_{\mu_3}^{\nu_3} & \delta_{\mu_4}^{\nu_3} \\
\delta_{\mu_1}^{\nu_4} & \delta_{\mu_2}^{\nu_4} & \delta_{\mu_3}^{\nu_4} & \delta_{\mu_4}^{\nu_4} \\
\end{pmatrix}, 
\end{equation}
while for dCS gravity \cite{dCS_01, dCS_02, dCS_03, QNM_dCS_01, QNM_dCS_02, QNM_dCS_03, QNM_dCS_04}, 
\begin{align}\label{eq:Amunu_dCS}
\mathscr{A}^{\mu \nu} & \equiv \left(\nabla_\sigma \vartheta \right) \epsilon^{\sigma \delta \alpha(\mu|} \nabla_\alpha R^{|\nu)}{}_{\delta} + \left(\nabla_\sigma \nabla_\delta \vartheta\right) {}^* R^{\delta (\mu \nu) \sigma}. 
\end{align}
$\mathscr{A}_{\vartheta}$ usually involves the product of the curvature tensor. 
For example, in sGB gravity, $\mathscr{A}_{\vartheta} = \mathscr{G}$, whereas in dCS gravity, $\mathscr{A}_{\vartheta} = \mathscr{P}$. 

\subsection{Background spacetime and scalar field}
\label{sec:bkgd}

Equation~\eqref{eq:field_eqs} allows us to construct a rotating BH spacetime (i.e.~a stationary, axisymmetric and vacuum spacetime) as an expansion in powers of the dimensionless spin $a$ in a given modified gravity theory. 
Since existing constraints indicate that $\zeta \ll 1 $~\cite{Nair:2019iur, Perkins:2021mhb}, we here focus on solutions within the small-coupling approximation.
We observe that, whereas Eq.~\eqref{eq:field_eqs} \textit{explicitly} depends on $\zeta$, Eq.~\eqref{eq:field_eqs_scalarfield} does not. 
Thus, we can solve Eq.~\eqref{eq:field_eqs} in the following iterative manner~\cite{Yunes:2009hc, Yagi:2012ya, Pani:2009wy, Ayzenberg:2014aka, Maselli:2015tta, McNees:2015srl, Cano_Ruiperez_2019}.
We first use the GR Kerr metric to compute $\mathscr{A}_{\vartheta}$, with which we solve the second equation of Eq.~\eqref{eq:field_eqs} for $\vartheta$. 
Since it is very difficult to solve the scalar-field equation exactly, one can instead solve for $\vartheta$ as a power series of $a$. Doing so, one finds that the solution takes the form~\cite{Yunes:2009hc,Yagi:2012ya,Cano_Ruiperez_2019} 
\begin{equation}\label{eq:vartheta_fns}
\vartheta (r, \chi) = \sum_{k=0}^{} \sum_{p=0}^{N_{r}(K)} \sum_{q=0}^{N_{\chi}(K)} \vartheta_{i, k, p, q} \frac{a^{k} \chi^{q} }{r^p}, 
\end{equation}
where we recall that $\chi = \cos{\theta}$ and where ${\vartheta}_{i, k, p, q}$ are constant, while $N_{r}(K)$ and $N_{\chi}(K)$ are also constants that depend on the order in $a$ to which one expands. 

Then, we use $\vartheta$ in the form of Eq.~\eqref{eq:vartheta_fns} to solve the first equation of Eq.~\eqref{eq:field_eqs} for the background metric.
Specifically, we solve the metric in the Boyer-Lindquist coordinates of the following form  \cite{Cano_Ruiperez_2019, Cano:2023jbk}, 
\begin{equation}\label{eq:metric}
\begin{split}
ds^2 &= g_{\mu \nu}^{(0)} dx^{\mu}  dx^{\nu} \\
& = - \left( 1-\frac{2 M r}{\Sigma} - \zeta H_1(r, \chi)\right) dt^2 \\
& - \left[ 1 + \zeta H_2(r, \chi) \right] \frac{4 M^2 a r}{\Sigma} (1 - \chi^2) d \phi dt \\
& \quad + \left[ 1 + \zeta H_3(r, \chi) \right] \left( \frac{\Sigma}{\Delta} dr^2 + \frac{\Sigma}{1 - \chi^2} d \chi^2 \right) \\
& \quad + \left[ 1 + \zeta H_4(r, \chi) \right](1-\chi^2) \\
& \quad \quad \times \left[r^{2} + M^2 a^{2}+\frac{2 M^3 a^{2} r}{\Sigma} (1 - \chi^2)\right] d\phi^2, \\
\end{split}
\end{equation}
where
\begin{equation}\label{eq:metric_quantities}
\begin{split}
\Sigma &= r^2 + M^2 a^2 \chi^2, \\
\Delta &= (r-r_+)(r-r_-), \\
r_{\pm} &= M(1 \pm b), \\
b &= \sqrt{1-a^2},
\end{split}
\end{equation}
and where $r_{\pm}$ are the GR radial location of the inner and outer event horizons respectively. The parameters $M$ and $0\leq a<1$\footnote{Note that the extremal limit for rotating BHs in this theory will be corrected from that in GR, as pointed out by \cite{Cano:2023jbk}. However, since $\zeta \ll 1$, we expect these corrections to be $\mathcal{O}(\zeta)$. Therefore, the modified extremal limit is not strictly relevant for parameter estimation of astrophysical ringdown signals. }
are the observable BH mass, which will be set to unity when we numerically compute the QNM frequencies (see Sec.~\ref{sec:Application_sGB}), and the observable dimensionless spin parameter, related to the observable spin angular momentum through $J = M^2 a$.
We denote the Kerr-BH metric in GR by $g^{\rm (GR)}_{\mu \nu} = g^{\rm (0)}_{\mu \nu}( \zeta = 0)$, and thus, $H_{i=1, ..., 4}(r, \chi)$ are the corrections to the Kerr background due to modified gravity, which can similarly be solved iteratively as series in $a$~\cite{Cano_Ruiperez_2019, Cano:2023jbk}, with the boundary condition that $g_{\mu \nu}^{(0)}$ is asymptotically flat at spatial infinity (i.e. $r \rightarrow + \infty$). Doing so, one finds that the solution takes the form
\begin{equation}\label{eq:H_fns}
H_i (r, \chi) = \sum_{k=0}^{K} \sum_{p=0}^{N_{r}'(K)} \sum_{q=0}^{N_{\chi}'(K)} h_{i, k, p, q} \frac{a^{k} \chi^{q} }{r^p}, 
\end{equation}
where $h_{i, k, p, q}$ are again all numbers, while  $N_{r}'(k)$ and $N_{\chi}'(k)$ 
are also numbers that depend on the truncation order of the series in $a$ (that is, on $K$). 
Note that within this parametrization, $H_{i=1, ..., 4}(r, \chi)$ do not explicitly depend on $\zeta$ or $\alpha$. 
One crucial advantage of this form of metric is that the radial coordinate of the outer and inner event horizons, corresponding to the radial roots of 
\begin{equation}
g^{rr} = \frac{1}{1+\zeta H_3 (r, \chi)}\frac{\Delta}{\Sigma} = 0, 
\end{equation}
are not changed by the modifications to gravity, 
\begin{equation}
r_{\pm} = M\left( 1 \pm b \right). 
\end{equation}

Two quantities related to the BH background that we will often use in our work are the horizon angular velocity, $\Omega_{\rm H}$, which can be computed using the $t \phi$ and $\phi \phi$ components of the metric via
\begin{equation}
\Omega_{\rm H} = - \frac{g_{t\phi}}{g_{\phi \phi}} \Big|_{r = r_+}, 
\end{equation}
and the surface gravity of the horizon, $\kappa$, which can be computed via 
\begin{equation}
\kappa =\lim_{r \rightarrow r_+} \frac{\partial_r \xi^2}{\sqrt{\xi^2/g^{rr}}},  
\end{equation}
where $\xi^2$ is the modulus square of the Killing vector of the BH horizon \cite{Cano_Ruiperez_2019}, which will be defined in more detail in Sec.~\ref{sec:null_geodesic_approach}, but for now we define via
\begin{equation}
\begin{split}
\lim_{r \rightarrow r_+} \xi^2 & = \left[g_{tt}+2\Omega_H g_{t \phi} + \Omega_{H}^2 g_{\phi \phi}\right]_{r=r_+} \\
& = \left(g_{tt} - \frac{g_{t \phi}^2}{g_{\phi \phi}}\right)_{r=r_+}\,. 
\end{split}
\end{equation}
Both can be written as 
\begin{equation}\label{eq:Omega_H_and_kappa}
\begin{split}
\Omega_{\rm H} & = \Omega^{(0)}_{\rm H} + \zeta \Omega^{(1)}_{\rm H}, \\
\kappa & = \kappa^{(0)} + \zeta \kappa^{(1)}, 
\end{split}
\end{equation}
where $\Omega^{(0)}_{\rm H}$ and $\kappa^{(0)}$ are, respectively, the GR horizon angular velocity and surface gravity 
\begin{equation}
\begin{split}
\Omega^{(0)}_{\rm H} & = \frac{a}{2Mb},  \qquad
\kappa^{(0)} = \frac{b}{2M(1+b)}, 
\end{split}
\end{equation}
and $\Omega^{(1)}_{\rm H}$ and $\kappa^{(1)}$ are the leading order in $\zeta$ correction to the horizon angular velocity and surface gravity respectively, each of which is a series in $a$ \cite{Cano_Ruiperez_2019}. 
For the reference of the reader, the explicit power series of $\Omega^{(1)}_{\rm H}$, $\kappa^{(1)}$, $\vartheta$ and $H_{i=1,...,4}$ as a power series up to the 40th order of $a$ are given in a Mathematica notebook as Supplemental Material. 

\subsection{Perturbations of fields}
\label{sec:metpert}

We now consider both metric and scalar perturbations of a BH in modified gravity theory. 
The linear metric perturbations can be written as 
\be \label{eq:metpert}
g_{\mu\nu} = g_{\mu\nu}^{(0)} + \epsilon \; h_{\mu\nu} \,,
\ee
where $g_{\mu\nu}^{(0)}$ is the background metric of Eq.~\eqref{eq:metric}, $h_{\mu\nu}$ is the metric perturbations, and $\epsilon$ is a bookkeeping parameter for the perturbations. 
The metric perturbations $h_{\mu\nu}$ can be written as 
\begin{widetext}
\begin{equation}\label{eq:metpert}
h_{\mu\nu} (t,r,\chi,\phi) = e^{im\phi-i\omega t}
\begin{pmatrix}
    h_1(r,\chi) & h_2(r,\chi) & -im(1-\chi^2)^{-1} h_5(r,\chi) &  (1-\chi^2) \partial_\chi h_5(r,\chi) \\
    * & h_3(r,\chi) & -im(1-\chi^2)^{-1} h_6(r,\chi) &  (1-\chi^2) \partial_\chi h_6(r,\chi) \\
	* & * & \left( 1 - \chi^2 \right)^{-1} h_4(r,\chi) & 0  \\
	* & * & * & \left( 1 - \chi^2 \right) h_4(r,\chi).  
\end{pmatrix}
\end{equation}
\end{widetext}
Here we have made use of the Regge-Wheeler gauge~\cite{Regge:PhysRev.108.1063, Berti_02}, a gauge that many modified gravity theories have sufficient residual degrees of freedom to enforce \cite{QNM_dCS_01, QNM_dCS_02, QNM_dCS_03, QNM_dCS_04, QNM_EdGB_01, QNM_EdGB_02, QNM_EdGB_03}.
Apart from the metric perturbations, we also need to consider perturbations of the scalar field to which the BH couples, 
\begin{equation}\label{eq:scalar_pert}
\vartheta(r, \chi) = \vartheta^{(0)}(r, \chi)  + \epsilon e^{im\phi-i\omega t} h_7 (r, \chi), 
\end{equation}
where $\vartheta^{(0)}(r, \chi)$ is the unperturbed rescaled scalar field (which is just Eq.~\eqref{eq:vartheta_fns}). 

Let us briefly discuss some of the variables introduced in this decomposition of the metric and scalar perturbations. 
The $\omega$ in $e^{-i \omega t}$ that multiplies $h_{i=1,...,7}$ in Eqs.~\eqref{eq:metpert} and~\eqref{eq:scalar_pert} is the same quantity, but this does not mean that we are imposing isospectrality, i.e.~that axial, polar, and scalar perturbations have the same frequency. 
As we shall see in the later sections, we will develop an algorithm to isolate axial modes, the polar modes and the scalar modes. That is, we will develop a method (essentially by picking the right ``parity-isolating'' initial guesses) that will ensure that only the $h_i$ of a given parity or scalar/metric type are turned on, while the others are suppressed. Doing so ensures that $\omega$ we solve for corresponds to the parity or type we have intended to isolate.  
Thus, Eqs.~\eqref{eq:metpert} and~\eqref{eq:vartheta_fns} are still general enough for the computation of the QNM frequencies when isospectrality is not preserved, even though the frequency of all types of perturbations is labeled by the same $\omega$. 

Before proceeding, let us discuss the parity content of Eq.~\eqref{eq:metpert}. 
The axial (also known as ``odd'' or ``magnetic'') metric perturbations are defined to satisfy \cite{Li:2023ulk, Nichols:2012jn}
\begin{equation}
\hat{P} \left[h^{\rm (A)}_{\mu \nu}\right] = -(-1)^{l} h^{\rm (A)}_{\mu \nu}. 
\end{equation}
Here $\hat{P} $ is the parity reversal operator [$\hat{P}f(\chi,\phi) = f(-\chi, \pi + \phi)$]
and $l$ is the azimuthal mode number of the QNM. 
The polar (also known as ``even'' and ``electric'') metric perturbations are defined to satisfy 
\begin{equation}
\hat{P} \left[h^{\rm (P)}_{\mu \nu}\right] = (-1)^{l} h^{\rm (P)}_{\mu \nu}. 
\end{equation}
These two conditions do not, in general, imply that $h_{i=1, ..., 4}$ are polar and $h_{i=5, 6}$ are axial. 
This statement is only true when $a=0$.
To illustrate this important observation, let us use the associated Legendre polynomials as an example angular spectral basis for the angular representation of the metric perturbations. These polynomials obey 
\begin{equation}\label{eq:Pellm_parity}
\hat{P} \left[P_{\ell}^{m}(\chi)\right] = (-1)^{\ell + m} P_{\ell}^{m}(\chi)\,,
\end{equation}
where $\ell$ and $m$ are the degree and order of the associated Legendre polynomial (and note that the degree of the Legendre polynomial $\ell$ is not to be confused with the QNM number $l$). 
When $a=0$, the perturbations of different $\ell$ decouple (so that $\ell$ and $l$ are the same), but when $a > 0 $, the perturbations of different $\ell$ are coupled. 
For spinning BHs then, the metric perturbations $h_{i=1, ..., 4}$ and $h_{i=5, 6}$ will ``take turns'' being of polar and axial type. 
For example, for the $l = 2$ modes, the metric perturbations $h_{i=1, ..., 4}$ are polar and $h_{i=5, 6}$ are axial when $\ell = 2, 4, 6, ... $; however, when $\ell = 3, 5, 7, ... $, because of Eq.~\eqref{eq:Pellm_parity}, the metric perturbations $h_{i=1, ..., 4}$ are axial and $h_{i=5, 6}$ are polar. Therefore, one cannot simply set $h_{i=5,6} = 0$ to study polar metric perturbations or $h_{i=1,\ldots,4}=0$ to study axial metric perturbations for BH backgrounds that are spinning.  

\section{Asymptotic behavior of the perturbations of the fields}
\label{sec:ansatz}

Having specified our ansatz of field perturbations around a BH in modified gravity, in this section, we determine the asymptotic behavior of the field perturbations near spatial infinity and the event horizon. 
In particular, using Boyer-Lindquist coordinates, we derive the asymptotic behavior of a massless field (i.e.~the wavefront of the field travels along a null geodesic) near the Killing horizon of a stationary and axisymmetric BH (in or outside GR). 

\subsection{Behavior near spatial infinity}

The asymptotic behavior of $h_i(r, \chi)$ can be determined by studying the asymptotic form of the background spacetime near spatial infinity. 
Near spatial infinity, the background spacetime (Eq.~\eqref{eq:metric}) is asymptotic to 
\begin{equation}
\begin{split}
ds^2 \sim & - \left[1-\frac{2M}{r} \left( 1 - \frac{1}{2} \zeta H_3^{(0)} \right) \right] dt^2 \\
& - \frac{4Ma}{r} \left(1 - \frac{1}{2} \zeta H_3^{(0)} \right) (1-\chi^2) dt d \phi \\
& + \left[ 1 + \zeta H_3^{(0)} \left( 1 - \frac{M}{r}\right) \right] \frac{d r^2}{1 - \frac{2M}{r}} \\
& + \left( 1 + \zeta H_3^{(0)} \right) \left[ r^2 \frac{d \chi^2}{1-\chi^2} + r^2 (1-\chi^2) d \phi^2 \right], 
\end{split}
\end{equation}
where 
\begin{equation}
\begin{split}
H_3^{(0)} & = \lim_{r \rightarrow + \infty} H_3 (r). 
\end{split}
\end{equation}

This asymptotic form of the background metric allows us to derive the asymptotic limit of the Eddington-Finkelstien coordinate $r_*$. Setting $d \chi = d \phi = 0 $, we can write the asymptotic form of the line element near spatial infinity as
\begin{equation}\label{eq:EF_coord_inf}
\begin{split}
ds^2 & \sim  \left[1-\frac{2M}{r} \left( 1 - \frac{1}{2} \zeta H_3^{(0)} \right) \right] \left( dt^2 - d r_*^2 \right), 
\end{split}
\end{equation}
where $r_*$ has been defined by 
\begin{equation}
dr_* = \left[ \frac{1 + \zeta H_3^{(0)} \left( 1 - \frac{M}{r} \right)}{\left[1-\frac{2M}{r} \left( 1 - \frac{1}{2} \zeta H_3^{(0)} \right) \right] \left(1-\frac{2M}{r}\right)} \right]^{\frac{1}{2}} dr. 
\end{equation}
This choice of $r_*$ is such that the null cone has slope of unity near spatial infinity, i.e.~$dt/dr_* = 1$ when $ds = 0$. 
Expanding the defining equation for $r_*$ to first order in $\zeta$ and integrating it, we obtain the asymptotic form of $r_*$ near spatial infinity as a function of $r$, namely
\begin{equation}
r_* \sim \left( 1 + \frac{1}{2} \zeta H_3^{(0)} \right) r + 2 M \log \left( r - 2M \right). 
\end{equation}

With this in hand, outgoing null waves near spatial infinity must travel along null cones in $(t,r_*)$ coordinates, which means that null waves must be proportional to
\begin{equation}
e^{i \omega r_*} \sim e^{i M (1+\frac{1}{2} \zeta H_3^{(0)}) \omega r} r^{2iM\omega}. 
\end{equation}
Observe that metric perturbations do not need to propagate at the speed of light near spatial infinity.
This is because the radial Boyer-Lindquist coordinate in modified gravity does not coincide with the usual radial coordinate, $R$, which measures the surface area of the two-sphere of a constant radius \cite{Cano_Ruiperez_2019}. 
Instead, the modified gravity Boyer-Lindquist coordinate $r$ and the GR Boyer-Lindquist coordinate $R$ near spatial infinity are related by 
\begin{equation}\label{eq:radial_coordinate_limit}
\begin{split}
r & \sim R \left(1 - \frac{1}{2} \zeta H_3^{(0)} \right) -  \frac{1}{2} \zeta H_3^{(1)}\,,  
\end{split}
\end{equation}
where $H_3^{(0)}$ and $H_3^{(1)}$ are defined by the asymptotic behavior of $H_3$ near spatial infinity, 
\begin{equation}
H_3 = H_3^{(0)} + \frac{H_3^{(1)}}{r} + \mathcal{O} \left( \frac{1}{r^2}\right). 
\end{equation}

Thus, the asymptotic behavior of both the metric and scalar-field perturbations near spatial infinity can be written as
\begin{equation}
\label{eq:asymptotic_limits3}
\begin{split}
h_k (r, \chi)  = ~ & e^{i \left(1 + \frac{1}{2} \zeta H_3^{(0)} \right)\omega r} r^{2 i M \omega  + \rho_{\infty}^{(k)}}  \sum^{\infty}_{p=0}\frac{a_{p}}{r^p},  
\end{split}
\end{equation}
where $a_p$ are constants, and $\rho_{\infty}^{(k)}$ is another $k$-dependent parameter controlling the divergent behavior of $h_{k}$ near spatial infinity (which will be specified later, in Eq.~\eqref{eq:rhos}).

\subsection{Behavior near the event horizon}

The asymptotic behavior of $h_k(r, \chi)$ near the event horizon can be determined using the properties of the horizon. 
Since the derivation of the asymptotic behavior near the horizon is more involved than that near spatial infinity, we will derive it through two different sets of arguments and then cross-check the conclusions to make sure the final result is correct. Anticipating the final result, we will find that the asymptotic behavior of $h_k$ near the event horizon, for a fixed $\chi$, must be 
\begin{equation}
\label{eq:asymptotic_limits1}
\begin{split}
\lim \limits_{r \rightarrow r_+} h_k (r, \chi) & \propto \lim \limits_{r \rightarrow r_+}  e^{- i \omega v + i m \varphi}\\
& \sim \left( r-r_+ \right)^{-i\frac{\omega - m \Omega_{\rm H}}{2 \kappa} - \rho_H^{(k)}}\sum_{p=0}^{\infty} b_{p} (r-r_+)^{p}, 
\end{split}
\end{equation}
where $b_{p}$ are constants and $\rho_{H}^{(k)}$ is a $k$-dependent parameter controlling the divergent behavior of $h_{k}$ at $r=r_+$ \cite{Chung:2023zdq}.

Before presenting the details of the derivation of the above result, let us make two remarks. 
First, having the same asymptotic boundary condition for all metric perturbations seems to contradict the Petrov type of sGB BHs. 
An sGB BH is of Petrov-type I, meaning that there are four distinct principal null directions, two of which are ingoing and the remaining two are outgoing \cite{Yagi:2012ya,Owen:2021eez}. 
However, we notice that the sGB corrections to the Kerr principal null directions found in~\cite{Owen:2021eez} are finite, and thus, they are suppressed relative to the Kerr principal null directions, which diverge at the event horizon. 
In other words, the divergent part of the Kerr principal null directions still dominates the asymptotic behavior of massless field perturbations. 
Second, the asymptotic form of $d v$ and $d \varphi$ that we will derive below may be useful to perform field quantization near the event horizon, thereby facilitating the study of the Hawking radiation of rotating BHs in modified gravity in the future.  

\subsubsection{Null geodesic approach}
\label{sec:null_geodesic_approach}

Near the event horizon, massless perturbations propagate along null geodesics that are ingoing at the event horizon \footnote{In many of the theories we consider here, like in sGB gravity and dCS gravity~\cite{Wagle:2019mdq}, there are only two propagating degrees of freedoms for metric perturbations. In the Regge-Wheeler gauge, metric perturbations are characterized by six functions. 
Thus, there are four degrees of freedom in the Regge-Wheeler gauge that are just gauge waves that follow null geodesics. 
The asymptotic behaviour of these four degrees of freedom is fixed by the (perturbed) Hamiltonian and momentum constraints, and, in turn, partially determined by the asymptotic behavior of the propagating degrees of freedom.
}. 
These geodesics are more suitably described by ingoing Kerr null coordinates $(v, r, \theta, \varphi)$, where
\begin{equation}
\begin{split}
v & = t + r_*, \\
\varphi & = \phi + \tilde{r}. 
\end{split}
\end{equation}
Here $v$ is the advanced time, and $r_*$ and $\tilde{r}$ are two coordinates. 
At the event horizon, these coordinates are the coordinates traced by a congruence generated by a Killing vector at the event horizon (see below). 
For GR BHs, $r_*$ and $\tilde{r}$ can be determined as a function of $r$, valid for $r \in [r_+, + \infty)$, by studying principal null geodesics of the spacetime. 
But for BHs in modified gravity, $r_*$ and $\tilde{r}$ cannot be determined in this way because some constants of the equations of motions, such as the Carter constant, may not exist, and the (natively second-order) geodesic equations cannot be decoupled into a set of first-order differential equations. 
Nonetheless, if the asymptotic dependence of $v$ and $\varphi$ on $r$ as $r \rightarrow r_+$ is known, we can still determine the asymptotic behavior of $h_k(r, \chi)$ as $r \rightarrow r_+$. 

One way to determine the asymptotic form of $v$ as a function of $r$ is using an argument that relies on Killing vectors. 
First, we note that $\tilde{\xi}^{\mu} = t^{\mu} + \Omega_{\rm H} \phi^{\mu}$ is a Killing vector in stationary and axisymmetric spacetimes, where $t^{\mu}$ and $\phi^{\mu}$ are the Killing vectors associated with stationarity and axisymmetry. Naturally, then, $\xi^{\mu} = - \tilde{\xi}^{\mu} = -\left( t^{\mu} + \Omega_{\rm H} \phi^{\mu} \right)$ is also a Killing vector, and geometrically, since $\xi^{\mu}$ is null, tangent, and normal to the event horizon, so is $-\xi^{\mu}$. 
Analytically, the defining equations of a Killing vector, $\mathcal{L}_{{\xi}^{\mu}} g_{\alpha \beta}=0$ (where $\mathcal{L}_{{\xi}^{\mu}}$ is the Lie derivative along ${{\xi}^{\mu}}$) or $\nabla_{(\alpha} \xi_{\beta)} = 0$, are invariant under sign reversal. 

With this at hand, we then consider the infinitesimal displacement of a congruence whose four-velocity (defined with respect to advance time $v$) is the Killing vector $\xi^{\mu}$, (see, e.g. page 204 of \cite{Poisson:2009pwt}), 
\begin{equation}\label{eq:adv_time_def}
dx^{\mu} = \xi^{\mu} d v. 
\end{equation}
To obtain a differential equation that defines $v$, we perform the inner product of $dx^{\mu}$ with $\xi_{\mu}$, 
\begin{equation}\label{eq:Killing_displacement}
\xi_{\mu} dx^{\mu} = \xi_{\mu} \xi^{\mu} d v = \xi^2 d v, 
\end{equation}
where $\xi^2 = \xi_{\mu} \xi^{\mu} $ is the modulus square of the Killing vector at the event horizon, which is [see, e.g.~Eq.~(4.7) of \cite{Cano_Ruiperez_2019}]
\begin{equation} \label{eq:xi_sq}
\begin{split}
\xi^2 & = \frac{g_{tt} g_{\phi \phi} - g_{t \phi}^2}{g_{\phi \phi}} = \mathcal{O}(r-r_+),  
\end{split}
\end{equation}
where the second equality holds near the event horizon. 

Next, let us consider the defining equation for the BH surface gravity, 
\begin{equation}\label{eq:kappa}
\xi^{\mu}\nabla_{\mu} \xi^{\nu} = - \kappa \xi^{\nu}, 
\end{equation}
where $\kappa > 0 $ is the surface gravity of the BH. Using this equation together with the Killing equation $\nabla_{(\mu} \xi_{\nu)} = 0$, we then have that [see, e.g.,~Eq.~(5.78) of \cite{Poisson:2009pwt} and Eq.~(4.13) of \cite{Cano_Ruiperez_2019}]
\begin{equation}\label{eq:Killing_vector_derivative}
\xi_{\mu} = \frac{1}{2 \kappa} \partial_{\mu} \xi^2. 
\end{equation}
Note that this equation, strictly speaking, is valid only on the event horizon because Eq.~\eqref{eq:kappa} is only valid at the event horizon [see Sec~4 of \cite{Cano_Ruiperez_2019}]. 
Note also that Eqs.~\eqref{eq:kappa} and~\eqref{eq:Killing_vector_derivative} differ from those in the literature (e.g., Eq.~(5.78) of \cite{Poisson:2009pwt} and Eq.~(4.13) of \cite{Cano_Ruiperez_2019}) only by a minus sign, because we have chosen to work with $\xi^{\mu}$ instead of $\tilde{\xi}^{\mu}$ and because $\kappa > 0$. 

With all of this background, we can now proceed to evaluate the asymptotic form of the Killing vector $\xi^{\mu}$. First, we note that since the background metric is a function of $\chi$ and $r$ only, so is $\xi^2$. 
Thus, at the horizon, $ \xi_{t} = \xi_{\phi} = 0$, and  explicit calculations show that $ \xi_{\chi} = 0$. 
Hence, near the event horizon, we must have that
\begin{equation}
\begin{split}
\xi_{t}, \xi_{\chi}, \xi_{\phi} ~~ \text{at most} ~ \sim \mathcal{O}(r-r_+).
\end{split}
\end{equation}
Technically, the falloff away from the horizon need not be an integer, but the metric of the background spacetime contains only rational functions of $r$ and $\chi$, so noninteger values will not arise.
As for $\xi_{r}$, we use Eq.~\eqref{eq:Killing_vector_derivative} to find that near the event horizon
\begin{equation}
\begin{split}
\xi_{r} & = \frac{1}{2 \kappa} \partial_r \xi^2 + \mathcal{O}(r-r_+). \\
\end{split}
\end{equation}
Note that the argument that $\xi^r = 0$ implies $\xi_r = g_{r \alpha} \xi^{\alpha} = 0 $ does not apply at the event horizon because $g_{rr}$ is singular there.

With all of this preliminary work behind us, we can now finally derive an expression for advanced time as a function of the modified Boyer-Lindquist coordinate $r$. We begin by substituting the above equation into Eq.~\eqref{eq:Killing_displacement} and using Eq.~\eqref{eq:adv_time_def} to find
\begin{equation}
 \xi^2 d v = \xi_r \left( \xi^r dv \right) = \xi_r dr = \frac{1}{2 \kappa} \partial_{r} \xi^2 dr + \mathcal{O}(r-r_+). 
\end{equation}
Rearranging, we have 
\begin{equation}
\begin{split}
d v & = \frac{1}{2 \kappa} \frac{1}{\xi^2} \partial_{r} \xi^2 dr + \mathcal{O}\Big( (r-r_+)^0 \Big) \\
& = \frac{1}{2 \kappa} \partial_{r} \log \xi^2 dr + \mathcal{O}\Big( (r-r_+)^0 \Big). 
\end{split}
\end{equation}
From Eq.~\eqref{eq:xi_sq}, we note that $\xi^2$ has a nonrepeated zero at $r=r_+$, which means we can write $\xi^2 = (r-r_+) \times \delta$, where $\delta \neq 0 $ at $r=r_+$. 
Hence, asymptotically, at a constant-$t$ hypersurface (i.e.~$dt = 0$) and as $r \rightarrow r_+$, we have 
\begin{equation}\label{eq:v_asymptotic}
v - v_0 \sim \frac{1}{2 \kappa} \log(r-r_+) + \mathcal{O}\Big(r-r_+\Big), 
\end{equation}
Note that the coordinate time $t$ has thus far not appeared in our calculations because we are working on a constant-time hypersurface, so that $dt = 0$. 
But even if $t$ were not a constant, its contribution to the asymptotic behavior will be suppressed relative to the logarithm divergence of $\log (r-r_+)$. where we have absorbed the $\delta$-dependent term from expanding the logarithm into $v_0$, which will then contribute a constant phase factor to $h_k$.
Note that while $r_+$ has not modified gravity correction (per the choice of coordinate system adopted to find the modified BH metric), the surface gravity $\kappa$ does admit a $\zeta$-dependent correction, which should be taken into account as we compute the QNM frequencies. 
Equation~\eqref{eq:v_asymptotic} indeed gives the correct asymptotic form of the advanced time of the GR Schwarzschild, Reissner-Nordstr\"{o}m, Kerr and Kerr-Newman BHs \cite{Carter:1968rr} when $\zeta = 0$, and it is consistent with the discussion in~\cite{Franzen:2020gke} (see page 28 in that reference). 

To find the relation between $\varphi$ and the modified Boyer-Lindquist coordinate $r$, we first recall that, in the GR Kerr background, $\varphi$ is defined as 
\begin{equation}
d\varphi = d\phi + \frac{a}{\Delta} dr. 
\end{equation}
Near the event horizon, $d\varphi$ is then asymptotic to 
\begin{equation}\label{eq:dvphi_asymptotic_GR}
d \varphi \sim d\phi + \frac{\Omega_{\rm H}}{2 \kappa} \frac{dr}{r-r_+}, 
\end{equation}
which is a differential equation relating $\phi$ and $r$. 
We seek a similar equation for a rotating BH in modified gravity, so we go back to Eq.~\eqref{eq:Killing_displacement}, take $\mu = \phi$, and recall that we are choosing $\xi^{\mu} = -\left(t^{\mu} + \Omega_{\rm H} \phi^{\mu}\right)$ to obtain
\begin{equation}
d \phi = \xi^{\phi} dv = - \Omega_{\rm H} dv \sim - \frac{\Omega_{\rm H}}{2 \kappa} \frac{dr}{r-r_+},  
\end{equation}
where we have made use of the asymptotic behavior of $dv$ in Eq.~\eqref{eq:v_asymptotic}. 
The asymptotic behavior of $d \phi$ suggests that 
\begin{equation}\label{eq:dvphi_asymptotic}
d \phi + \frac{\Omega_{\rm H}}{2 \kappa} \frac{dr}{r-r_+} \sim \mathcal{O}\Big[(r-r_+)^p \Big], 
\end{equation}
where $p > - 1$. 
In principle, $p$ can assume a noninteger value between $-1$ and 0. 
However, as the BH background metric (Eq.~\eqref{eq:metric}) contains only a rational function of $r$, $p$ can only be a non-negative integer. 
This implies that $d \phi + [{\Omega_{\rm H}}/({2 \kappa})] [{dr}/({r-r_+})]$ is finite (and thus regular) as $r \rightarrow r_+$, and one may define a $\varphi$ coordinate in exactly the same way as done in Eq.~\eqref{eq:dvphi_asymptotic_GR}. 
When $\zeta = 0$, Eq.~\eqref{eq:dvphi_asymptotic} reduces to the $d \varphi$ of the GR Kerr BH (Eq.~\eqref{eq:dvphi_asymptotic_GR}). 

Using the asymptotic form of $v$ and $\varphi$, we can determine the asymptotic behavior of $h_k$ near the horizon to be
\begin{equation}\label{eq:asymptotic_behaviour_EH_adv_time}
h_k (r \rightarrow r_+) \propto e^{-i \omega v + i m \varphi} \propto (r-r_+)^{-i\frac{\omega - m \Omega_{\rm H}}{2 \kappa}} e^{i m \phi - i \omega t}\,, 
\end{equation}
which is exactly what we anticipated in Eq.~\eqref{eq:asymptotic_limits1}.

\subsubsection{d'Alembertian operator approach}

Another way to determine the asymptotic behavior of $h_k(r, \chi) $ at the event horizon is to carry out an asymptotic analysis of the wave equation of a massless scalar field near the event horizon. 
Analytically, metric perturbations obey massless wave equations, which are similar to the wave equation governing a massless scalar field. 
Physically, both the metric and scalar perturbations of a BH follow null geodesics into the BH. 
Thus, deriving the asymptotic behavior of scalar-field perturbations near the horizon allows us to learn about the leading-order divergent behavior of the perturbations along the radial coordinate, offering us insight to fully determine the asymptotic behavior of metric perturbations. 

The wave equation governing massless scalar-field perturbations is simply
\begin{equation}
\Box \Phi = 0, 
\end{equation}
where $\Box$ is the d'Alembertian operator in curved spacetime, 
\begin{equation}
\Box \Phi = \frac{1}{\sqrt{-g}} \partial_{\mu} \left( \sqrt{-g} g^{\mu \nu} \partial_{\nu} \Phi \right) = 0,  
\end{equation}
and $\Phi$ is a scalar field. 
Let us rewrite this equation as 
\begin{equation}
\Box \Phi = \partial_{\mu} \left( g^{\mu \nu}  \partial_{\nu} \Phi \right) + \left( \partial_{\mu} \log \sqrt{-g} \right) g^{\mu \nu} \partial_{\nu} \Phi\,,
\label{Boxeq-new-rewrite}
\end{equation}
and consider the first and second contractions separately. 

In Boyer-Lindquist coordinates, the first contraction can be written as 
\begin{equation}
\begin{split}
& \partial_{\mu} \left( g^{\mu \nu}  \partial_{\nu} \Phi \right) \\
& = g^{tt} \frac{\partial^2 \Phi}{\partial t^2} + 2 g^{t\phi} \frac{\partial^2 \Phi}{\partial t \partial \phi}  + g^{\phi \phi} \frac{\partial^2 \Phi}{\partial \phi^2} \\
& \quad + \frac{\partial}{\partial r} \left( g^{rr} \frac{\partial \Phi}{\partial r} \right) + \frac{\partial}{\partial \chi} \left( g^{\chi \chi} \frac{\partial \Phi}{\partial \chi} \right), 
\end{split}
\end{equation}
where we recall that the explicit expressions for the components of the inverse of the metric $g_{\mu \nu} $ are
\begin{align}
g^{tt} & = \frac{g_{\phi \phi}}{g_{tt} g_{\phi \phi} - g_{t \phi}^2}, \qquad
g^{t\phi}  = - \frac{g_{t \phi}}{g_{tt} g_{\phi \phi} - g_{t \phi}^2}, \\
g^{\phi \phi} & = \frac{g_{tt}}{g_{tt} g_{\phi \phi} - g_{t \phi}^2}, \qquad
g^{rr}  = \frac{1}{g_{rr}}, \qquad
g^{\chi \chi}  = \frac{1}{g_{\chi \chi}}. 
\end{align}
Using Eq.~\eqref{eq:xi_sq} evaluated at the event horizon, 
\begin{equation}
\begin{split}
& \left[ g_{tt} g_{\phi \phi} - g_{t \phi}^2 \right]_{r=r_+} = g_{\phi \phi}(r=r_+) \xi^2\,,
\end{split}
\end{equation}
and since $\xi^2 \sim \mathcal{O}(r-r_+)$ near the event horizon, we then have that
\begin{equation}
g_{tt}(r=r_+) = \left[ \frac{g_{t \phi}^2}{g_{\phi \phi}}\right]_{r=r_+}\,. 
\end{equation}
We can now use this expression on the components of the inverse metric to find that  
\begin{equation}
\begin{split}
g^{tt}(r=r_+) & = \frac{1}{\xi^2}, \\
g^{t\phi}(r=r_+) & = - \left[\frac{g_{t \phi}}{g_{\phi \phi}}\right]_{r=r_+} \frac{1}{\xi^2} = \frac{\Omega_{\rm H}}{\xi^2}, \\
g^{\phi \phi}(r=r_+) & = \left[\frac{g_{t t}}{g_{\phi \phi}}\right]_{r=r_+} \frac{1}{\xi^2} = \left[ \frac{g_{t \phi}^2}{g_{\phi \phi}^2}\right]_{r=r_+} \frac{1}{\xi^2} =  \frac{\Omega_{\rm H}^2}{\xi^2},  
\end{split}
\end{equation}
where we have assumed that $\Omega_{\rm H} \geq 0 $, since $\Omega_{\rm H} = - g_{t \phi}(r=r_+) / g_{\phi \phi}(r=r_+) $. 
Using what we have learned about the behavior of the components of the inverse metric near the event horizon, and multiplying both sides of $\Box \Phi = 0$ by $\xi^2$, we can transform this equation into the following form: 
\begin{equation}
\begin{split}
& \frac{\partial^2 \Phi}{\partial t^2} + 2 \Omega_{\rm H} \frac{\partial^2 \Phi}{\partial t \partial \phi}  + \Omega_{\rm H}^2 \frac{\partial^2 \Phi}{\partial \phi^2} + \xi^2 \frac{\partial}{\partial r} \left( g^{rr} \frac{\partial \Phi}{\partial r} \right) \\
& + \xi^2 \frac{\partial}{\partial \chi} \left( g^{\chi \chi} \frac{\partial \Phi}{\partial \chi} \right) + \xi^2 \left( \partial_{\mu} \log \sqrt{-g} \right) g^{\mu \nu} \partial_{\nu} \Phi \\
& = 0. 
\end{split}
\end{equation}

Let us now focus on the second contraction of Eq.~\eqref{Boxeq-new-rewrite}, namely $\left( \partial_{\mu} \log \sqrt{-g} \right) g^{\mu \nu} \partial_{\nu} \Phi$. Expanding this contraction, we have that
\begin{equation}
\begin{split}
& \left( \partial_{\mu} \log \sqrt{-g} \right) g^{\mu \nu} \partial_{\nu} \Phi \\
& = \left( \partial_{r} \log \sqrt{-g} \right) g^{r r} \partial_{r} \Phi + \left( \partial_{\chi} \log \sqrt{-g} \right) g^{\chi \chi} \partial_{\chi} \Phi. 
\end{split}
\end{equation}
Now, note that $ g^{\chi \chi} $ and $\partial_{\mu} \log \sqrt{-g}$ are at most $\sim \mathcal{O}(1)$ near the event horizon.  
Thus, $\left( \partial_{\chi} \log \sqrt{-g} \right) g^{\chi \chi} \partial_{\chi} \Phi$ is also at most $\sim \mathcal{O}(\Phi)$. 
As for $\left( \partial_{r} \log \sqrt{-g} \right) g^{r r} \partial_{r} \Phi$, near the event horizon we have that 
\begin{equation}
\begin{split}
& g^{r r} \sim \mathcal{O}(r-r_+), \\
& \frac{\partial \Phi}{\partial r} \text{ at most } \sim \mathcal{O}\left[ (r-r_+)^{-1} \Phi \right]. 
\end{split}
\end{equation}
Hence $\left( \partial_{r} \log \sqrt{-g} \right) g^{r r} \partial_{r} \Phi$ is also at most $\sim \mathcal{O}(\Phi)$.

With this in hand, we can now combine what we have learned about the two contractions that define Eq.~\eqref{Boxeq-new-rewrite}. From the above arguments, we have that
\begin{equation}
\begin{split}
& \xi^2 \frac{\partial}{\partial \chi} \left( g^{\chi \chi} \frac{\partial \Phi}{\partial \chi} \right) + \xi^2 \left( \partial_{\mu} \log \sqrt{-g} \right) g^{\mu \nu} \partial_{\nu} \Phi \\
& \sim \mathcal{O}\left[ (r-r_+) \Phi \right]. 
\end{split}
\end{equation}
Hence, as $r \rightarrow r_+ $, the equation $\Box \Phi = 0 $ is asymptotic to 
\begin{equation}\label{eq:asymptotic_dAlembertian_01}
\frac{\partial^2 \Phi}{\partial t^2} + 2 \Omega_{\rm H} \frac{\partial^2 \Phi}{\partial t \partial \phi}  + \Omega_{\rm H}^2 \frac{\partial^2 \Phi}{\partial \phi^2} + \xi^2 \frac{\partial}{\partial r} \left( g^{rr} \frac{\partial \Phi}{\partial r} \right) = 0. 
\end{equation}
This equation can be further simplified using the asymptotic limit of $g^{rr}$ at the horizon, which can be computed using  Eq.~\eqref{eq:Killing_vector_derivative} and taking $\mu = r$ \cite{Cano_Ruiperez_2019}, 
\begin{equation}
\xi_r = \frac{1}{2 \kappa} \partial_r \xi^2. 
\end{equation}
Since $\xi_t = \xi_{\chi} = \xi_{\phi} = 0 $ at $r=r_+$, $\xi^2 $ can be computed using just $\xi_r$ and $g^{rr}$ at the event horizon,
\begin{equation}
\left.\xi^2\right|_{r_+} = - g^{rr} \left( \xi_r \right)^2 = - g^{rr} \left( \frac{\partial_r \xi^2}{2 \kappa}  \right)^2. 
\end{equation}
We include the negative sign here because by Eq.~\eqref{eq:xi_sq}, 
\begin{equation}
\lim_{r-r_+ \rightarrow 0^{+}} \xi^2 = 0^{-}, 
\end{equation}
whereas $g^{rr} > 0 $ for $r>r_+$. 
Rearranging, we have the asymptotic value of $g^{rr}$ at the event horizon, 
\begin{equation}
\left.g^{rr}\right|_{r_+} = -\xi^2 \left( \frac{2 \kappa }{\partial_r \xi^2}  \right)^2. 
\end{equation}
Using this asymptotic limit of $g^{rr}$ at the event horizon, we then have that
\begin{align}
 \xi^2 \frac{\partial}{\partial r} \left( g^{rr} \frac{\partial \Phi}{\partial r} \right)  &= - \xi^2 \frac{\partial}{\partial r} \left[ \xi^2 \left( \frac{2 \kappa }{\partial_r \xi^2}  \right)^2 \frac{\partial \Phi}{\partial r} \right] \\
& = - \frac{2 \kappa\xi^2}{\partial_r \xi^2} \frac{\partial}{\partial r} \left( \frac{2 \kappa \xi^2 }{\partial_r \xi^2} \frac{\partial \Phi}{\partial r} \right) 
\nonumber \\
 & \quad - \xi^2 \left( \frac{\partial}{\partial r} \frac{2 \kappa}{\partial_r \xi^2}  \right) \frac{2 \kappa \xi^2 }{\partial_r \xi^2} \frac{\partial \Phi}{\partial r}\,,
\end{align}
where in the second line we used the product rule.
As $r \rightarrow r_+ $, $\xi^2 \sim \mathcal{O}(r-r_+) $, $\partial_r \Phi$ is at most $\mathcal{O}\left[\left(r-r_+\right)^{-1} \Phi \right]$, and 
\begin{equation}
\begin{split}
& \frac{\xi^2}{\partial_r \xi^2} = \frac{1}{\partial_r \log \xi^2} = \frac{1}{\partial_r \log (r-r_+)} \sim \mathcal{O}(r-r_+), \\
& \frac{\partial}{\partial r} \frac{1}{\partial_r \xi^2} \text{ at most }\sim \mathcal{O}(1), \\
& \Rightarrow \xi^2 \left( \frac{\partial}{\partial r} \frac{2 \kappa}{\partial_r \xi^2}  \right) \frac{2 \kappa \xi^2 }{\partial_r \xi^2} \frac{\partial \Phi}{\partial r} \sim \mathcal{O}[(r-r_+) \Phi]. 
\end{split}
\end{equation}
Hence, asymptotically, the last term can be discarded, so that $\Box \Phi = 0$ is asymptotic to, 
\begin{equation}
\label{eq:asymptotic_dAlembertian_02}
\frac{\partial^2 \Phi}{\partial t^2} + 2 \Omega_{\rm H} \frac{\partial^2 \Phi}{\partial t \partial \phi}  + \Omega_{\rm H}^2 \frac{\partial^2 \Phi}{\partial \phi^2}  - \frac{2 \kappa\xi^2}{\partial_r \xi^2} \frac{\partial}{\partial r} \left( \frac{2 \kappa \xi^2 }{\partial_r \xi^2} \frac{\partial \Phi}{\partial r} \right) = 0. 
\end{equation}

Equation~\eqref{eq:asymptotic_dAlembertian_02} motivates us to define a new radial coordinate, $r_*$, whose asymptotic form as $r \rightarrow r_+ $ is 
\begin{equation}
\frac{\partial}{\partial r_*} = \frac{2 \kappa\xi^2}{\partial_r \xi^2} \frac{\partial}{\partial r} = \frac{2 \kappa}{\partial_r \log \xi^2} \frac{\partial}{\partial r}.  
\end{equation}
Since $\xi^2 \sim \mathcal{O}(r-r_+)$, integrating this differential equation yields
\begin{equation}
r_* \sim \frac{1}{2 \kappa} \log(r-r_+), 
\end{equation}
as $r \rightarrow r_+ $. Note that this is exactly the same expression we found in Eq.~\eqref{eq:v_asymptotic} when we considered the derivation in terms of coordinates adapted to ingoing null geodesics.  

In terms of $r_*$, Eq.~\eqref{eq:asymptotic_dAlembertian_02} becomes 
\begin{equation}\label{eq:asymptotic_dAlembertian_03}
\begin{split}
& \frac{\partial^2 \Phi}{\partial t^2} + 2 \Omega_{\rm H} \frac{\partial^2 \Phi}{\partial t \partial \phi}  + \Omega_{\rm H}^2 \frac{\partial^2 \Phi}{\partial \phi^2} - \frac{\partial^2 \Phi}{\partial r_*^2 } = 0.
\end{split}
\end{equation}
If $\Phi \propto e^{i m \phi - i \omega t}$, Eq.~\eqref{eq:asymptotic_dAlembertian_03} becomes 
\begin{equation}
\frac{\partial^2 \Phi}{\partial r_*^2 } + \left( \omega - m \Omega_{\rm H}\right)^2 \Phi = 0, 
\end{equation}
which implies that the $\Phi$ that is ingoing at the horizon is asymptotic to
\begin{equation}\label{eq:asymptotic_behavior_horizon}
\begin{split}
\Phi & \propto e^{ - i \left( \omega - m \Omega_{\rm H} \right) r_* + i m \phi - i \omega t} \\
& \sim (r-r_+)^{-i\frac{\omega - m \Omega_{\rm H}}{2 \kappa}} e^{i m \phi - i \omega t}, 
\end{split}
\end{equation}
which is the asymptotic behavior we anticipated in Eq.~\eqref{eq:asymptotic_limits1}.

\subsection{Ansatz for the metric and scalar perturbations}
\label{sec:ansatz_summary}

The results above motivate us to "resum and peel off" the asymptotic behaviors of all of the $h_k(r, \chi)$ perturbations through the following product decomposition
\be \label{eq:radspec}
h_k(r, \chi) = A_k(r)  u_k(r, \chi) \,,
\ee
where $u_k(r,\chi)$ is a finite and bounded ``correction factor,'' while $A_k(r)$ is an ``asymptotic factor.'' The latter is chosen to be
\begin{equation}\label{eq:asym_prefactor}
\begin{split}
A_k(r) = & e^{i \left(1 + \frac{1}{2} \zeta H_3^{(0)} \right)\omega r} r^{2 i M \omega  + \rho_{\infty}^{(k)}} \left( \frac{r-r_+}{r}\right)^{-i\frac{\omega - m \Omega_{\rm H}}{2 \kappa}- \rho_H^{(k)}}\,,
\end{split}
\end{equation}
so that the correction factor goes to unity both at the event horizon and at spatial infinity. The parameters $\rho_{H}^{(k)}$ and $\rho_{\infty}^{(k)}$ in Eq.~\eqref{eq:asym_prefactor} control the divergence of the metric function at the event horizon and spatial infinity. 
In principle, these parameters could depend on the index $k$, the dimensionless spin $a$, and the coupling constant $\zeta$.
For the time being, we shall assume that $\rho_{H}^{(k)}$ and $\rho_{\infty}^{(k)}$ are not modified by $a$ and $\zeta$, so that they assume the values for a GR Kerr BH, 
\begin{equation}\label{eq:rhos}
\begin{split}
\rho_H^{(k)} & = 
\begin{cases}
2, ~~ \text{for $k = 3$, }\\
1, ~~ \text{for $k = 2\text{  or  } 6$, }\\
0, ~~ \text{otherwise,}
\end{cases} \\
\rho_{\infty}^{(k)} & = 
\begin{cases}
2, ~~ \text{for $k \neq 4$ nor 7,}\\
1, ~~ \text{for $k = 4$,} \\
-1, ~~ \text{for $k = 7$.} \\
\end{cases}
\end{split} 
\end{equation}
We determined $\rho_{H}^{(7)}$ and $\rho_{\infty}^{(7)}$ by inspecting the asymptotic behavior of the solutions to the scalar D'Alembertian equation in a Kerr background \cite{Teukolsky_01_PRL, Teukolsky:1973ha}. 
In Sec.~\ref{sec:022_a_01}, we shall justify the correctness of these $\rho_{H}^{(k)}$ and $\rho_{\infty}^{(k)}$ choices through the exponential convergence of the QNM frequencies we calculate. 

Let us now use that the correction function $u_k(r, \chi)$ is bounded and finite to represent it as a spectral expansion. In order to achieve this, we must change coordinates so that they all have a finite domain. The radial coordinate, in particular, is semi-infinite, as its domain is $r \in [r_+,\infty]$. 
Following \cite{Langlois:2021xzq, Jansen:2017oag, Chung:2023zdq}, let us then define the compactified spatial coordinate
\begin{equation}\label{eq:z}
z(r) = \frac{2r_+}{r} - 1, 
\end{equation}
which maps the semi-infinite domain of $r$ to the finite domain $z \in [-1,1]$. 
As in \cite{Chung:2023wkd, Chung:2023zdq}, let us then expand $u_k(r, \chi)$ as a linear combination of the product of Chebyshev and associated Legendre polynomials, 
\begin{equation}
\label{eq:spectral_decoposition_correction_factor}
u_k (z, \chi) = \sum_{n=0}^{\infty} \sum_{\ell=|m|}^{\infty} v_k^{n \ell}  T_{n}(z) P^{|m|}_{\ell}(\chi)\,,
\end{equation}
where $ v_k^{n \ell}$ are \textit{constant} coefficients.
As pointed out in \cite{Chung:2023zdq, Chung:2023wkd}, we could have chosen a different basis for the spectral expansion, as long as the basis is complete and orthogonal. 
As we will show below, the product of Chebyshev and associated Legendre polynomials suffices for our purposes.  

Let us then compose the ansatz for the metric and field perturbations to obtain
\begin{equation}\label{eq:spectral_decoposition_factorized}
h_k (z, \chi) = A_k(z) \sum_{n=0}^{\infty} \sum_{\ell=|m|}^{\infty} v_k^{n \ell}  T_{n}(z) P^{|m|}_{\ell}(\chi). 
\end{equation}
This equation provides a formal expression for the full spectral expansion of all field perturbations in the angular coordinate $\chi$ and the compactified spatial coordinate $z$ as a double infinite series over the chosen basis functions. 
In practice, infinite series cannot be used in numerical computations, so we truncate them at some $\mathcal{N}_z$ for the Chebyshev sum and some $\mathcal{N}_{\chi}$ for the associated Legendre sum, leading to
\begin{equation}\label{eq:spectral_decoposition_factorized_finite}
h_k (r, \chi) = A_k(r) \sum_{n=0}^{\mathcal{N}_z} \sum_{\ell=|m|}^{\mathcal{N}_{\chi}+|m|} v_k^{n \ell}  \; T_{n}\left[z(r)\right] \; P^{|m|}_{\ell}(\chi)\,,
\end{equation}
Moreover, we can also, in principle, choose $\mathcal{N}_z$ and $\mathcal{N}_{\chi}$ to be different from each other, but we shall further choose to set $\mathcal{N}_z = \mathcal{N}_{\chi} = N$, so that
\begin{equation}\label{eq:spectral_decoposition_factorized_finite}
h_k (r, \chi) = A_k(r) \sum_{n=0}^{N} \sum_{\ell=|m|}^{N+|m|} v_k^{n \ell}  \; T_{n}\left[z(r)\right] \; P^{|m|}_{\ell}(\chi)\,.
\end{equation}
When numerically calculating the QNM frequencies of BHs in modified gravity, we will use this last truncated spectral expansion. 

\section{Linearized field equations about the background spacetime as an algebraic problem in the METRICS approach}
\label{sec:METRICS}

In this section, we will derive the linearized field equations with the field perturbation ansatz described in the previous section. 
We will then convert these equations into an algebraic problem. 

\subsection{The linearized field equations}
\label{sec:Linearized_EFEs}

We begin by substituting the perturbed metric and scalar field of Eq.~\eqref{eq:spectral_decoposition_factorized} into the field equations  [Eq.~\eqref{eq:field_eqs}]. 
Linearizing the latter, one finds a system of 11 coupled, partial differential equations for the seven unknown functions $h_k(r,\chi)$. 
The process of linearization, however, can be simplified slightly as follows. 
When substituting the asymptotic factor into the field equations for the metric, i.e.  $R_{\mu}{}^{\nu} + \zeta \left( \mathscr{A}_{\mu}{}^{\nu} - T_{\mu}{}^{\nu} \right) = 0$, we need to include the \textit{explicit} $\zeta$ modifications to $\Omega_{\rm H}$ and $\kappa$ in the asymptotic factor, because the equation is satisfied up to first order in $\zeta$. 
However, when considering the scalar-field equation, i.e. $\square \vartheta + \mathscr{A}_{\vartheta} = 0$, we can set $\zeta$ to zero when computing $\Omega_{\rm H}$ and $\kappa$, because $\mathscr{A}_{\vartheta}$ is computed in the GR Kerr background. 
Nonetheless, this is not to say that the scalar-field equation is not modified by $\zeta$, because the d'Alambertian depends on $h_k(r,\chi)$, and thus, the equation is coupled to the equations for the metric perturbations. 

We can now cast the linearized field equations into a form that can also be spectrally expanded. 
We note that, even with the corrections due to modified gravity, the components of the background metric tensor $g_{\mu\nu}^{(0)}$ in Boyer-Lindquist coordinates are still rational functions of $r$ and $\chi$ [c.f. Eq.~\eqref{eq:metric}]. 
Therefore, the coefficient functions multiplying the field perturbations $h_k$ in the linearized field equations must also be rational functions of $r$ and $\chi$, since they can only depend on background quantities and their derivatives \cite{Chung:2023wkd, Chung:2023zdq}. 
Thus, we can always cast the $k$-th linearized field equation, after factorization and multiplication through common denominators, in the following form: 
\begin{align}\label{eq:pertFE-1}
& \sum_{j=1}^{7} \sum_{\eta = 0}^{1} \sum_{\alpha, \beta = 0} \sum_{\gamma=0} \sum_{\delta=0}^{d_{r}} \sum_{\sigma=0}^{d_{\chi}} \mathcal{G}_{k, \eta, \gamma, \delta, \sigma, \alpha, \beta, j} \zeta^{\eta} \omega^\gamma r^{\delta} \chi^{\sigma} \partial_{r}^{\alpha} \partial_{\chi}^{\beta} h_j 
\nonumber \\
&= 0 \,,
\end{align}
where $\mathcal{G}_{k, \eta, \gamma, \delta, \sigma, \alpha, \beta, j}$ is a complex function of $M$, $m$, and $a$ only.
The upper limit of the sums over $\alpha+\beta$ and over $\gamma$ depends on the nature of the modified gravity theory considered. 
For sGB gravity, $\alpha+\beta \leq 3 $ and $\gamma \leq 2 $ because $\mathscr{A}_{\mu} {}^{\nu}$ involves at most second-order derivatives of the metric and the scalar field [c.f. Eq.~\eqref{eq:Amunu_sGB}]. 
On the other hand, for dCS gravity, $\alpha+\beta \leq 4 $ and $\gamma \leq 3 $ because derivatives of the Ricci tensor are involved [c.f. Eq.~\eqref{eq:Amunu_dCS}]. 
Note that the upper limit of the sum over $j$ has been increased to 7 to accommodate the scalar-field perturbation. 
The constants $d_r$ and $d_\chi$ are the degree in $r$ and $\chi$ of the coefficient of a given term in the linearized field equations respectively, which depend on the component of the linearized field equation we are focusing on, and, thus, on the summation indices $(\alpha,\, \beta, \, k, \, j)$. 
We sum only from $\eta = 0 $ to 1, because the metric modifications and the field equations, and thus, all calculations that follow, are valid up to first order in $\zeta$ only. 
After factorizing each of the linearized field equations to obtain common denominators, we find that there are also prefactors, such as powers of $1-\chi^2$, $\Delta$, and $\Sigma$, multiplying the equations.
These prefactors contain no metric perturbation functions and are nonzero except at $r = r_+$, and $\chi = \pm 1$. 
As these common factors are nonzero in the computational domain (except at the boundaries), we divide the equations by them to simplify them and improve their numerical stability.
When $\zeta = 0$, Eq.~\eqref{eq:pertFE-1} reduces to the equations that correspond to the metric and scalar perturbations of a Kerr BH in GR. 

Equation~\eqref{eq:pertFE-1} represents a system of coupled and (two-dimensional) partial differential equations.  
As observed in \cite{Chung:2023wkd, Chung:2023zdq}, the modulus of $\mathcal{G}_{k, \eta, \gamma, \delta, \sigma, \alpha, \beta, j}$ can change by $\sim 10$ orders of magnitude across different $\gamma, \delta, \sigma, \alpha, \beta$, and $j$ in one equation, and the largest modulus of $\mathcal{G}_{k, \eta, \gamma, \delta, \sigma, \alpha, \beta, j}$ of different equations can also change by $\sim 20$ orders of magnitude. 
As in \cite{Chung:2023wkd, Chung:2023zdq}, we prevent overflow by normalizing every partial differential equation, such that the largest modulus of the coefficient of each equation is one. 
This is allowed by the homogeneity of the linearized field equations.

We can now substitute the truncated spectral expansion of the field perturbation functions (Eq.~\eqref{eq:spectral_decoposition_factorized}) into Eq.~\eqref{eq:pertFE-1} to transform the latter into a system of linear algebraic equations. 
Since $r$ is a rational function of $z$ (see Eq.~\eqref{eq:z}), the coefficient functions of the linearized equations of $u_k$ can only be rational functions of $z$ \cite{Chung:2023wkd}. 
Therefore, when we substitute Eq.~\eqref{eq:radspec} into Eq.~\eqref{eq:pertFE-1}, 
we can factorize the $k$-th partial differential equation as
\begin{equation}\label{eq:system_3}
\begin{split}
& \sum_{j=1}^{7} \sum_{\eta = 0}^{1} \sum_{\alpha, \beta = 0} \sum_{\gamma=0} \sum_{\delta=0}^{d_{z}} \sum_{\sigma=0}^{d_{\chi}} \mathcal{K}_{k, \eta, \gamma, \delta, \sigma, \alpha, \beta, j} \zeta^{\eta} \omega^\gamma z^{\delta} \chi^{\sigma} \partial_{z}^{\alpha} \partial_{\chi}^{\beta} u_j \\
& = 0 \,, 
\end{split}
\end{equation}
where $d_z$ and $d_{\chi}$ are the degree of $z$ and $\chi$ of the coefficient of the partial derivative $\partial_{z}^{\alpha} \partial_{\chi}^{\beta} \{...\} $ in the equations respectively, while $\mathcal{K}_{k, \eta, \gamma, \delta, \sigma, \alpha, \beta, j}$ are complex functions of $M, m$, $a$, $\zeta$, $\rho_H^{(k)}$ and $\rho_{\infty}^{(k)}$ only (for every value of the summation indices $ \alpha, \beta, \gamma, \delta, \sigma$, and $j$). 

\subsection{Converting the linearized field equations into algebraic equations}
\label{sec:Conversion}

Let us now convert the linearized field equations into an algebraic system of equations using our spectral expansion. 
We first substitute the truncated spectral expansion of the $u_k$ functions into Eq.~\eqref{eq:system_3}, 
\begin{equation}\label{eq:system_4}
\begin{split}
& \sum_{j=1}^{7} \sum_{\eta = 0}^{1} \sum_{\alpha, \beta = 0} \sum_{\gamma=0} \sum_{\delta=0}^{d_{z}} \sum_{\sigma=0}^{d_{\chi}} \mathcal{K}_{k, \eta, \gamma, \delta, \sigma, \alpha, \beta, j} \zeta^{\eta}  \omega^\gamma z^{\delta} \chi^{\sigma} \\
& \quad \quad \quad \times \partial_{z}^{\alpha} \partial_{\chi}^{\beta} \Bigg\{\sum_{n=0}^{\mathcal{N}_z} \sum_{\ell=|m|}^{\mathcal{N}_{\chi}+|m|} v_j^{n \ell} T_{n}(z) P^{|m|}_{\ell}(\chi) \Bigg\} = 0.  
\end{split}
\end{equation}
These equations can be further simplified by using the defining equations for the Chebyshev polynomials and associated Legendre polynomials, namely
\begin{equation}
\begin{split}
\frac{d^2 T_n}{d z^2} & = \frac{1}{1-z^2} \left( z \frac{d T_n}{dz} - n^2 T_n \right), \\
\frac{d^2 P_{\ell}^{|m|}}{d \chi^2} & = \frac{1}{1-\chi^2} \Big( 2 \chi \frac{d P_{\ell}^{|m|}}{d \chi} - \ell(\ell+1) P_{\ell}^{|m|} \\
& \quad \quad \quad \quad \quad - \frac{m^2}{1-\chi^2} P_{\ell}^{|m|} \Big). 
\end{split}
\end{equation}
The above allows us to obtain more factors of $1-\chi^2, 1-z$, or $1+z$ when factorizing the linearized field equations, further simplifying Eq.~\eqref{eq:system_3}. 

We then express the left-hand side of Eq.~\eqref{eq:system_3} as a linear combination of the Chebyshev and associated Legendre polynomials, 
\begin{equation}\label{eq:elliptic_eqn_v2}
\sum_{n=0}^{\mathcal{N}_z} \sum_{\ell=|m|}^{\mathcal{N}_{\chi}+|m|} w_k^{n \ell} T_{n}(z) P^{|m|}_{\ell}(\chi) = 0\,, 
\end{equation}
where $w_k^{n \ell}$ are independent of $z$ and $\chi$, but depend on $M$, $a$, $n$, $\ell$, $m$, and\footnote{Recall that $n$ and $\ell$ here do not denote the overtone and azimuthal-mode number of the QNM frequency. Rather, $n$, and $\ell$ are the order of the Chebyshev and the degree of the associated Legendre polynomials.} $\omega$, and $k \in [1,11]$. 
By the orthogonality of $T_{n}(z)$ and of $P^{|m|}_{\ell}(\chi)$, the linearized field equations are satisfied when $ w_k^{n \ell} = 0 $ for every $k, n$ and $\ell$ because they are homogenous. 
Comparing Eq.~\eqref{eq:system_3} to Eq.~\eqref{eq:system_4}, we notice that $w_k^{n \ell}$ depends on $v_k^{n \ell}$ linearly, because
\begin{equation}
\label{eq:pertFE-2}
\begin{split}
w_k^{n \ell} = \sum_{j=1}^{7} \sum_{n'=0}^{\mathcal{N}_z} \sum_{\ell'=|m|}^{\mathcal{N}_{\chi}+|m|} \left[ \mathbb{D}_{n \ell, n' \ell'}(\omega) \right]_{kj} v_j^{n' \ell'}, 
\end{split}
\end{equation}
where $\mathbb{D}_{n \ell, n' \ell'}(\omega)$ are $11 \times 7$ matrices, whose elements are polynomials in $\omega$ and can be obtained by evaluating the inner product given by Eq. (41) of \cite{Chung:2023zdq}.  

Let us now introduce some notation to simplify the resulting expressions. As in \cite{Chung:2023zdq}, we introduce the following (Euclidean) vectors~\cite{Chung:2023zdq}, 
\begin{equation}\label{eq:vector_arrangement}
\begin{split}
& \textbf{v}_{n \ell} = \left( v_1^{n \ell}, v_2^{n \ell}, v_3^{n \ell}, v_4^{n \ell}, v_5^{n \ell}, v_6^{n \ell}, v_7^{n \ell} \right)^{\rm T} \,, \\
& \textbf{w}_{n \ell} = \left( w_1^{n \ell}, w_2^{n \ell}, ..., w_{11}^{n \ell} \right)^{\rm T} \,,
\end{split}
\end{equation}
so that Eq.~\eqref{eq:pertFE-2} can be written as 
\begin{equation}\label{eq:vector_equations}
\textbf{w}_{n \ell} = \sum_{n'=0}^{\mathcal{N}_z} \sum_{\ell'=|m|}^{\mathcal{N}_{\chi}+|m|} \mathbb{D}_{n \ell, n' \ell'}(\omega) \textbf{v}_{n' \ell'} = 0\,,
\end{equation}
where the $\mathbb{D}_{n \ell, n' \ell'}$ matrix is now ``dotted'' into $\textbf{v}_{n' \ell'}$ (with a flat, Euclidean metric). 
Let us further define the vectors $\textbf{v}$ and $\textbf{w}$ as
\begin{equation}
\begin{split}
\textbf{v} & = \left\{ \textbf{v}_{00}^{\rm T}, \textbf{v}_{01}^{\rm T}, ..., \textbf{v}_{0 \mathcal{N}_{\chi}}^{\rm T}, ..., \textbf{v}_{1 \mathcal{N}_{\chi}}^{\rm T}, ...,  \textbf{v}_{(\mathcal{N}_{z}+1)(\mathcal{N}_{\chi}+1)}^{\rm T} \right\}^{\rm T}, \\
\textbf{w} & = \left\{ \textbf{w}_{00}^{\rm T}, \textbf{w}_{01}^{\rm T}, ..., \textbf{w}_{0 \mathcal{N}_{\chi}}^{\rm T}, ..., \textbf{w}_{1 \mathcal{N}_{\chi}}^{\rm T}, ...,  \textbf{w}_{(\mathcal{N}_{z}+1)(\mathcal{N}_{\chi}+1)}^{\rm T} \right\}^{\rm T}\,,
\end{split}
\end{equation}
which store all $\textbf{v}_{n \ell}$ and $\textbf{w}_{n \ell}$  respectively.
Note that $\textbf{v}$ is a $7(\mathcal{N}_{z}+1)(\mathcal{N}_{\chi}+1)$-dimensional vector, whereas $\textbf{w}$ is a $11(\mathcal{N}_{z}+1)(\mathcal{N}_{\chi}+1)$-dimensional vector.
Then, Eq.~\eqref{eq:pertFE-2} can be more compactly written as 
\begin{equation}\label{eq:augmented_matrix_01}
\begin{split}
\textbf{w} = \tilde{\mathbb{D}} (\omega) \textbf{v} = \left[ \sum_{\gamma=0} \tilde{\mathbb{D}}_{\gamma} \omega^\gamma \right] \textbf{v} = \textbf{0} \,,
\end{split}
\end{equation}
where the $\tilde{\mathbb{D}}_{\gamma = 0,1,2, ....}$ matrices are constant, $ 11 (\mathcal{N}_z + 1)(\mathcal{N}_{\chi} + 1) \times 7 (\mathcal{N}_z + 1)(\mathcal{N}_{\chi} + 1) $ rectangular matrices, which are all linear in $\zeta$. 
The QNM frequencies of the modified BH in modified gravity correspond to the $\omega$ such that Eq.~\eqref{eq:augmented_matrix_01} admits a nontrivial solution $\textbf{v}$. 

\section{First $\zeta$-order eigenvalue perturbation in the METRICS Approach}
\label{sec:eigenvalue_pert}

The modified gravity theories we are considering are effective, and thus, their defining actions ought to be understood as curvature (or derivative) expansions. The field equations that result from varying such an approximate action are therefore also approximate. Any solutions of these approximate field equations are, of course, also approximate by construction. More concretely for the problem at hand, the background metric (Eq.~\eqref{eq:metric}) satisfies the field equations only up to the first order in $\zeta$, because the field equations are only valid to this order. For consistency, then, we now compute the first-order-in-$\zeta$ modification to the QNM frequencies. 
To this end, we develop an eigenvalue-perturbation scheme, which allows us to estimate the first $\zeta$-order modifications to the QNM frequencies. 

We begin by expanding $\mathbb{D}_k, \omega$, and $\textbf{v}$ as a power series in $\zeta$, 
\begin{equation}\label{eq:eigenperturbation_01}
\begin{split}
\mathbb{D}_k & = \sum_{j=0} \zeta^j \mathbb{D}^{(j)}_k, \quad
\omega  = \sum_{j=0} \zeta^j \omega^{(j)}, 
\quad
\textbf{v} = \sum_{j=0} \zeta^j \textbf{v}^{(j)}, 
\end{split}
\end{equation}
and then truncate the series at $j=1$, 
\begin{equation}\label{eq:eigenperturbation_02}
\begin{split}
\mathbb{D}_k & = \mathbb{D}^{(0)}_k + \zeta \mathbb{D}^{(1)}_k, \\
\omega & = \omega^{(0)} + \zeta \omega^{(1)}, \\
\textbf{v} & = \textbf{v}^{(0)} + \zeta \textbf{v}^{(1)}.  
\end{split}
\end{equation}
Substituting these perturbed quantities into Eq.~\eqref{eq:augmented_matrix_01}, we have the zeroth-order-in-$\zeta$ equation, 
\begin{equation}\label{eq:zeroth-zeta-order}
\tilde{\mathbb{D}}^{(0)} (\omega^{(0)}) \textbf{v}^{(0)} = \textbf{0}, 
\end{equation}
which is the same as the METRICS equation for the QNM frequencies for Kerr black holes in GR \cite{Chung:2023wkd}. The first-order parts in $\zeta$ of Eq.~\eqref{eq:augmented_matrix_01} is
\begin{equation}\label{eq:first-zeta-order}
\begin{split}
& \omega^{(1)} \frac{\partial \tilde{\mathbb{D}}^{(0)} (\omega)}{\partial \omega}\Big|_{\omega = \omega^{(0)}} \cdot \textbf{v}^{(0)} + \tilde{\mathbb{D}}^{(0)} (\omega^{(0)}) \cdot \textbf{v}^{(1)} \\
& + \tilde{\mathbb{D}}^{(1)} (\omega^{(0)}) \cdot  \textbf{v}^{(0)} = \textbf{0}, 
\end{split}
\end{equation}
where we have defined 
\begin{equation}
\tilde{\mathbb{D}}^{(p)} (\omega^{(q)}) = \sum_{k=0} \tilde{\mathbb{D}}^{(p)}_k \left[\omega^{(q)}\right]^k. 
\end{equation}
Recall that these matrices are power series in $\omega$, and the explicit upper limit of summation depends on the modified theory.

To explain the further steps for solving these equations, let us focus on the polar perturbations as an example. 
For the polar perturbations, we impose the following conditions to break the homogeneity or linear-scaling invariance of Eq.~\eqref{eq:first-zeta-order}, 
\begin{equation}
\begin{split}
&\left[ \textbf{v}^{(0)}\right]_{k=1}^{n=0,\ell=|m|} = 1, \\
&\left[ \textbf{v}^{(1)}\right]_{k=1}^{n=0,\ell=|m|} = 0. 
\end{split}
\end{equation}
These conditions are allowed by the homogeneity of Eq.~\eqref{eq:first-zeta-order}. 
If $\left[ \textbf{v}^{(1)}\right]_{k=1}^{n=0,\ell=|m|} \neq 0$, we can always divide the whole $\textbf{v}$ by $\left[ \textbf{v}^{(0)} + \zeta \textbf{v}^{(1)} \right]_{k=1}^{n=0,\ell=|m|}$ to meet the above conditions. 

The zeroth-order-in-$\zeta$ equation ~\eqref{eq:zeroth-zeta-order} can be solved using the Newton-Raphson algorithms discussed in \cite{Chung:2023wkd}. 
After numerically obtaining $\textbf{v}^{(0)}$ and $\omega^{(0)}$, we can use them to solve for $\textbf{v}^{(1)}$ and $\omega^{(1)}$. 
Since we have set $\left[ \textbf{v}^{(1)}\right]_{k=1}^{n=0,\ell=|m|} = 0$, we are left with the following unknowns
\begin{equation}
\textbf{x}^{(1)} = \left\{ \left[ \textbf{v}^{(1)}\right]_{k\neq1}^{n\neq0,\ell\neq|m|}, \omega^{(1)}\right\}. 
\end{equation}
We can solve Eq.\eqref{eq:first-zeta-order} for $\textbf{x}^{(1)}$ by computing the Moore-Penrose inverse of the following Jacobian matrix 
\begin{equation}
[\mathbb{J}]_{ij} = \frac{\partial \left[\mathbb{D}^{(0)}(\omega)\cdot \textbf{v}^{(1)} \right]}{\partial [\textbf{x}^{(1)}]_{j}} \Bigg|_{\textbf{x} = \textbf{x}_{(n)}}, 
\end{equation}
such that 
\begin{equation}\label{eq:x_perturbation}
\textbf{x}^{(1)} = - [\mathbb{J}]^{-1} \cdot \left[\tilde{\mathbb{D}}^{(1)} (\omega^{(0)}) \cdot \textbf{v}^{(0)} \right]. 
\end{equation}
This Jacobian matrix is exactly the same as that defined by Eq. (41) of \cite{Chung:2023wkd}, which was evaluated at the GR $\textbf{x}$ (thus, we here denote them by the same symbol $\mathbb{J}$). 
In actual numerical computations, we only need to save the inverse of the Jacobian matrix and $\textbf{x}$ at the last Newton-Raphson step when solving the linearized Einstein equations in GR, and then use them to compute $\textbf{x}^{(1)}$ by Eq.~\eqref{eq:x_perturbation}. 
Equation~\eqref{eq:x_perturbation} is a key result of this work. 

The above procedure can also be similarly performed for the axial perturbations using the following initial guess, 
\begin{equation}
\begin{split}
&\left[ \textbf{v}^{(0)}\right]_{k=5}^{n=0,\ell=|m|} = 1, \\
&\left[ \textbf{v}^{(1)}\right]_{k=5}^{n=0,\ell=|m|} = 0, 
\end{split}
\end{equation}
and for scalar-led perturbations using the following initial guess, 
\begin{equation}
\begin{split}
&\left[ \textbf{v}^{(0)}\right]_{k=7}^{n=0,\ell=|m|} = 1, \\
&\left[ \textbf{v}^{(1)}\right]_{k=7}^{n=0,\ell=|m|} = 0. 
\end{split}
\end{equation}
Since the steps are identical to the polar case, we will not present them here. 

From Eq.~\eqref{eq:x_perturbation}, we can immediately conclude that isospectrality may not persevere in modified gravity. 
The $\textbf{v}^{(0)}$ of the axial, polar, and scalar-led perturbations is different. 
Given a gravity theory, even though $\tilde{\mathbb{D}}^{(1)}(\omega)$ is the same,,
$\tilde{\mathbb{D}}^{(1)} (\omega^{(0)}) \cdot \textbf{v}^{(0)}$ can be different because $\textbf{v}^{(0)}$ of the axial, polar and scalar perturbations is different. 
Thus, in general, $\omega^{(1)}$ of perturbations led by different sectors is different, which implies the departure from the isospectrality in GR. 

The above is not the only way to perturbatively solve for the QNM frequencies through a spectral-expansion method. Recall that the metric perturbation was earlier decomposed into the product of an asymptotic factor $A_k(r)$ and a correction factor $u_k(r,\chi)$ [see Eq.~\eqref{eq:radspec}]. We could have therefore worked directly with the correction factor; in fact, the above procedure is equivalent to writing 
\begin{equation}
u_j = u^{(0)}_j + \zeta u^{(1)}_j, 
\end{equation}
then first solving 
\begin{equation}
\begin{split}
& \sum_{j=1}^{7} \sum_{\alpha, \beta = 0}^{\alpha+\beta \leq 3} \sum_{\gamma=0}^{2} \sum_{\delta=0}^{d_{z}} \sum_{\sigma=0}^{d_{\chi}} \mathcal{K}_{k, \eta = 0, \gamma, \delta, \sigma, \alpha, \beta, j} \omega^\gamma z^{\delta} \chi^{\sigma} \partial_{z}^{\alpha} \partial_{\chi}^{\beta} u^{(0)}_j \\
& = 0 \,, 
\end{split}
\end{equation}
for $u^{(0)}_j$, and then using them to solve 
\begin{equation}
\begin{split}
& \sum_{j=1}^{7} \sum_{\alpha, \beta = 0}^{\alpha+\beta \leq 3} \sum_{\gamma=0}^{2} \sum_{\delta=0}^{d_{z}} \sum_{\sigma=0}^{d_{\chi}} \mathcal{K}_{k, \eta = 0, \gamma, \delta, \sigma, \alpha, \beta, j} \omega^\gamma z^{\delta} \chi^{\sigma} \partial_{z}^{\alpha} \partial_{\chi}^{\beta} u^{(1)}_j \\
& = - \sum_{j=1}^{7} \sum_{\alpha, \beta = 0} \sum_{\gamma=0} \sum_{\delta=0}^{d_{z}} \sum_{\sigma=0}^{d_{\chi}} \mathcal{K}_{k, \eta = 1, \gamma, \delta, \sigma, \alpha, \beta, j} \zeta \omega^\gamma z^{\delta} \chi^{\sigma} \partial_{z}^{\alpha} \partial_{\chi}^{\beta} u_j^{(0)} 
\end{split}
\end{equation}
for $u^{(1)}_j$. This, of course, is more complicated than the method we introduced above because would then have to solve a system of coupled partial differential equations for the correction factor. 
Having said that, the above procedure is not the same as writing 
\begin{equation}
h_j = h^{(0)}_j + \zeta h^{(1)}_j, 
\end{equation}
and then first solving 
\begin{equation}
\begin{split}
& \sum_{j=1}^{7} \sum_{\alpha, \beta = 0}^{\alpha+\beta \leq 3} \sum_{\gamma=0}^{2} \sum_{\delta=0}^{d_{z}} \sum_{\sigma=0}^{d_{\chi}} \mathcal{G}_{k, \eta = 0, \gamma, \delta, \sigma, \alpha, \beta, j} \omega^\gamma z^{\delta} \chi^{\sigma} \partial_{z}^{\alpha} \partial_{\chi}^{\beta} h^{(0)}_j \\
& = 0 \,, 
\end{split}
\end{equation}
and then solving
\begin{equation}
\begin{split}
& \sum_{j=1}^{7} \sum_{\alpha, \beta = 0}^{\alpha+\beta \leq 3} \sum_{\gamma=0}^{2} \sum_{\delta=0}^{d_{z}} \sum_{\sigma=0}^{d_{\chi}} \mathcal{K}_{k, \eta = 0, \gamma, \delta, \sigma, \alpha, \beta, j} \omega^\gamma z^{\delta} \chi^{\sigma} \partial_{z}^{\alpha} \partial_{\chi}^{\beta} h^{(1)}_j \\
& = - \sum_{j=1}^{7} \sum_{\alpha, \beta = 0} \sum_{\gamma=0} \sum_{\delta=0}^{d_{z}} \sum_{\sigma=0}^{d_{\chi}} \mathcal{K}_{k, \eta = 1, \gamma, \delta, \sigma, \alpha, \beta, j} \zeta \omega^\gamma z^{\delta} \chi^{\sigma} \partial_{z}^{\alpha} \partial_{\chi}^{\beta} h_j^{(0)} 
\end{split}
\end{equation}
for $h^{(1)}_j$. 
In the latter approach, since the asymptotic factor of $h^{(0)}_j$ is a GR one, so is that of $h^{(1)}_j$, which means that the effects of modified gravity on the asymptotic factor (or behavior) of $h_j$ are ignored. 
On the other hand, by solving $u_j$ perturbatively, the effects of modified gravity on the asymptotic factor are also included. 

We conclude this section by drawing a parallel between the eigenvalue perturbation theory of METRICS and the time-independent perturbation theory of quantum mechanics. 
Equation~\eqref{eq:x_perturbation} is the black hole perturbation theory equivalent to the equations of the first-order eigenenergy shift of a quantum mechanical system with a slight time-independent modification to its potential. 
In particular, $\omega^{(0)}$ plays the role of the unperturbed eigenfrequency, $\textbf{v}^{(0)}$ the role of the wave function of an unperturbed eigenstate, $\mathbb{D}^{(1)}$ the role of the time-independent modifications to the potential of the quantum mechanical system, $[\mathbb{J}]^{-1}$ the role of normalization in quantum mechanics, and the dot products between $[\mathbb{J}]^{-1}, \mathbb{D}^{(1)}$ and $\textbf{v}^{(0)}$ the volume integral of the modulus square of the unperturbed wave function. 
In this work, we focus only on the leading-$\zeta$ order because we work on effective field theories that are only valid to this order. 
But also as in quantum mechanics, one can, in principle, further develop the eigenvalue perturbation theory in METRICS to an arbitrary order in $\zeta$, provided that the background BH spacetime in the gravity theory is also available up to the corresponding order in $\zeta$. 

\section{Application to rotating BHs in scalar-Gauss-Bonnet gravity}
\label{sec:Application_sGB}

To illustrate the application of METRICS to BHs in modified gravity, in this section, we apply the eigenvalue perturbation scheme developed above to BHs in sGB gravity, which belongs to a popular class of effective theory that has recently been studied extensively~\cite{EdGB_01, EdGB_02, EdGB_03, EdGB_04, QNM_EdGB_01, QNM_EdGB_02, QNM_EdGB_03, PhysRevD.47.5259}.
Specifically, we will apply METRICS to compute the QNM frequencies of the axial and polar perturbations of the 022, 033, and 021 modes. 
We focus on these $l$ and $m$ modes because there have been claims that suggest the presence of these QNMs in detected ringdown signals \footnote{We remind the reader that there is debate over whether multiple QNMs have been actually detected (see, e.g., \cite{Cotesta:2022pci, Isi:2022mhy, Carullo:2023gtf}). Although interesting, these debates do not concern us here. 
} \cite{Siegel:2023lxl, Gennari:2023gmx}. 
We focus on the fundamental modes ($n=0$) because they have the longest lifetimes. 
Finally, we focus more extensively on the axial and polar QNMs of the metric perturbation because these are the ones that can be detected with GW interferometers; sGB gravity only possesses the same two tensor polarizations as GR~\cite{Wagle:2019mdq}. Nonetheless, for completeness, we do present the scalar-mode frequencies of the 022 mode. 

\subsection{Numerical implementation}

We first numerically solve Eq.~\eqref{eq:zeroth-zeta-order} according to the procedure detailed in Sec. IV of \cite{Chung:2023wkd}. 
Throughout our numerical computations, $M$ is set to one (i.e. $M=1$), thus defining a system of units. 
To speed up the convergence of our numerical results, we initialize the Newton-Raphson iterations with the known values of the Kerr QNM frequencies. 
This is a justified procedure, given that we have already showed that METRICS can accurately compute the (known) values of the QNM frequencies of Kerr BHs in GR \cite{Chung:2023wkd}. 
The inverse of the Jacobian matrix is computed using the built-in \texttt{PseudoInverse} function of \textit{Mathematica} to double precision. 

\begin{figure*}[htp!]
\centering  
\subfloat{\includegraphics[width=0.47\linewidth]{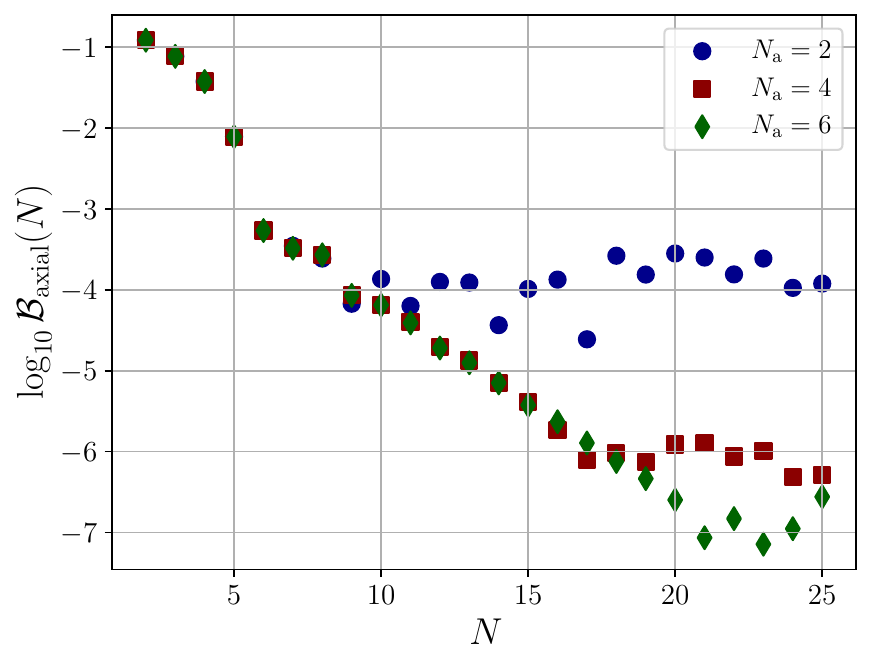}}
\subfloat{\includegraphics[width=0.47\linewidth]{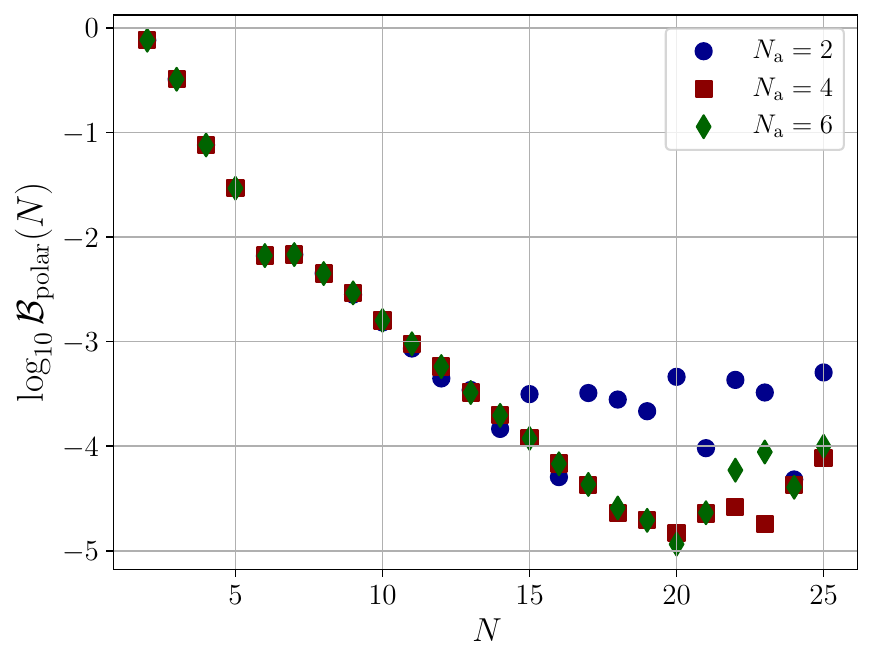}}
\caption{The backward modulus difference $\mathcal{B}(N) = |\omega^{(1)}(N)-\omega^{(1)}(N-1)|$, where $\omega^{(1)}$ is the leading-order modifications to the $nlm=$022-mode frequency (defined by Eq.~\eqref{eq:eigenperturbation_02}) of the axial (left) and polar (right) perturbations of a BH of $a=0.1$ in shift-symmetric Einstein dilaton Gauss-Bonnet (sGB) gravity as a function of the spectral order ($N$). 
The background metric of the rotating BH includes modifications due to shift symmetric sGB gravity up to $N_{\rm a} = 2$, 4, and 6 orders in $a$. 
We observe that for all $N_{\rm a}$, $\mathcal{B}(N)$ first decreases exponentially and then fluctuates around a constant after reaching a specific spectral order. 
Nonetheless, the constant about which $\mathcal{B}(N)$ fluctuates can be significantly reduced if we include the metric corrections of sGB of a larger order in $a$. 
This is reasonable because a background metric with sGB corrections of a higher order of $a$ satisfies the field equations better. 
The $\omega^{(1)}$ of the scalar-led perturbations shows a similar tendency, but in the interest of space, we omit the scalar-led $\omega^{(1)}$ from the figure. 
}
\label{fig:a_0.1}
\end{figure*}

Formally, $H_i (r, \chi)$, $\vartheta(r,\chi)$, $\Omega^{(1)}$ and $\kappa^{(1)}$ are an infinite series in $a$. 
In practice, it is neither possible nor necessary to include $H_i (r, \chi)$ and $\vartheta(r,\chi)$ to infinite order for actual computations that are sought to a desired precision. 
Instead, one only needs to truncate these quantities at a given order in $a$, such that the inclusion of higher-order terms do not affect the calculation of our frequencies beyond a desired precision. 
Through this work, we will truncate $H_i (r, \chi)$, $\vartheta(r,\chi)$, their derivatives,
$\Omega^{(1)}$, and $\kappa^{(1)}$ at the same order in $a$, 
\begin{equation}\label{eq:H_fns_truncated}
\begin{split}
H_i (r, \chi) & \approx \sum_{k=0}^{N_{\rm a}} \sum_{p=0}^{N_{r}(k)} \sum_{q=0}^{N_{\chi}(k)} h_{i, k, p, q} \frac{a^{k} \chi^{q} }{r^p}, \\
\vartheta (r, \chi) & \approx  \sum_{k=0}^{N_{\rm a}} \sum_{p=0}^{N_{r}(k)} \sum_{q=0}^{N_{\chi}(k)} \vartheta_{i, k, p, q} \frac{a^{k} \chi^{q} }{r^p}, 
\end{split}
\end{equation}
where $N_{\rm a}$ is the truncation order in $a$ (yet to be specified). 
We have checked that the $H_i (r, \chi)$ and $\vartheta(r, \chi)$ used here indeed satisfy the field equations~\eqref{eq:field_eqs} up to the $N_{\rm a}$th order of $a$ by substituting directly the corrected metric into the field equations. 

\subsection{Validation Example: the 022 mode at $a=0.1$}
\label{sec:022_a_01}

\begin{figure*}[htp!]
\centering  
\subfloat{\includegraphics[width=0.47\linewidth]{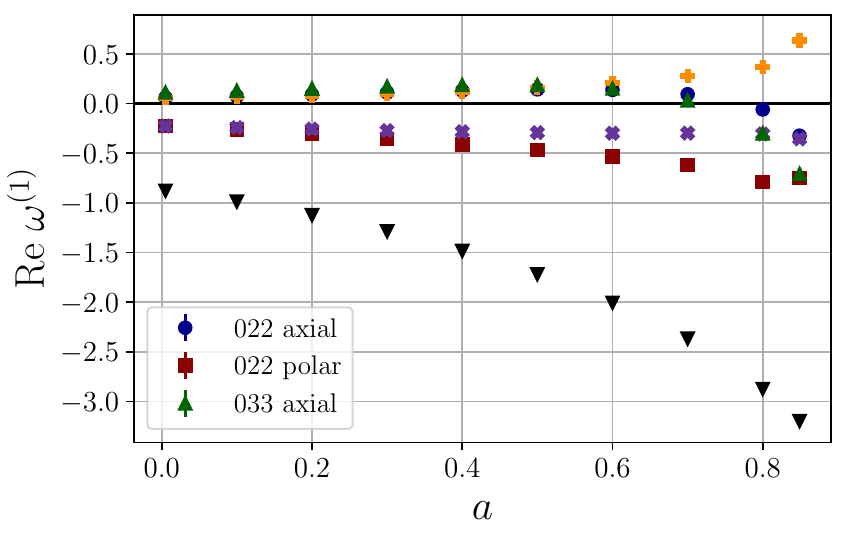}}
\subfloat{\includegraphics[width=0.47\linewidth]{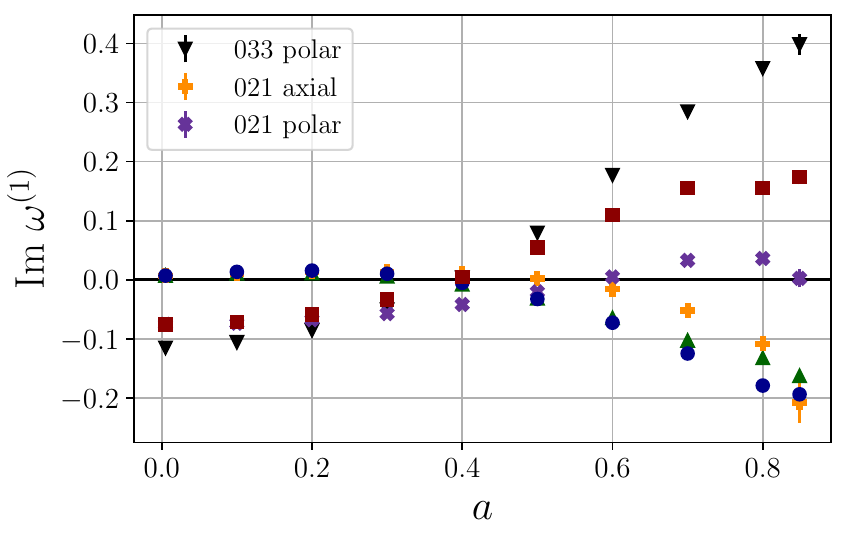}}
\caption{The real (left) and imaginary (right) parts of $\omega^{(1)}$  of the axial (blue dots) and polar (red triangles) metric perturbations of the $nlm = 022$, 033 and 021 modes as a function of dimensionless spin.  
We observe that $\omega^{(1)}$ of the axial and polar perturbations does not coincide, which indicates that the isospectrality is broken in sGB gravity. 
Moreover, we observe that $|\omega^{(1)}|$ is smaller for the axial perturbations compared to that of the polar perturbations, and for $a \lesssim 0.3 $, $\text{Re}\omega^{(1)} $ and $\text{Im}\omega^{(1)}$ for the polar perturbations. 
These two features are consistent with the previous studies regarding slowly rotating BHs in sGB gravity. 
Finally, we observe that, although we have included the numerical uncertainty (defined by Eq.~\eqref{eq:optimal_spectral_order}) of our frequency computations as error bars, the error bars for $a \leq 0.8$ are too small to be seen in the figure and are visualized in Fig.~\ref{fig:delta}. 
The small uncertainties indicate that the $\omega^{(1)}$ we computed is accurate. 
}
\label{fig:omega_1_022}
\end{figure*}

We start by studying the 022 mode of sGB BHs with $a=0.1$, whose results already reflect the general QNM features that we will observe for higher $a$ BHs. 
The blue, red and green symbols in Fig.~\ref{fig:a_0.1} represent the base-10 logarithm of the backward modulus difference of $\omega^{(1)}$ of the axial (left panel) and polar perturbations (right panel), 
\begin{equation}\label{eq:BWD}
\mathcal{B} (N) = |\omega^{(1)}(N) - \omega^{(1)}(N-1)|, 
\end{equation}
as a function of the spectral order $N$, for $N_{\rm a}=2$, $4$ and $6$, respectively.
Observe that, for both parities, $ \mathcal{B}(N)$ first decreases approximately exponentially, until reaching a certain spectral order, at which point $ \mathcal{B}(N)$ fluctuates around an approximate constant. 

The saturation of the backward modulus difference is related to the fact that the background metric satisfies the field equations only up to a chosen $N_{\rm a}$-th order in $a$. 
At small $N$ spectral order, the error in the corrected metric is not well resolved, since the error induced by the truncation of the spectral expansion dominates.
But as the spectral order is gradually increased, the error due to the $N_{\rm a}$ truncation becomes more important and begins to affect the computation. 
From Fig.~\ref{fig:a_0.1}, observe that $\mathcal{B} (N)$ decreases exponentially to a significantly smaller value and begins to fluctuate around a constant value at a larger spectral order as $N_{\rm a}$ is increased. 

Observe that, at relatively small spectral order (e.g., for $N \leq 8$ for the axial perturbations and $N \leq 11$ for the polar perturbations), the backward modulus difference is approximately independent of $N_{\rm a}$. 
This is because, at these small spectral orders, the $\omega^{(1)}$ computed with any $N_{\rm a}$ is the same. 
This feature indicates that increasing $N_{\rm a}$ can help improve the ``convergence'' of $\omega^{(1)}$ with respect to the spin expansion. 

The results shown in Fig.~\ref{fig:a_0.1} also offer a guideline to choose $N_{\rm a}$ such that the resulting $\omega^{(1)}$ is of the desired accuracy. 
As estimated in \cite{Maselli:2023khq}, the relative measurement uncertainty of the real and imaginary parts of the QNM frequencies combined across $\mathcal{O}(10^3)$ ringdown signals detected by the next-generation detectors is $\mathcal{O}(10^{-4})$. 
Since $\zeta$ must be small [i.e.~smaller than $\mathcal{O}(0.1)$], we should ensure the uncertainty in $\omega^{(1)}$ is smaller than $\mathcal{O}(10^{-3})$. 
Figure~\ref{fig:a_0.1} suggests that to achieve such an accuracy, we need to select $N_{\rm a}$ such that $a^{N_{\rm a}} \leq 10^{-2} $ at least. This implies that if $a = 0.1$ then $N_{\rm a} > 2$, but if $a=0.3$ then $N_{\rm a}>4$, and if $a = 0.7$ then $N_{\rm a} > 13$.  
Finally, the exponential convergence of $\omega^{(1)}$ also justifies the validity of the asymptotic behavior derived in Sec.~\ref{sec:ansatz_summary} and our choice of $\rho_H^{(k)}$ and $\rho_{\infty}^{(k)}$. 
If any of these details were not correct, $\omega^{(1)}$ would not converge exponentially. 

\subsection{More rapidly rotating BHs}
\label{sec:Rotating_BHs_results}

We apply METRICS to compute the QNM frequencies of more rapidly rotating BHs in sGB gravity. 
For the purpose of the computations, we have recalculated the sGB metric corrections to 40th order in $a$, and this is the BH background we employ in our calculations. 
For $a \leq 0.7$, we include metric modifications up to $N_{\rm a}$th order in $a$, such that $N_{\rm a}$ is the least even integer that satisfies $a^{N_{\rm a}} < 10^{-4}$ (we chose an even $N_{\rm a}$ order because the $H_i (r, \chi)$ functions are a power series in $a^2$).
We will terminate our computations at $a=0.85$ because $(0.85)^{40} \sim 1.5 \times 10^{-3}$, smaller than $10^{-2}$, a necessary criterion that we noted in Sec.~\ref{sec:022_a_01}. If one wishes to consider BHs that are more rapidly spinning, one would need to use a background metric to higher than 40th order in a small-spin expansion (or an appropriate resummation or numerical background), which is not impossible but is computationally expensive.
The QNM frequency for $a > 0.85$ will be extrapolated by constructing a fitting expression for the QNM frequency of $a \leq 0.85 $ (see Eq.~\eqref{eq:omega_1_fitted}). 
Nevertheless, we expect our QNM frequencies are still useful for the analyses of the detected ringdown signals because the spin of many remnant BHs is $< 0.85$ \cite{LIGOScientific:2016vlm, LIGO_07, LIGO_11}. 

At a given $a$ and $N_{\rm a}$, we compute $\omega^{(1)}$ from $N=1$ to 25. 
We stop at 25 because we find that the backward modulus difference is usually saturated or minimized from $N = 20$ to $N=25$. 
Then, we select the optimal spectral order, defined as
\begin{equation}\label{eq:optimal_spectral_order}
N_{\rm opt} = \text{arg}\min_{N} \mathcal{B}(N), 
\end{equation}
which allows us to define $\omega^{(1)}(N_{\rm opt})$ as the QNM frequency perturbation and 
\begin{equation}\label{eq:numerical_uncertainty}
\delta = \mathcal{B}(N_{\rm opt})
\end{equation}
as the numerical uncertainty in $\omega^{(1)}$.

\begin{figure}[htp!]
\centering  
\subfloat{\includegraphics[width=\columnwidth]{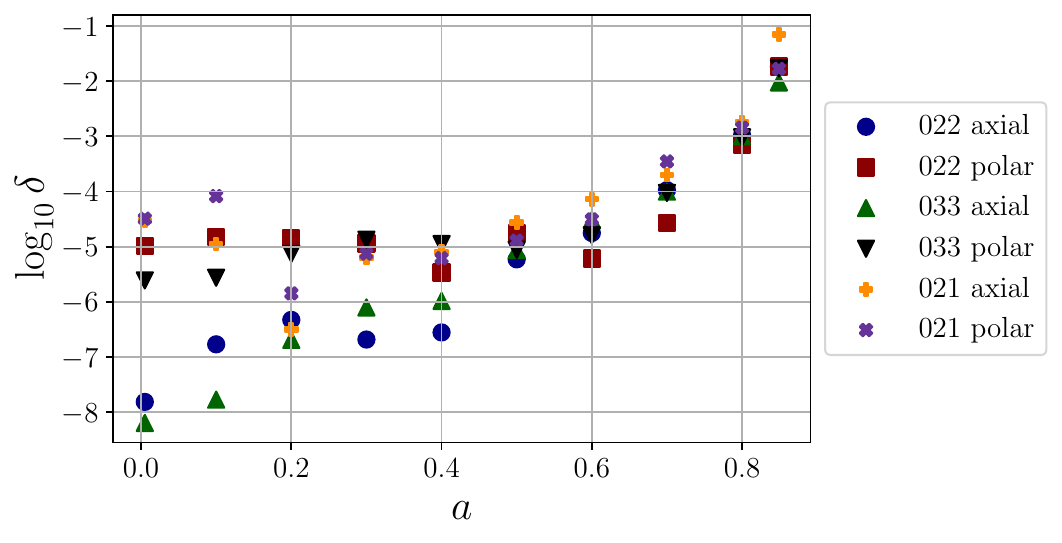}}
\caption{
The base-10 logarithms of the numerical uncertainty, the minimum backward modulus difference of $\omega^{(1)}$ for $N \leq 25$. 
We observe that the uncertainty increases with $a$ in general. 
This is reasonable because the background metric that we constructed only satisfies the field equations to a finite order of dimensionless spin. 
The error of the background increases with the dimensionless spin even we have adjusted the spin-truncation order of the background metric which we keep for QNM frequency computations. 
Nonetheless, for $a \leq 0.7 $, we can keep the numerical uncertainty of all modes below $\sim 10^{-4}$. 
Thus, the $\omega^{(1)}$ that we computed using METRICS is accurate enough (see Sec.~\ref{sec:022_a_01} for the definition of "desired" accuracy) to be applied to analyze astrophysical ringdown signals. 
}
\label{fig:delta}
\end{figure}

Figure~\ref{fig:omega_1_022} shows the real and imaginary parts of $\omega^{(1)}$ of the 022-, 033- and 021-mode frequency of the axial and polar perturbations, together with their uncertainties (which, for the most part, cannot be seen in the figure). 
Tables~\ref{tab:omega_1_022},~\ref{tab:omega_1_033} and~\ref{tab:omega_1_021} of Appendix~\ref{sec:Appendix_B} in show the numerical value of real and imaginary parts of $\omega^{(1)}$ of the 022-, 033-, and 021-mode frequency of both parities. 
From the figure, we immediately observe that  isospectrality is broken, which is expected in modified gravity theories in general \cite{QNM_EdGB_01, QNM_EdGB_02, QNM_EdGB_03, QNM_dCS_01, QNM_dCS_02, QNM_dCS_03, QNM_dCS_04, isospectrality-paper}. 
Explicitly, we observe that the frequencies of the polar perturbations change more significantly compared to that of the axial perturbations.  
This is reasonable because the axial perturbations couple more weakly to sGB terms \cite{Blazquez-Salcedo:2016enn, Pierini:2021jxd, Pierini:2022eim} (also see Sec.~\ref{sec:SF_role}). 
We also observe that, for slowly rotating BHs (e.g. $a \lesssim 0.3$), the real and imaginary parts of the polar 022-mode frequency both decrease as sGB gravity comes into effect, i.e.~$\text{Re}(\omega^{(1)}) < 0 > \text{Im}(\omega^{(1)})$ when $a \lesssim 0.3$ for the polar $022$ mode. 
Both of these features are consistent with previous studies of the QNM frequencies of slowly rotating BHs in sGB gravity  \cite{Blazquez-Salcedo:2016enn, Pierini:2021jxd, Pierini:2022eim}\footnote{We will quantitatively show that our frequencies are consistent with those computed in \cite{Blazquez-Salcedo:2016enn, Pierini:2021jxd, Pierini:2022eim} in Sec.~\ref{sec:Reslts_comparison}, after constructing a fitting polynomial of our frequencies [see Eq.~\eqref{eq:omega_1_fitted}].}.
Apart from the broken isospectrality, observe that the numerical uncertainty (defined by the minimal backward modulus difference across a range of spectral orders) is included in Fig.~\ref{fig:omega_1_022} as error bars. 
Nonetheless, for $a \leq 0.8 $, the error bars are too small to be visualized while keeping the symbols a reasonable size. 
This indicates that the numerical uncertainty of $\omega^{(1)}$ is tiny.

Figure~\ref{fig:delta} plots the numerical uncertainty as a function of $a$, which clearly shows that the uncertainty, in general, increases with $a$.
Observe also that the uncertainty of the axial and polar frequencies is different, which is because the $\omega^{(1)}$ of the axial and polar modes spans different numerical ranges.
The growth of the uncertainty with $a$ is reasonable because the background metric that we constructed satisfies the field equations only to an $N_{\rm a}$th order in $a$, which means that the error of our background metric will at least be of $(N_{\rm a}+1)$th order. 
However, $a^{N_{\rm a}+1}$ still grows with $a$ even with $N_{\rm a}$ adjusted such that $a^{N_{\rm a}} < 10^{-4}$. 
For example, at $a = 0.005$, $N_{\rm a} = 4$ and an order-of-magnitude estimate of the error is $a^{N_{\rm a}+1} \sim 10^{-10}$, whereas at $a=0.85$, $N_{\rm a} = 40$, and $a^{N_{\rm a}+1} \sim 10^{-3}$.  
These levels of uncertainty in $\omega^{(1)}$ are well within the desired accuracy defined in Sec.~\ref{sec:022_a_01} for the analysis of astrophysical ringdown signals. 
Moreover, the numerical uncertainty for the relatively large $a$ cases can always be reduced if the metric corrections are kept to a higher order in $a$.

\begin{figure*}[htp!]
\centering  
\subfloat{\includegraphics[width=0.47\linewidth]{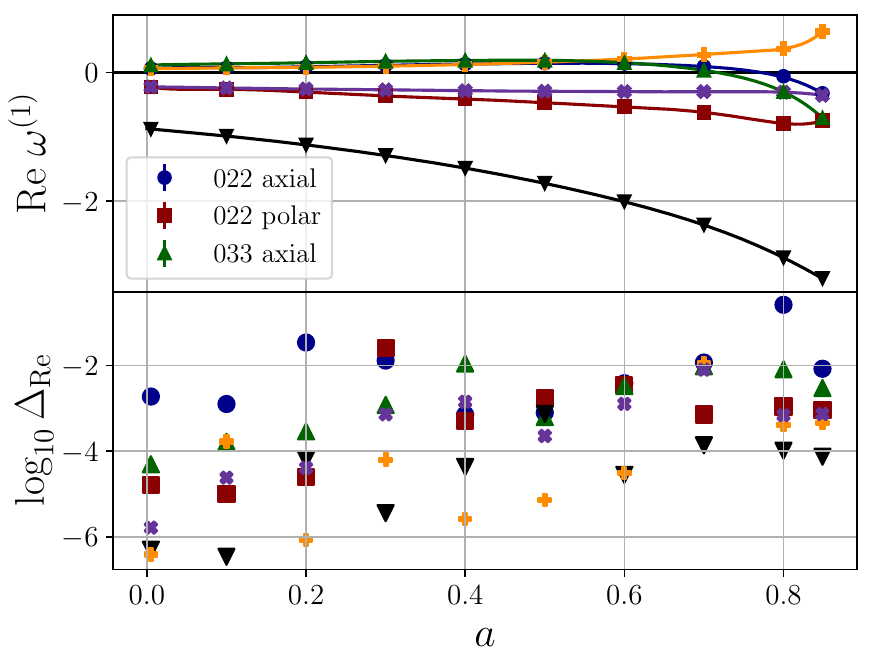}}
\subfloat{\includegraphics[width=0.47\linewidth]{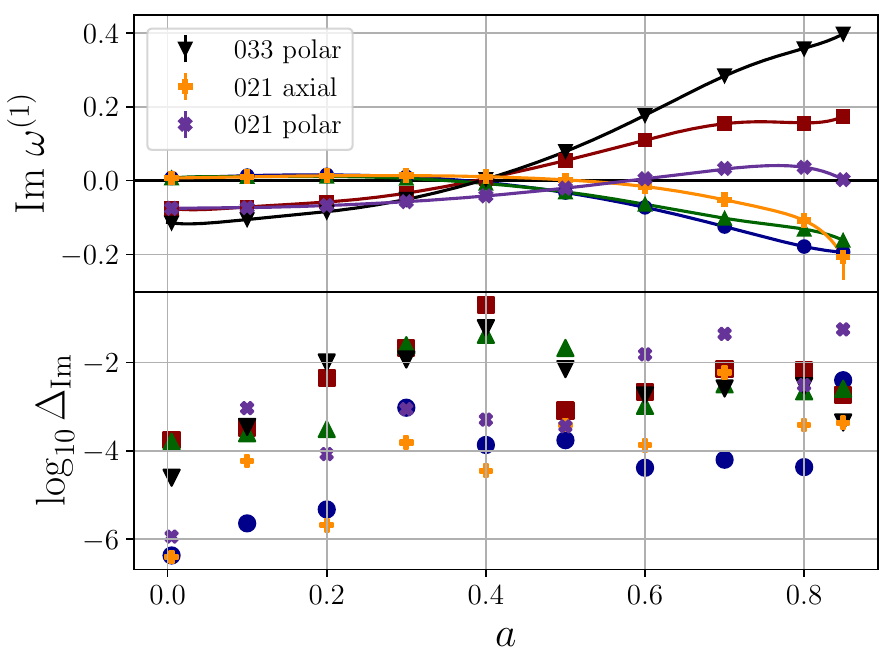}}
\caption{Real (top left) and imaginary (top right) parts of $\omega^{(1)}$ for the 022, 033 and 021 modes of both parities, computed using METRICS (symbols) and using the optimal degree-8 fitting polynomials (curves), as a function of $a \in [0, 0.85]$. 
The relative fractional errors of the real (bottom left) and imaginary (bottom left) parts of $\omega^{(1)}$, computed using the fitting polynomials and the METRICS data, show the accuracy of the fit. 
Observe that the optimal fitting polynomials are excellent approximations to the METRICS $\omega^{(1)}$, with an error of at most $\sim 10 \%$. 
}
\label{fig:fitting_polynomials_check}
\end{figure*}

Using the METRICS frequencies we computed at different spins, we now fit their real and imaginary parts of all modes except the 021 axial mode with a degree-8 polynomial, 
\begin{equation}\label{eq:omega_1_fitted}
\begin{split}
\omega^{(1)} & = \sum_{j=0}^{8} w_{j} a^j\,,
\end{split}
\end{equation}
where in this fit we have set $M=1$ and the $w_j$ are complex constants. 
We fit these modes as a degree-8 polynomial because we find this degree minimizes the fitting loss functions. 
As for the frequency of the 021 axial mode, we fit its real and imaginary parts as a degree 14 polynomial because we observed unphysical oscillations for small $a$ if we fit the frequency with a polynomial of a smaller degree. 
The numerical value of $w_j$ is obtained using the built-in \texttt{NonLinearModelFit} function in \textit{Mathematica}, with numerical uncertainty included as the error of the real and imaginary parts of $\omega^{(1)}$. 
The explicit numerical value and uncertainty of $w_j$ are given in Tables~\ref{tab:poly_fit_coeffs} and \ref{tab:poly_fit_coeffs_uncertainty} respectively. 
For the convenience of the reader, below we provide the fitting polynomials truncated at $a^4$: 
\begin{widetext}
\begin{align}
\omega_{022,\rm A} = & (0.055241 + 0.00684399i) + (0.985294+0.0688128i) a 
+ (- 18.4902 + 0.207235i) a^2 
\nonumber \\
&+ (157.802- 3.3872i) a^3 
+ (- 682.224 + 14.8323i) a^4 + \ldots\,, \\
\omega_{022,\rm P} = & (-0.215202 -0.0734094i) 
+(- 2.51816 - 0.411031i)a 
+ (48.4659 + 9.59898i)a^2 
\nonumber \\
& +(- 431.428 - 82.4765i)a^3 
+(1935.01 + 378.832i)a^4 + \ldots\,, \\
\label{eq:METRICS_fitting_polynomials}
\omega_{033,\rm A} = & (0.109706 + 0.0061152i) 
+ (0.685561+0.231546i) a 
+ (- 11.6733 - 4.20702i) a^2 
\nonumber \\
& + (108.698 + 37.6761i) a^3 
+ (-510.61- 178.723i) a^4 + \ldots\,, 
\\
\omega_{033,\rm P} = & (-0.872789 -0.113506i) 
+ (- 1.20198- 0.339974i) a 
+ (2.82484 + 9.09087i) a^2 
\nonumber \\
& + (- 35.6024 - 75.7708i) a^3
+ (165.435 + 341.913i) a^4 + \ldots\,, \\
\omega_{021,\rm A} = & (0.0665864 + 0.00342921i) 
+ (- 1.59526+0.789588i) a 
+ (38.1384 - 16.7089i) a^2 
\nonumber \\
& + (- 347.625 + 147.414i) a^3 
+ (1622.8 - 663.824i) a^4 + \ldots\,, \\
\omega_{021,\rm P} = & (-0.22612 -0.0754879i) 
+ (0.0933968 + 0.0504285i)a 
+(-5.91521 - 0.968127i)a^2 
\nonumber \\
& +(54.1801 + 10.6838i)a^3 
+(-252.83 - 51.1785i)a^4 + \ldots\,,
\end{align}
\end{widetext}

To check that the fitting polynomials obtained can accurately estimate the $\omega^{(1)}$ computed using METRICS, we compute $\omega^{(1)}$ using the fitting polynomials and compare it with the $\omega^{(1)}$ using METRICS. The top panels of Fig.~\ref{fig:fitting_polynomials_check} show the real (left) and imaginary (right) parts of $\omega^{(1)}$ for the 022, 033 and 021 modes of both parities, computed using the degree-8 fitting polynomials and as a function of $a \in [0, 0.85]$. 
By visual inspection, the fitting polynomials pass through the $\omega^{(1)}$ data computed with METRICS perfectly, indicating that these polynomials can indeed accurately compute $\omega^{(1)}$. 
To further qualitatively gauge the accuracy of the fitting polynomials, the bottom panels of Fig.~\ref{fig:fitting_polynomials_check} show the relative fractional errors of the real (bottom, left) and imaginary (bottom, right) parts of $\omega^{(1)}$, computed using the fitting polynomials and the METRICS data. 
Observe that the relative fractional error is at most $\sim 10 \%$, consistent with our observation that the fitting polynomials accurately estimate the METRICS $\omega^{(1)}$.
The QNM frequencies could also be fitted using other functional forms, such as a rational function in $a$ through a Pad\'e function. 
However, we experimented with this idea and found that fitting the frequencies with a polynomial gives the most accurate estimate of the QNM frequencies at the computed $a$. This result makes sense since the $\omega^{(1)}$ we presented in Fig.~\ref{fig:omega_1_022} clearly show no poles, and thus, a Pad\'e resummation is not guaranteed to increase accuracy. 

\subsection{Comparisons with the previous results in slow-rotation expansions}
\label{sec:Reslts_comparison}

Fitting the QNM frequencies as a polynomial in $a$ also allows us to compare our results to those in \cite{Pierini:2022eim}. 
In \cite{Pierini:2022eim}, $f(\Phi) = \exp(\Phi)$, but in the small-coupling limit, $\Phi \ll 1 $, we can  approximate the coupling function well as $f(\Phi) \approx 1 + \Phi $, which reduces to the coupling function that we considered in this work, as we have explained in the Introduction. 
Hence, we can compare the $w_j$ of our fitting polynomials with the fitting coefficients of the leading-coupling terms in \cite{Pierini:2022eim}. 
By reading Eqs.~(40) and (47) and Tables I--III of \cite{Pierini:2022eim}, we find that the leading $\zeta$-order modifications to the polar-mode frequencies in \cite{Pierini:2022eim} are given by \footnote{Note that our $\alpha$ is $\alpha/4$ in \cite{Pierini:2022eim}. 
Hence, the $\zeta$ of our paper is $16 \zeta^2 $ in \cite{Pierini:2022eim}. }
\begin{align}
\omega^{(1), \rm PG}_{022\rm P} & = (-0.22496 -0.0752i) \nonumber \\
& \quad + (-0.33536+0.00064i)a \nonumber \\
& \quad + (-0.32+0.30432i)a^2, \nonumber \\
\label{eq:freq_Gultierai}
\omega^{(1), \rm PG}_{033\rm P} & = (-0.87248 -0.11504i) \nonumber \\
& \quad + (-1.03488-0.05664i)a \\
& \quad + (-0.96224+0.39792i)a^2, \nonumber \\
\omega^{(1), \rm PG}_{021\rm P} & = (-0.22496 -0.0752i) \nonumber \\
& \quad + (-0.16768+0.00032i)a \nonumber \\
& \quad + (0.08176+0.15408i)a^2. \nonumber \\
\nonumber 
\end{align}
By comparing the above expressions and Eq.~\eqref{eq:METRICS_fitting_polynomials}, 
we find that the $w_{j=0,1,2}$ of our fitting polynomials are close to the coefficient shown above. 
We identify several possible causes for the small differences, including but not limited to the following.  
First, the order in dimensionless spin and $\zeta$ that is used in the two studies is different. 
In this work, at $a=0.1$, we use $N_{\rm a} \geq 4$, whereas in \cite{Pierini:2022eim} $N_{\rm a} = 2$ throughout their computations. Moreover, in this work we compute only the leading $\zeta$ modifications to the frequencies, whereas terms of higher degree in $\zeta$ are also included in the fitting expression of \cite{Pierini:2022eim}. 
Second, the two studies used different numerical methods to compute the QNM frequencies. 
We compute the QNM frequencies using METRICS, whereas in \cite{Pierini:2022eim} the shooting method is used, which is can be sensitive to the accuracy of the boundary conditions and the shooting point. 
Third, in this work $f(\Phi) = \Phi$, whereas in \cite{Pierini:2022eim} $f(\Phi) = \exp(\Phi)$. 
When applying the shooting method, there might be regions where $\Phi$ is beyond the small-coupling limit, rendering the comparison not entirely fair. 
All these discrepancies can lead to differences between the QNM frequencies in the two studies.  

\begin{figure}[tp!]
\centering  
\subfloat{\includegraphics[width=\columnwidth]{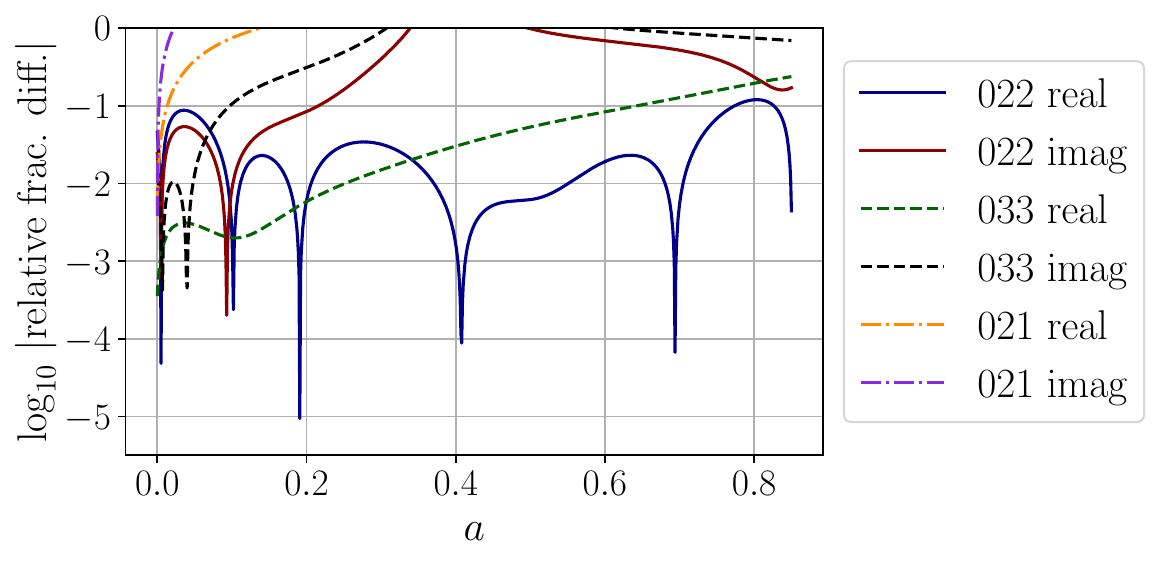}}
\caption{The relative fractional difference [defined by Eq.~\eqref{eq:results_comparison}] between fitting polynomials of the real and imaginary parts of the polar frequency of the 022 (solid), 033 (dashed) and 021 (dash-dotted) modes obtained using METRICS and that in \cite{Pierini:2022eim} through the small-spin expansions. 
}
\label{fig:Comparison}
\end{figure}

In spite of all these possible discrepancies, our fitting polynomials are actually consistent with the results of \cite{Pierini:2022eim}, allowing us therefore to estimate the spin at which the second-order-in-spin approximation becomes highly inaccurate. 
Let us then compare the frequencies computed using Eq.~\eqref{eq:freq_Gultierai} and that computed using our fitting polynomials by computing the relative fractional differences between them, 
\begin{equation}\label{eq:results_comparison}
\text{relative frac. diff.} = \frac{\omega^{(1)}_{\rm Re / Im}(a) - \omega^{(1), \rm PG}_{\rm Re / Im}(a)}{\omega^{(1)}_{\rm Re / Im}(a)}.
\end{equation}
Figure~\ref{fig:Comparison} shows the base-10 logarithms of the absolute value of the relative fractional differences of the real and imaginary parts of the polar frequency of the polar 022 (solid lines), 021 (dashed lines) and 033 (dash-dotted lines) modes. 
This comparison indicates that our polar frequencies agree well with those computed in \cite{Pierini:2022eim} using the small-spin expansions for small spins. However, the imaginary part of the dominant mode (022) computed in \cite{Pierini:2022eim} presents an error larger than 30 \% relative to the METRICS frequencies above $a = 0.3$; for other modes [like the (021) mode], the small spin expansion becomes inaccurate for much small spins. As $a$ increases, the frequencies computed using the small-spin expansions become increasingly inaccurate, and the METRICS frequencies should be used instead. 

\subsection{Parity content of metric perturbations}
\label{sec:PD}

One great advantage of METRICS is its convenience in reconstructing metric perturbations, which requires only the reading and rearrangement of the elements of the eigenvector. 
This advantage allows us to examine the parity content of the metric perturbations, which can be put to good use to check the sanity of our calculations. 
Since we compute the modification to the QNM frequency of the axial and polar perturbations separately, we expect that the corresponding $\textbf{v}^{(1)}$ should also be solely axial and polar for consistency. 

To quantify the parity content of the metric perturbations, we define the parity dominance (PD) in the following way. 
\begin{equation}\label{eq:PD_01}
\text{PD} = \frac{\rm Amp(P)}{\rm Amp(A)}, 
\end{equation}
where 
\begin{align}\label{eq:PD_02}
\left[\text{Amp(P)}\right]^2 &= \displaystyle \sum_{k=1}^{4} \sum_{n=0}^{N} \sum_{\ell = \rm even} \frac{(\ell+m)!}{(2\ell+1)(\ell-m)!} |v_k^{(1), n \ell}(N)|^2 
\nonumber \\
&+ \sum_{k=5,6} \sum_{n=0}^{N} \sum_{\ell = \rm odd} \frac{(\ell+m)!}{(2\ell+1)(\ell-m)!} |v_k^{(1), n \ell}(N)|^2, \\
\left[\text{Amp(A)}\right]^2 & = \displaystyle \sum_{k=5, 6} \sum_{n=0}^{N} \sum_{\ell = \rm even} \frac{(\ell+m)!}{(2\ell+1)(\ell-m)!} |v_k^{(1), n \ell}(N)|^2
\nonumber \\
&+\sum_{k=1}^{4} \sum_{n=0}^{N} \sum_{\ell = \rm even} \frac{(\ell+m)!}{(2\ell+1)(\ell-m)!} |v_k^{(1), n \ell}(N)|^2. 
\end{align}
Here $\sum_{\ell = \rm even}$ and $\sum_{\ell = \rm odd}$, respectively, stand for the sum of over all $\ell$ even and odd integers between $|m|$ and $N+|m|$. 
Heuristically, the PD gives a rough estimate of the ratio between the amplitude of the polar perturbations to that of the axial ones. 
Note that unlike the PD which we defined in \cite{Chung:2023wkd}, when estimating the amplitude of the polar perturbations (i.e. the numerator), we do not only take $u_{i=1,...,4}$, but also the $u_{i=5,6}$, in accordance with the definition of the axial and the polar parity that we developed in this paper. 
Recall from Sec.~\ref{sec:metpert} that the polar perturbations satisfy $ \hat{P} h^{\rm (P)}_{\mu \nu} = (-1)^{l} h^{\rm (P)}_{\mu \nu} $, where $\hat{P}$ is the parity operator. 
Since we are using associated Legendre polynomials as our angular spectral basis, which obey Eq.~\eqref{eq:Pellm_parity} upon parity transformation, $h_{i=1, ..., 4}$ and $h_{i=5, 6}$ will take turns being of polar and axial type.  
For example, for the $l = 2$ modes, when $\ell = 2$, $h_{i=1, ..., 4}$ are polar and $h_{i=5, 6}$ are axial; however, when $\ell = 3$, because of the parity of the associated Legendre polynomials (see Eq.~\eqref{eq:Pellm_parity}), $h_{i=1, ..., 4}$ are axial and $h_{i=5, 6}$ are polar. 

\begin{figure}[tp!]
\centering  
\subfloat{\includegraphics[width=\columnwidth]{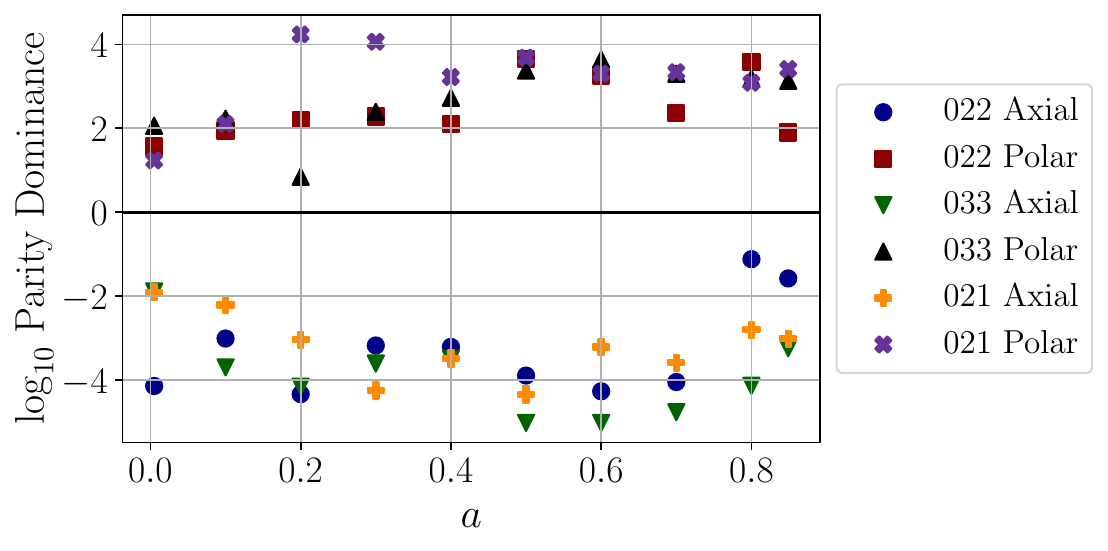}}
\caption{
The base-10 logarithm of the parity dominance (see Eq.~\eqref{eq:PD_02}) of the modification to the axial and polar perturbations as a function of dimensionless spin $(a)$. 
PD gives a rough estimate of the ratio between the amplitude of the polar perturbations to that of the axial ones. 
PD $\gg 1$ indicates that the perturbations are dominantly polar, and PD $\ll 1$ indicates that the perturbations are dominantly axial. 
We observe that the PD is much larger than 1 (marked by the solid horizontal line in black) for the polar perturbations and much smaller than 1 for the axial perturbations. 
This feature indicates that the $\textbf{v}^{(1)}$ (defined by Eq.~\eqref{eq:eigenperturbation_02}) of the polar perturbations remain dominantly polar; and that of the axial perturbations remain mostly axial. 
}
 \label{fig:PD}
\end{figure}

With this definition of PD, the PD of purely polar perturbations is $\infty$, and that of axial perturbations is $0$. 
However, due to numerical error, the PD of $\textbf{v}^{(1)}$ which we obtain will never be $\infty$ or 0. 
Thus, a $\rm PD \gg 1 $ indicates that the perturbations are dominantly polar, while $\rm PD \ll 1$ indicates that the perturbations are dominantly axial. 
Figure~\ref{fig:PD} shows the PD of the axial and polar perturbations as a function of $a$. 
We see that for all $a$ and the QNMs for which we compute the QNM frequencies, PD $\gtrapprox 10^2 $ for the polar perturbations and $\rm PD \lesssim 10^{-3} $ otherwise, despite fluctuations. 
This indicates that $\textbf{v}^{(1)}$ of the polar perturbations remains mostly polar and that of the axial perturbations remains mostly axial for all $a$ explored.

The astute observer will notice that the parity dominance for some modes does not separate as cleanly as for other modes. For example, at a spin of $a = 0.2$, the 021 mode has a parity dominance of $\sim 10^4$ for the polar mode and $\sim 10^{-3}$ for the axial mode; however, the 033 mode has a dominance of $\sim 10$ for the polar mode and $\sim 10^{-4}$ for the axial mode. The reason that the polar mode has such a low parity dominance is a technical one. As explained in Sec.~\ref{sec:eigenvalue_pert}, when solving the linearized field equations for the modified gravity corrections, one must first solve the linearized Einstein equations. In this paper, when doing so, we did not optimize the solution to the linearized Einstein equations to maximize its accuracy; instead, we were content with GR frequencies that are accurate to better than $10^5$ relative to continuous fraction solutions to the Teukolsky equation. Skipping this optimization step (which can be remedied in the future, if desired) means that certain GR modes will be computed more accurately than others, with the 033 one being only accurate to $10^{-5}$ while others are accurate to $10^{-8}$. This difference in accuracy in the calculation of the GR modes contaminates the accuracy of the sGB corrections, leading to a lower parity dominance for some modes. Having said that, we have ensured that the inaccuracies in the GR modes are not contaminating the sGB corrections to the complex frequencies beyond what was quoted in this paper.  

\subsection{Understanding the role of the scalar field $\vartheta$}
\label{sec:SF_role}

In Sec.~\ref{sec:Rotating_BHs_results}, we observed that  $|\omega^{(1)}|$ is larger for the polar perturbations, and we attributed this feature to the fact that axial perturbations couple more weakly to the sGB terms. 
One way to justify this claim is to examine the amplitude of $\vartheta$, which can be directly read from the elements of $\textbf{v}^{(1)}$. 
Inspecting Eqs.~\eqref{eq:field_eqs} and~\eqref{eq:Amunu_sGB}, we realize that the linearized field equations contain no terms that involve the product of $\mathscr{G}$ and $h_{\mu \nu}$ (and their derivatives) in the small and shift-symmetric coupling.
Thus, $h_{\mu \nu}$ and their derivatives can only interact with $\mathscr{G}$ through $\vartheta$, which can be obtained by solving $\square \vartheta + \mathscr{A}_{\vartheta} = 0$, which nothing but $\square \vartheta + \mathscr{G} = 0$ for the shift-symmetric case. 
Moreover, $\mathscr{G}^{(1)}$ is dominated by the polar parity terms\footnote{Since the explicit expression of $\mathscr{G}^{(1)}$ is lengthy and not insightful, we include it in a \textit{Mathematica} notebook that is available upon reasonable request.}. 
Hence, the amplitude of $\vartheta$ indicates the strength of the interplay between the metric perturbations and the sGB terms in the field equations. 

Let us quantify the strength of $\vartheta$ explicitly by defining the scalar-field abundance (i.e.~the $\vartheta$ abundance) as the ratio between the $L^2$ norm of $\vartheta$ and the amplitude of the axial or polar perturbations. 
For the modes of even $l+|m|$, the $\vartheta$ abundance for the axial/polar perturbations is then defined as 
\begin{equation}\label{eq:SFA}
\begin{split}
& \vartheta~\text{abundance(A/P)} = \frac{\text{Amp}(\vartheta)}{\rm Amp(A/P)}, \\ 
\end{split}
\end{equation}
where 
\begin{equation}
\begin{split}
\text{Amp}(\vartheta) = \sum_{n=0}^{N} \sum_{\ell = |m|}^{N+|m|} \frac{(\ell+m)!}{(2\ell+1)(\ell-m)!} |v_7^{(1), n \ell}(N)|^2. 
\end{split}
\end{equation}
We divide $\text{Amp}(\vartheta)$ by Amp(A) or Amp(P) before comparison because the amplitude of the axial and polar perturbations may at first be different. 
Such a division can make the comparison more fair and informative. 

\begin{figure}[tp!]
\centering  
\subfloat{\includegraphics[width=\columnwidth]{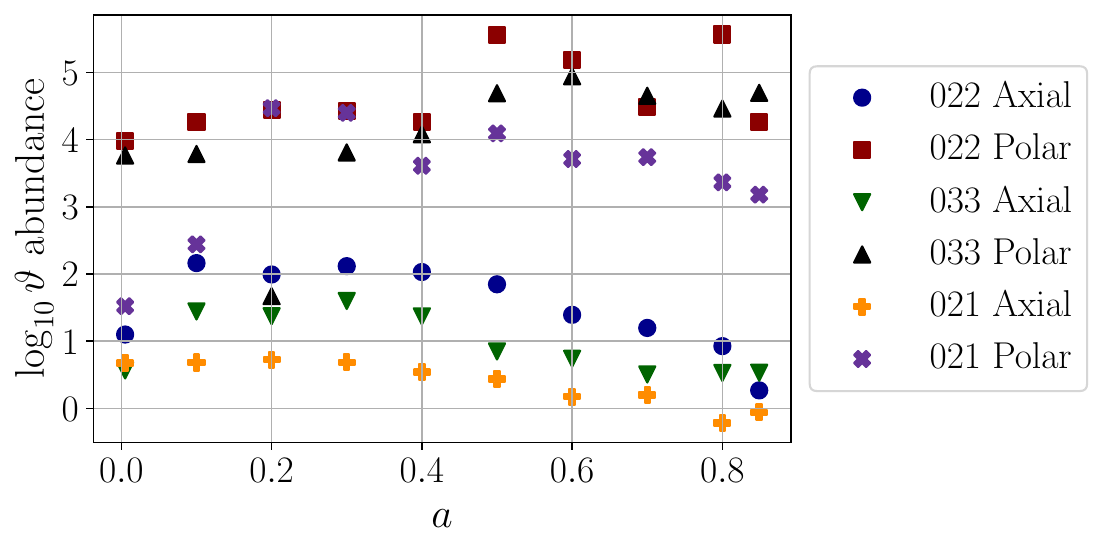}}
\caption{
The $\vartheta$ abundance (defined by Eq.~\eqref{eq:SFA}) of the axial and polar perturbations of different QNMs as a function of $a$. 
Heuristically, the $\vartheta$ abundance is the ratio between the amplitude of the scalar-field perturbations ($\vartheta^{(1)}$) and that of the axial or polar metric perturbations. 
We observed that the $\vartheta$ abundance for the polar perturbations is about 2 orders of magnitude stronger than that for the axial perturbations, indicating that the polar perturbations of different QNMs are always accompanied by a larger scalar field. 
In other words, the sGB terms exert stronger effects on the polar perturbations than on the axial perturbations, explaining why $|\omega^{(1)}|$ of the polar perturbations is larger (see Fig.~\ref{fig:omega_1_022}). 
}
\label{fig:SFA}
\end{figure}

Figure~\ref{fig:SFA} shows the $\vartheta$ abundance of the axial and polar perturbations of different QNMs as a function of $a$. 
Observe that the abundance of the polar perturbations is at least about 2 orders of magnitude larger than that of the axial perturbations, which indicates that the polar perturbations are accompanied by a larger $\vartheta$. 
In other words, the polar perturbations couple to the terms that stem from $\mathscr{G}$ more strongly than the axial perturbations do.

\section{Concluding remarks}
\label{sec:conclusion}

In this work, we extended METRICS to the study of gravitational perturbations of rotating BHs in modified gravity, where we have focused on sGB gravity as an example. 
Our work yielded three important results. 
The first is the asymptotic behavior of gravitational perturbations at the event horizon of a general and axially symmetric BH in modified gravity. 
The second is the eigenvalue perturbation theory of METRICS, which allows us to compute the modifications to the gravitational QNM frequencies of BHs in modified gravity. 
The third is the leading-order modifications to the gravitational QNM frequency of the $nlm = 022$, 033 and 021 modes as a function of dimensionless spins in sGB gravity. 
The numerical uncertainty of the frequency of these modes is $\lesssim 10^{-5}$ for $a \leq 0.6$, $\lesssim 10^{-4} $ for $0.6< a \leq 0.7$, and $\lesssim 10^{-3} $ for $0.7< a \leq 0.85$. 

Our work demonstrates that METRICS is an effective tool to study the perturbations of BHs in modified gravity. 
Even without decoupling and simplifying the linearized field equations, METRICS enables us to accurately compute the QNM frequencies in sGB gravity.  
To the best of our knowledge, our work is the first accurate computation of gravitational QNM frequencies of a rapidly rotating BH (of $a \sim 0.85$) coupled to a scalar field in modified gravity theories.
By simply reading off the eigenvectors of the METRICS solution, we can swiftly reconstruct the metric and scalar perturbations, which could be highly involved in other formalisms \cite{Ripley:2020xby, isospectrality-paper}. 
These results offer insight into how different types of perturbations affect the QNM frequencies of BHs in modified gravity theories. 
The successful application of METRICS to sGB gravity shows that METRICS has the high potential to unlock the QNM spectra and metric perturbations of BHs in other modified gravity theories.

METRICS also fundamentally alters the nature of black hole perturbation studies. 
As mentioned before, the conventional approach when computing QNM frequencies has traditionally focused on simplifying the linearized field equations into several master equations, which often requires special analytical transformations that have to be devised on a case-by-case basis. 
METRICS provides a unified framework to compute the QNM frequencies of BHs without these transformations in different modified gravity theories (provided that these theories reduce to GR continuously when the coupling constant vanishes). 
Our results show that, as long as sufficient computational resources are available, we can always compute the QNM spectra in a modified gravity theory to high accuracy.
Although in this work, as a proof of principle, we have computed the QNM frequencies up to $a = 0.85$ in sGB gravity, METRICS can, in principle, be applied to larger $a$ BHs with similar accuracy as long as a sufficiently accurate background metric is available. 

Future work could focus on refining the METRICS approach even further. 
Thus far, METRICS has only been applied within the Regge-Wheeler gauge. 
Although the Regge-Wheeler gauge is adequate for many modified gravity theories, it will also be beneficial to extend METRICS to other gauges, because other formalisms and theories may need them. 
For example, the modified Teukolsky formalism of~\cite{Li:2022pcy, Asad_Aaron_2022, Cano:2023tmv, Cano:2023jbk} uses the in-going and outgoing radiation gauges. 
Being able to compute the QNM frequencies and reconstruct the perturbations in other gauges would facilitate cross-checks and comparisons between different formalisms. 
Also, in this work, we have developed an eigenvalue perturbation theory approach only to leading order in the coupling parameter of the modified theory. 
In principle, one can further develop this approach to arbitrary order in the coupling parameter. 
Although for tests of GR, knowing the leading-order modifications to the QNM frequencies should be sufficient, knowing the next-to-leading-order modifications can consolidate our understanding of the effects of modified gravity theories on the QNM response and also further inform us about the stability of BHs in modified gravity.  

Apart from the refinements of METRICS, this framework also opens up several other lines of research. 
First, the fitting functions for $\omega^{(1)}$ (Eq.~\eqref{eq:omega_1_fitted}) that we obtained could be used to analyze detected astrophysical ringdown signals and place the first (sGB) theory-specific constraints with ringdown data; the analyses are ongoing and results will be presented in a future paper. 
We also plan to extend the studies in this paper to other modified gravity theories, including sGB gravity with different potentials and coupling functions [e.g. $V(\Phi) = \frac{1}{2} \mu^2 \Phi^2$ or $f(\Phi) = \exp(\beta \Phi)$, for some real $\mu$ and $\beta$], and dynamical Chern-Simons gravity; this is also in progress and results will be reported elsewhere. 
Once we obtain the modifications to the QNM frequencies in these theories, we can construct fitting expressions and place these theories to the test with astrophysical ringdown data. 

Second, we can apply METRICS to model the waveforms of extreme-mass-ratio inspirals (EMRIs) around a Kerr BH.
Modeling EMRI waveforms usually involves self-force calculations, which require solving the inhomogenous linearized Einstein equations with source terms \cite{Pound:2021qin}. 
In this paper, we have demonstrated that METRICS is efficient in solving inhomogenous linearized field equations. 
Hence, METRICS could also, in principle, be useful when solving the inhomogenous linearized Einstein equations required in self-force calculations. 
Moreover, reconstructing the waveform of an EMRI through METRICS only requires reading the eigenvector at the end of the calculation, which spares us from the effort of reconstructing the metric perturbations from the Weyl scalars, the current standard approach in EMRI waveform modeling. 

Finally, the QNM frequencies presented in this paper can help establish an accurate waveform model for GWs emitted by binary BHs in sGB gravity, which can ultimately lead to even more tests of sGB gravity that include all coalescence  (inspiral, merger and ringdown) stages. 
Recent breakthroughs in numerical relativity make simulating binary black hole coalescence in sGB gravity possible \cite{AresteSalo:2022hua, AresteSalo:2023hcp, AresteSalo:2023hcp}. 
However, accurately extracting the QNM frequencies from numerical simulations in sGB gravity requires high resolutions and can be computationally challenging. 
Our QNM frequencies can spare the need for these high-resolution simulations for accurate QNM frequency extraction. 
 
\section*{Acknowledgements}

The authors acknowledge the support from the Simons Foundation through Award No. 896696, the NSF through award PHY-2207650 and NASA through Grant No. 80NSSC22K0806. 
The authors would like to thank Emanuele Berti, Alessandra Buonanno, Vitor Cardoso, Gregorio Carullo, Gregory Gabadadze, Leonardo Gualtieri, Dongjun Li, and Shinji Tsujikawa for insightful discussion about this work. 
N.Y. would like to thank Takahiro Tanaka for insightful discussion about this work. 
The calculations and results reported in this paper were produced using the computational resources of the Illinois Campus Cluster, a computing resource that is operated by the Illinois Campus Cluster Program (ICCP) in conjunction with National Center for Supercomputing Applications (NCSA), and is supported by funds from the University of Illinois at Urbana-Champaign, and used Delta at NCSA through allocation PHY240142 from the Advanced Cyberinfrastructure Coordination Ecosystem: Services $\&$ Support (ACCESS) program, which is supported by National Science Foundation grants 2138259, 2138286, 2138307, 2137603, and 2138296.
The author would like to specially thank the investors of the IlliniComputes initiatives and GravityTheory computational nodes for permitting the authors to execute runs related to this work using the relevant computational resources. 

\appendix

\section{Symbols}
\label{sec:Appendix_A}

The calculations presented in this paper involved numerous symbols. 
For the convenience of the reader, we provide a list of the symbols and their definitions in this appendix. 

\begin{itemize}
    \item $\alpha$, when not suscripted nor superscripted, is the dimensional coupling constant of modify, first defined in Eq.~\eqref{eq:Lagrangian}. 
    \item $a$ is the dimensionless spin of the BH, first defined in Eq.~\eqref{eq:metric}. 
    \item $ A_k (r) $ is the asymptotic prefactor of the $k$-th perturbation variable, first defined in Eq.~\eqref{eq:asym_prefactor}. 
    \item (A) is the superscript or supscript which denotes the quantity concerning the axial perturbations, first defined in Table.~\ref{tab:omega_1_022}. 
    \item $b = \sqrt{1-a^2}$, first defined in Eq.~\eqref{eq:metric_quantities}. 
    \item $\mathcal{B}(N)$ is the backward modulus difference of the QNM frequency, first defined in Eq.~\eqref{eq:BWD}. 
    \item $d_{r}$ is the degree of $r$ of the coefficient of the partial derivative of the linearized field equations, first defined in Eq.~\eqref{eq:pertFE-1}. 
    \item $d_{\chi}$ is the degree of $\chi$ of the coefficient of the partial derivative of the linearized field equations, first defined in Eq.~\eqref{eq:pertFE-1}. 
    \item $d_{z}$ is the degree of $z$ of the coefficient of the partial derivative of the compactified linearized field equations, first defined in Eq.~\eqref{eq:system_3}. 
    \item $\mathbb{D}(\omega) $ is the coefficient matrix of spectral expansion, from one particular basis to another, first defined in Eq.~\eqref{eq:pertFE-2}. 
    \item $\delta $ is the numerical uncertainty of the METRICS frequencies, first defined in Eq.~\eqref{eq:numerical_uncertainty}. 
    \item $\Delta = (r-r_+)(r-r_-)$, first defined in Eq.~\eqref{eq:metric_quantities}. 
    \item $\mathcal{G}_{k, \eta, \gamma, \delta, \sigma, \alpha, \beta, j} $ is the coefficient of $\omega^\gamma r^{\delta} \chi^{\sigma} \partial_{r}^{\alpha} \partial_{\chi}^{\beta} h_j$ of the linearized field equations of $h_j$, first defined in Eqs.~\eqref{eq:pertFE-1}. 
    \item $h_k(r, \chi)$ is the functions of metric perturbations, first defined in Eqs.~\eqref{eq:metpert} and \eqref{eq:scalar_pert}. 
    \item $H_k(r, \chi)$ is corrections to the Kerr metric due to modified gravity, first defined in Eqs.~\eqref{eq:metric}. 
    \item $i = \sqrt{-1}$ is the imaginary unit. 
    \item $k$ in the subscript is the component of the metric perturbation functions and $k= 1, ..., 7$, first defined in Eqs.~\eqref{eq:metpert} and \eqref{eq:scalar_pert}.
    \item $\kappa$ is the surface gravity on BH horizon, first defined in Eqs.~\eqref{eq:Omega_H_and_kappa}.
    \item $\mathcal{K}_{k, \alpha, \beta, \gamma, \delta, \sigma, j} $ is the coefficient of $\omega^\gamma z^{\delta} \chi^{\sigma} \partial_{z}^{\alpha} \partial_{\chi}^{\beta}(...)$ of the linearized field equations in $z$ and $\chi$, first defined in Eq.~\eqref{eq:system_3}. 
    \item $l$ is the azimuthal mode number of the gravitational QNMs, first defined in Sec.~\ref{sec:intro}. 
    \item $\ell$ is the degree of associate Legendre polynomial used in spectral expansion, first defined in Eq.~\eqref{eq:spectral_decoposition_correction_factor}.
    \item $M$ is the BH mass, which is taken to be $M=1$ throughout this work, first defined in Eq.~\eqref{eq:metric}. 
    \item $m$ is the azimuthal number of the metric perturbations, first defined in Eqs.~\eqref{eq:metpert} and \eqref{eq:scalar_pert}. 
    \item $N$ is the number of the Chebyshev and associated Legendre polynomials used in the full spectral expansion, first defined below  Eq.~\eqref{eq:spectral_decoposition_factorized_finite}. 
    \item $\mathcal{N}_{\chi}$ is the number of the associated Legendre polynomials included in the spectral expansion, first defined in Eq.~\eqref{eq:spectral_decoposition_factorized_finite}. 
    \item $\mathcal{N}_{z}$ is the number of the Chebyshev polynomials included in the spectral expansion, first defined in Eq.~\eqref{eq:spectral_decoposition_factorized_finite}. 
    \item (P) is the superscript which denotes the quantity concerning the parity-led perturbations, first defined in , first defined in Table.~\ref{tab:omega_1_022}.
    \item PD is the parity dominance, which characterizes the parity content of metric perturbations, first defined in Eq.~\eqref{eq:PD_01}.
    \item $r_\pm = M(1\pm\sqrt{1-a^2})$ is the radial coordinate of the position of the event horizon of the Kerr BH, first defined below Eq.~\eqref{eq:metric_quantities}. 
    \item $r_*$ is the tortoise coordinate, first defined in Eq.~\eqref{eq:EF_coord_inf}. 
    \item $\rho_{\infty}^{(k)}$ and $\rho_H^{(k)}$ are the parameters that characterize the boundary conditions of $h_k$ in future null infinity and at the horizon, first defined in Eq.~\eqref{eq:asymptotic_limits1} and \eqref{eq:asymptotic_limits3}.
    \item $\Omega_{\rm H}$ is the angular velocity of BH horizon, first defined in Eq.~\eqref{eq:Omega_H_and_kappa}. 
    \item $\vartheta$ is the scaled scalar field in modified gravity theories, first defined above Eq.~\eqref{eq:Tmunu}. 
    \item $z = \frac{2 r_+}{r} - 1$ is the variable that maps $r$ into a finite domain, first defined in Eq.~\eqref{eq:z}. 
    \item $\zeta = \alpha^2 / M^4 $ is the dimensionless coupling parameter of modified gravity, first defined in Eq.~\eqref{eq:zeta}. 
\end{itemize}

\section{Additional Tables}
\label{sec:Appendix_B}

The following tables present additional details related to the results described in the main body of this paper. In particular, Tables~\ref{tab:omega_1_022}, \ref{tab:omega_1_033}, and \ref{tab:omega_1_021} present the sGB corrections to the quasinormal frequencies for the $022$, $033$ and $021$ modes respectively, for various choices of BH spin. Meanwhile, Table~\ref{tab:omega_1_022_scalar} presents the sGB quasinormal frequencies of the scalar mode. Tables~\ref{tab:poly_fit_coeffs} and~\ref{tab:poly_fit_coeffs_uncertainty} show the coefficients of the fitting polynomials, as well as their uncertainties, respectively.   

\begin{table*}[htb]
\resizebox{\textwidth}{!}{
\begin{tabular}{c|c|c|c|c|c|c|c}
\hline
$a$ & $N_{\rm a}$ & $N^{\rm (A)}_{\rm opt}$  & $\omega^{(1)}_{\rm (A)}$ & $\delta^{\rm (A)}$ & $N^{\rm (P)}_{\rm opt}$  & $\omega^{(1)}_{\rm (P)}$ & $\delta^{\rm (P)}$ \\ \hline
0.005 & 4 & 25 & $0.05983677380180108 + 0.007192825134851422i$ & $(0.251+1.61i) \times 10^{-8}$ & 21 & $-0.22663749127162247 -0.07524763567237269i$ & $(1.02+0.0390i) \times 10^{-5}$ \\
0.1 & 4 & 24 & $0.07281552745099619 + 0.013531763175821453i$ & $(1.36+0.596i) \times 10^{-6}$ & 20 & $-0.26227702246215134 - 0.07145798641074617i$ & $(1.11+0.984i) \times 10^{-5}$ \\
0.2 & 8 & 21 & $0.0907539924479579 + 0.015647215719244878i$ & $(0.0889+1.45i) \times 10^{-7}$ & 19 & $-0.3066709116903176 -0.05830502121797751i$ & $ (1.36+0.451i) \times 10^{-5}$ \\
0.3 & 10 & 22 & $0.11124804644377795 + 0.010330072870775808i$ & $(1.46+1.48i) \times 10^{-7}$ & 20 & $-0.3568577921282271 - 0.03334328807781678i$ & $(1.10+0.317i) \times 10^{-5}$ \\
0.4 & 14 & 21 & $0.13073382288624735 -0.0050047880088150976i$ & $(2.66 + 0.855i) \times 10^{-7}$ & 22 & $-0.4111938224678359 + 0.004668633581125168i$ & $(3.30+0.800i) \times 10^{-6}$ \\
0.5 & 16 & 19 & $0.14312291185082282 - 0.03237775518326025i$ & $(2.45+5.36i) \times 10^{-6}$ & 18 & $-0.4689206189969932 +0.05438911010968178i$ & $ (0.344+1.76i) \times 10^{-5}$ \\
0.6 & 22 & 14 & $0.13807501299240688  - 0.07248938044869968i$ & $(0.214+1.76i) \times 10^{-5}$ & 18 & $-0.5331370100327957 + 0.1099710619280786i$ & $(5.38+2.85i) \times 10^{-6}$ \\
0.7 & 30 & 15 & $0.09354320193472176-0.12455863075416929i$ & $ (1.05+0.0678i) \times 10^{-4}$ & 23 & $-0.6193307127940813 + 0.1550769221600632i$ & $ (0.433+3.86i) \times 10^{-4}$ \\
0.8 & 40 & 15 & $-0.05959665172558104 - 0.17878061231480724i$ & $ (9.07+4.91i) \times 10^{-4}$ & 17 & $-0.7918609719800095 + 0.15494665090331627i$ & $ (2.21+6.25i) \times 10^{-4}$ \\ 
0.849 & 40 & 15 & $-0.3236241048489177 - 0.19358591148846216i$ & $ (1.43+1.15i) \times 10^{-2}$ & 15 & $-0.7479793034686937 + 0.17364176377429175i$ & $ (1.36+1.23i) \times 10^{-2}$ \\ \hline
\end{tabular}
}
\caption{
\label{tab:omega_1_022}
$\omega^{(1)}$ of the $nlm=022$-mode gravitational perturbations of rotating BHs in scalar Gauss Bonnet (sGB) gravity at different dimensionless spins $a$ (first column). 
The superscripts (A) and (P) respectively stand for axial and polar perturbations. 
$N_{\rm opt}$ is the optimal spectral order, the spectral order which minimizes the backward modulus difference of $\omega^{(1)}$. 
$\delta$ is the numerical uncertainty of the frequency calculations, which is the respective minimal backward modulus difference of the real and imaginary parts of the frequency. 
The numerical value of the real and imaginary parts of $\omega^{(1)}$ is rounded to the nearest decade which is larger than the numerical uncertainty. 
}
\end{table*}

\begin{table*}[htb]
\resizebox{\textwidth}{!}{
\begin{tabular}{c|c|c|c|c|c|c|c}
\hline
$a$ & $N_{\rm a}$ & $N^{\rm (A)}_{\rm opt}$  & $\omega^{(1)}_{\rm (A)}$ & $\delta^{\rm (A)}$ & $N^{\rm (P)}_{\rm opt}$  & $\omega^{(1)}_{\rm (P)}$ & $\delta^{\rm (P)}$ \\ \hline
0.005 & 4 & 20 & $0.11286078919794015 + 0.007173507267031212i$ & $ (1.04+6.32i) \times 10^{-9}$ & 17 & $-0.8787330636118746-0.11499069888716668i$ & $ (2.18+1.12i) \times 10^{-6}$ \\
0.1 & 6 & 19 & $0.13066629176256211+0.011038421005643784i$ & $(1.06+1.32i) \times 10^{-8}$ & 16 & $-0.9875555961636844-0.10558386081665214i$ & $ (2.58+0.928i) \times 10^{-6}$ \\
0.2 & 8 & 17 & $0.1518073459136097+0.0115278223210461i$ & $ (1.32+1.57i) \times 10^{-7}$ & 16 & $-1.1245702806895668-0.08525138927080578i$ & $(5.15+5.32) \times 10^{-6}$ \\
0.3 & 10 & 23 & $0.17289901776399458+0.006099519382483931i$ & $ (4.61+6.45i) \times 10^{-7}$ & 18 & $-1.2887523667768863 - 0.050250615075774774i$ & $ (1.29+0.152i) \times 10^{-5}$ \\
0.4 & 14 & 19 & $0.18861427338653164-0.00755337816196977i$ & $ (7.41+7.32) \times 10^{-7}$ & 17 & $-1.485183229054195+0.003593110536591726i$ & $ (0.464+1.01i) \times 10^{-5}$ \\
0.5 & 16 & 13 & $0.18855622885326895 - 0.0310881937864389i$ & $ (1.11+8.61i) \times 10^{-6}$ & 17 & $-1.720738577576597 + 0.07952486536808223i$ & $ (8.72+0.892i) \times 10^{-6}$ \\
0.6 & 22 & 15 & $0.15216745480132232- 0.064254054959898i$ & $ (0.185+3.04i)\times 10^{-5}$ & 15 & $-2.00700863307933 + 0.1770630228719039i$ & $ (1.48+0.620i)\times 10^{-5}$ \\
0.7 & 30 & 15 & $0.033764760999829946-0.10255535266797444i$ & $(6.54+7.52i) \times 10^{-5}$ & 17 & $-2.367832028905901 + 0.2846211474620919i$ & $(8.349+4.37i) \times 10^{-5}$ \\ 
0.8 & 40 & 15 & $-0.2989143336916561 - 0.13192572629913388i$ & $ (9.42+3.48i) \times 10^{-4}$ & 14 & $-2.876137347093313 + 0.3576923047281628i$ & $ (6.77+6.78i) \times 10^{-4}$ \\ 
0.849 & 40 & 16 & $-0.706397474437324 - 0.16229155895183256i$ & $ (7.29+6.06i) \times 10^{-3}$ &  14 & $-3.1991817526896615 + 0.39805904301121964i$ & $ (0.595+1.84i) \times 10^{-2}$ \\ \hline
\end{tabular}
}
\caption{
\label{tab:omega_1_033}
Identical to Table ~\ref{tab:omega_1_022}, except that $nlm = 033$. 
}
\end{table*}

\begin{table*}[htb]
\resizebox{\textwidth}{!}{
\begin{tabular}{c|c|c|c|c|c|c|c}
\hline
$a$ & $N_{\rm a}$ & $N^{\rm (A)}_{\rm opt}$  & $\omega^{(1)}_{\rm (A)}$ & $\delta^{\rm (A)}$ & $N^{\rm (P)}_{\rm opt}$  & $\omega^{(1)}_{\rm (P)}$ & $\delta^{\rm (P)}$ \\ \hline
0.005 & 4 & 24 & $0.05948521200430079+0.007009731665666319i$ & $ (5.26+3.69i) \times 10^{-5}$ & 14 & $-0.22579391150054562-0.07525874375642161i$ & $ (2.96+0.228i) \times 10^{-5}$ \\
0.1 & 6 & 20 & $0.06674484236317868 + 0.010532065820924585i$ & $ (1.11+0.0622i) \times 10^{-5}$ & 18 & $-0.24134198879459592 - 0.07336129762739471i$ & $ (2.73+4.63i) \times 10^{-5}$ \\
0.2 & 8 & 21 & $0.0788014109220061+0.013126720850122453i$ & $ (2.53+1.90i) \times 10^{-7}$ & 20 & $-0.25715508043788304-0.06761134054016793i$ & $(1.39+0.303i) \times 10^{-6}$ \\
0.3 & 10 & 18 & $0.09681103316083872 + 0.01364389449901715i$ & $ (2.71+0.516i) \times 10^{-6}$ & 24 & $-0.2714569186336462-0.05712134162831006i$ & $ (4.37+6.09i) \times 10^{-6}$ \\
0.4 & 14 & 25 & $0.12257424673702388+0.010711734937522976i$ & $ (7.44+3.30i) \times 10^{-6}$ & 19 & $ -0.2839011173927588-0.04161437565704773i$ & $ (4.79+3.91i) \times 10^{-6}$ \\
0.5 & 16 & 24 & $0.15887915267623143+0.0022214694485569453i$ & $ (2.20+1.63i) \times 10^{-5}$ & 22 & $-0.29252740652330544-0.020361139691473795i$ & $ (1.14+0.603i) \times 10^{-5}$ \\
0.6 & 22 & 17 & $0.20824890018928244-0.01603064248271835i$ & $ (2.02+2.36i) \times 10^{-5}$ & 17 & $-0.2983421497190903+0.005031018346571159i$ & $ (5.01+5.21i)\times 10^{-5}$ \\
0.7 & 30 & 22 & $0.27867234154211573 - 0.052030059461859325i$ & $(6.76+1.89i) \times 10^{-4} $ & 15 & $-0.30057599044531 + 0.03140654347441085i$& $ (0.275+1.22i) \times 10^{-4} $ \\ 
0.8 & 40 & 11 & $0.36776266018908643 - 0.10807501119184693i$ & $ (0.106+1.83i) \times 10^{-3}$ & 14 & $-0.306345246631091 + 0.036180734446990215i$ & $ (0.573+1.32i) \times 10^{-3}$ \\ 
0.849 & 40 & 10 & $0.6352503982110855 - 0.20749661006414685i$ & $ (6.10+3.45i) \times 10^{-2}$ & 11 & $-0.35540935973252025 + 0.002487636563500928i$ & $ (0.421+1.51i)  \times 10^{-2}$ \\ \hline
\end{tabular}
}
\caption{
\label{tab:omega_1_021}
Identical to Table ~\ref{tab:omega_1_022}, except that $nlm = 021$. 
}
\end{table*}

\begin{table*}[htb]
\begin{tabular}{c|c|c|c|c}
\hline
$a$ & $N_{\rm a}$ & $N^{\rm (S)}_{\rm opt}$  & $\omega^{(1)}_{\rm (S)}$ & $\delta^{\rm (S)}$ \\ \hline
0.005 & 4 & 19 & $0.6045798545323886 + 0.0581423873432243i$ & $(1.26-0.614i) \times 10^{-5}$  \\
0.1 & 4 & 17 & $0.6520081777182252 + 0.049164102037905326i$ & $(0.269+2.80i) \times 10^{-5}$ \\
0.2 & 8 & 18 & $0.7046912702423356 + 0.03603026646139007i$ & $(0.908+3.96i) \times 10^{-6}$ \\
0.3 & 10 & 17 & $0.7589595233768467 + 0.01842926109601173i$ & $(0.346+7.60i) \times 10^{-6}$ \\
0.4 & 14 & 16 & $0.8120444787616149 - 0.00414803677523237i$ & $(1.93+0.173i) \times 10^{-5}$ \\
0.5 & 16 & 17 & $0.8582013985710617 - 0.03173757333700589i$ & $(4.60+1.28i) \times 10^{-6}$ \\
0.6 & 22 & 19 & $0.8852904646574657 - 0.06275621308225965i$ & $(5.83+1.08i) \times 10^{-6}$ \\
0.7 & 30 & 16 & $0.8657377111656217 - 0.09150362347338614i$ & $ (0.747+1.06i) \times 10^{-5}$ \\
0.8 & 40 & 20 & $0.7292910025639827 - 0.09881446399022309i$ & $ (1.88+0.497i) \times 10^{-5}$ \\ 
0.849 & 40 & 20 & $0.566457285841918 - 0.07810396025985997i$ & $ (0.518+3.23i) \times 10^{-5}$ \\ \hline
\end{tabular}
\caption{
\label{tab:omega_1_022_scalar} 
Identical to Table ~\ref{tab:omega_1_022}, except that the mode is the scalar mode. 
}
\end{table*}

\begin{table*}[]
\resizebox{\textwidth}{!}{
\begin{tabular}{c|cc|cc|cc|}
\hline
\multirow{2}{*}{$w_j$} & \multicolumn{2}{c|}{022}                                                     & \multicolumn{2}{c|}{033}                                                 & \multicolumn{2}{c|}{021}                                                   \\ \cline{2-7} 
                       & \multicolumn{1}{c|}{(A)}                         & (P)                       & \multicolumn{1}{c|}{(A)}                      & (P)                      & \multicolumn{1}{c|}{(A)}                      & (P)                        \\ \hline
$w_0$                  & \multicolumn{1}{c|}{$ 0.055241 + 0.00686713 i $} & $ -0.215202-0.0734094i $  & \multicolumn{1}{c|}{$ 0.109706+0.0061152 i $} & $-0.872789-0.113506i $  & \multicolumn{1}{c|}{$0.0590468+0.00690883i$}   & $ -0.22612 -0.0754879 i $  \\
$w_1$                  & \multicolumn{1}{c|}{$ 0.985294+0.0636233i $}     & $ -2.51816-0.411031i $ & \multicolumn{1}{c|}{$ 0.685561+0.231546i$} & $-1.20198-0.339974 i $  & \multicolumn{1}{c|}{$0.0899258+0.0179754i$}   & $ 0.0933968 + 0.0504285i $ \\
$w_2$                  & \multicolumn{1}{c|}{$ -18.4902 + 0.325569i $}     & $ 48.4659+9.59898i $  & \multicolumn{1}{c|}{$ -11.6733-4.20702i $}  & $2.82484 +9.09087 i $ & \multicolumn{1}{c|}{$- 0.470918+0.460483i$}    & $ - 5.91521-0.968127i $    \\
$w_3$                  & \multicolumn{1}{c|}{$ 157.802 -4.46276i $}     & $ -431.428-82.4765i $   & \multicolumn{1}{c|}{$ 108.698+37.6761i $}   & $-35.6024-75.7708i $   & \multicolumn{1}{c|}{$4.30802- 3.94314i $}    & $ 54.1801+10.6838i $       \\
$w_4$                  & \multicolumn{1}{c|}{$-682.224 +19.8274i $}       & $ 1935.01+378.832i $      & \multicolumn{1}{c|}{$-510.61-178.723i $}    & $165.435 + 341.913 i $   & \multicolumn{1}{c|}{$ - 8.45233+ 13.3649i $} & $ - 252.83-51.1785i $       \\
$w_5$                  & \multicolumn{1}{c|}{$ 1660.08  -55.4163i $}      & $ -4831.15-965.058i $     & \multicolumn{1}{c|}{$1334.38 +467.415i $}   & $-434.981-857.681 i$     & \multicolumn{1}{c|}{$- 8.21766 - 16.1916i$}    & $  659.362+139.483i $       \\
$w_6$                  & \multicolumn{1}{c|}{$ -2307.78 + 87.744i $}      & $ 6808.99 +1386.33i $      & \multicolumn{1}{c|}{$-1970.88-687.899i $}     & $638.737+1219.16i$     & \multicolumn{1}{c|}{$29.5539 - 13.5289i $}   & $ - 969-215.727i $      \\
$w_7$                  & \multicolumn{1}{c|}{$ 1711.32 -73.0696 i $}      & $ -5065.1-1050.91i $     & \multicolumn{1}{c|}{$1536.55+532.083i $}   & $-488.58-915.499i $    & \multicolumn{1}{c|}{$53.3926+26.2421i $}    & $ 749.935 + 176.707i $     \\ 
$w_8$                  & \multicolumn{1}{c|}{$-526.516 + 25.0324  i $}      & $ 1544.32+325.905i $     & \multicolumn{1}{c|}{$-493.701-167.701i $}   & $150.043+280.837i $    & \multicolumn{1}{c|}{$- 63.872+ 36.1966i $}    & $ - 237.614 - 59.776i $     \\
$w_9$                  & \multicolumn{1}{c|}{$-$}      & $-$     & \multicolumn{1}{c|}{$-$}   & $- $    & \multicolumn{1}{c|}{$ - 262.086 - 15.5455i $}    & $ -$     \\ 
$w_{10}$                  & \multicolumn{1}{c|}{$-$}      & $-$     & \multicolumn{1}{c|}{$-$}   & $- $    & \multicolumn{1}{c|}{$ 21.8749 - 77.7481$}    & $-$     \\ 
$w_{11}$                  & \multicolumn{1}{c|}{$-$}      & $-$     & \multicolumn{1}{c|}{$-$}   & $- $    & \multicolumn{1}{c|}{$799.931 - 56.6442i $}    & $-$     \\ 
$w_{12}$                  & \multicolumn{1}{c|}{$-$}      & $-$     & \multicolumn{1}{c|}{$-$}   & $- $    & \multicolumn{1}{c|}{$225.947 + 84.4078i $}    & $- $     \\ 
$w_{13}$                  & \multicolumn{1}{c|}{$-$}      & $-$     & \multicolumn{1}{c|}{$-$}   & $- $    & \multicolumn{1}{c|}{$- 1942.69 +198.31i $}    & $- $     \\ 
$w_{14}$                  & \multicolumn{1}{c|}{$-$}      & $-$     & \multicolumn{1}{c|}{$-$}   & $- $    & \multicolumn{1}{c|}{$ 1174.3- 182.029i$}    & $ - $     \\ \hline
\end{tabular}
}
\caption{
\label{tab:poly_fit_coeffs}
The coefficients $w_j$ of the fitting polynomial (c.f. Eq.~\eqref{eq:omega_1_fitted}) to the axial and polar frequencies of the $nlm=022, 033$ and 021.}
\end{table*}

\begin{table*}[]
\resizebox{\textwidth}{!}{
\begin{tabular}{c|cc|cc|cc|}
\hline
\multirow{2}{*}{$w_j$} & \multicolumn{2}{c|}{022}                                                     & \multicolumn{2}{c|}{033}                                                 & \multicolumn{2}{c|}{021}                                                   \\ \cline{2-7} 
                       & \multicolumn{1}{c|}{(A)}                         & (P)                       & \multicolumn{1}{c|}{(A)}                      & (P)                      & \multicolumn{1}{c|}{(A)}                      & (P)                        \\ \hline
$w_0$                  & \multicolumn{1}{c|}{$ 9.53 \times 10^{-6} + 0.00686713 i $} & $ 0.00457066+0.00233301i $  & \multicolumn{1}{c|}{$ 0.0017747+0.000565499i $} & $0.000425088+0.00185872i $  $- $ & \multicolumn{1}{c|}{$-$} & {$0.0075939+0.00411396i$}    \\
$w_1$                  & \multicolumn{1}{c|}{$0.891167 + 0.0013926i $}     & $ 0.892279+0.271754i $ & \multicolumn{1}{c|}{$ 0.202205+0.078463i$} & $0.0944822+0.350078i $  & \multicolumn{1}{c|}{ $- $} & {$0.557172+0.362529i$} \\
$w_2$                  & \multicolumn{1}{c|}{$18.1278 + 0.029502i $}     & $ 19.3688+5.6713i $  & \multicolumn{1}{c|}{$ 4.25712+1.67866i $}  & $1.97781+ 7.1689i $ & \multicolumn{1}{c|}{$- $} & {$11.9661+6.57228i$}    \\
$w_3$                  & \multicolumn{1}{c|}{$146.535+0.250497i $}     & $ 163.664+47.2939i $   & \multicolumn{1}{c|}{$ 36.4784+14.2138i $}   & $16.1772+57.2912i $   & \multicolumn{1}{c|}{$- $} &{$106.628+50.186i $}       \\
$w_4$                  & \multicolumn{1}{c|}{$609.58+1.09047i $}       & $ 692.595+202.145i $      & \multicolumn{1}{c|}{$160.602+60.7758i $}    & $68.233+234.331i $   & \multicolumn{1}{c|}{$- $} &{$485.328+199.799i $}       \\
$w_5$                  & \multicolumn{1}{c|}{$1429.63+2.65319i $}      & $ 1614.95+484.046i $     & \multicolumn{1}{c|}{$394.953 + 143.826i $}   & $162.737+538.798i$     & \multicolumn{1}{c|}{$- $} & {$1221.93 + 448.967i$}       \\
$w_6$                  & \multicolumn{1}{c|}{$1910.19+3.64319i $}      & $ 2108.76 + 656.174i $      & \multicolumn{1}{c|}{$547.518+191.188i $}     & $221.279+704.804i$     & \multicolumn{1}{c|}{$- $} & {$1718.49 + 574.311i $}    \\
$w_7$                  & \multicolumn{1}{c|}{$1356.36+2.63502i$}      & $ 1445.54+470.048i $     & \multicolumn{1}{c|}{$399.431+133.72i $}   & $159.596+489.772i $    & \multicolumn{1}{c|}{$- $} &  {$1262.3+390.343i $}      \\ 
$w_8$                  & \multicolumn{1}{c|}{$396.962+0.779457i$}      & $ 405.09+138.139i $     & \multicolumn{1}{c|}{$119.119+38.2928i $}   & $47.2759+140.335i $    & \multicolumn{1}{c|}{$- $} &  {$376.818 +109.465i $}    \\ \hline
\end{tabular}
}
\caption{
\label{tab:poly_fit_coeffs_uncertainty}
The uncertainty of the real and imaginary parts of $w_j$ of the fitting polynomial (c.f. Eq.~\eqref{eq:omega_1_fitted}) to the axial and polar frequencies of the $nlm=022, 033$ and 021.
Note that the uncertainty of the fitting polynomial of the 021 axial mode is undefined because the frequency is actually overfitted to avoid unphysical oscillations of the frequency for small $a$ (see discussion around Eq.~\eqref{eq:omega_1_fitted}). }
\end{table*}

\bibliography{ref}

\end{document}